\documentclass[epj,final]{svjour}

\usepackage[dvips]{graphicx}
\usepackage{pstricks}
\usepackage[english]{babel}
\usepackage{longtable}
\usepackage{amssymb,latexsym,amsmath}
\usepackage{cite}
\usepackage{dsfont}
\usepackage{psfrag}
\usepackage{bm}
\usepackage{multirow}
\usepackage{booktabs}
\usepackage{units}

\def\Id{{\rm 1\kern-.3em I}}

\begin{document}

\definecolor{darkblue}{rgb}{0,0,.5}
\definecolor{darkgreen}{rgb}{0,.5,0}
\definecolor{darkred}{rgb}{.5,0,0}
\newcommand{\corr}[1]{{\color{black}{#1}}}

\title{
  Effects of a spin-flavour dependent interaction on 
  light-flavoured baryon helicity amplitudes
}
\author{
  Michael Ronniger\thanks{e-mail: \texttt{ronniger@hiskp.uni-bonn.de}} 
  \and 
  Bernard Ch. Metsch
}

\titlerunning{
  Effects of a spin-flavour dependent interaction on helicity amplitudes
}
\authorrunning{
  M. Ronniger \emph{et al.}
}
\institute{
  Helmholtz Institute f\"ur Strahlen-- und Kernphysik (Theorie), 
  Universit\"at Bonn, 
  Nu\ss allee 14-16, 
  D-53115 Bonn,
  Germany
}
\date{\today}

%
\abstract{
This paper is a continuation of previous work about the effects of a
phenomenological flavour dependent force in a relativistically covariant
constituent quark model based on the Salpeter equation on the structure of
light-flavoured baryon resonances. Here the longitudinal and transverse
helicity amplitudes as studied experimentally in the electro-excitation of
nucleon- and $\Delta$ resonances are calculated.  In particular the amplitudes
for the excitation of three and four star resonances as calculated in a
previous model $\mathcal A$ are compared to those of the novel model $\mathcal
C$ as well as to existing and partially new experimental data such as
\textit{e.g.} determined by the CB-ELSA collaboration. A brief discussion of
some improvements to model $\mathcal C$ is given after the introduction.
\PACS{
      {11.10.St}{Bound and unstable states; Bethe-Salpeter equations}  \and
      {12.39.Ki}{Relativistic quark model} \and
      {13.40.Gp}{Electromagnetic form factors} \and
      {13.40.Hq}{Electromagnetic decays}
     } 
} 
\maketitle

\renewcommand{\figurename}{Fig.} 


\section{Introduction\label{intro}}

This paper is a continuation of our previous work~\cite{Ronniger} on the
description of baryon resonances in a covariant Bethe-Salpeter framework.
While in~\cite{Ronniger} we concentrated on the description of the mass
spectrum in the present paper we discuss the results on electromagnetic
transition amplitudes obtained on the basis of the Salpeter amplitudes
determined previously. The dynamical ingredients of the relativistically
covariant quark model used are instantaneous interaction kernels describing
confinement, a spin-flavour dependent interactions kernel motivated by
instanton effects as well as in addition a phenomenologically introduced
spin-flavour dependent interaction. The latter was found to improve the description
obtained previously in~\cite{LoeMePe1,LoeMePe2,LoeMePe3} in particular of
Roper-like scalar excitations as well as the position of some negative parity
$\Delta$-resonances slightly below 2 GeV. With instantaneous interaction
kernels the Bethe-Salpeter equation reduces to the more tractable Salpeter
equation which can be cast in the form of an eigenvalue equation for the
masses and the Salpeter amplitudes. These in turn determine the vertex
functions for any on-shell momentum of the baryons which then enter the
electromagnetic current matrix elements. The details of this procedure can be
found in~\cite{Merten}\,.

Spin-flavour dependent effective quark-quark interactions have also been studied
previously by the Graz
group~\cite{%
Glozman1996,Glozman1997,Glozman1998_1,%
Glozman1998_2,Theussl,Glantschnig,Melde07,Melde08,%
Plessas}
which obtained very satisfactory results for the mass spectra up to 1.7 GeV as
well as for the corresponding nucleon form factors on the basis of a truncated
pseudo-vector coupled Yukawa potential, where the tensor force terms were
neglected. In~\cite{Ronniger} we found a phenomenological \textit{Ansatz}
which includes a short ranged flavour-singlet and a flavour-octet exchange
with pseudoscalar-like coupling of Gaussian form to be most effective for the
improvements mentioned above. This newly introduced interaction kernel
increased the number of parameters of seven in the former model (Model
$\mathcal{A}$, see~\cite{LoeMePe2,LoeMePe3}) to ten in the new version (Model
$\mathcal{C}$, see~\cite{Ronniger}), which we still consider to be acceptable
in view of the multitude of baryon masses described accurately in this
manner. For the values of the parameters we refer to table~\ref{tab:par},
where we listed an improved set of parameters for the interaction kernels of
model $\mathcal C$ together with the values used in~\cite{Ronniger} displayed
in brackets. Likewise the parameters for the interaction kernels of model
$\mathcal{A}$ obtained from calculations within a larger model space, see
also~\cite{Ronniger}\, are listed along with the values (in brackets) as
determined in \cite{LoeMePe2,LoeMePe3} in smaller model spaces.
\begin{table}[!htb]
\centering
\caption{
  Model parameter values for the novel model $\mathcal{C}$ in comparison to
  those of model $\mathcal{A}$~\cite{LoeMePe2,LoeMePe3}. The bracketed numbers
  in the column for model $\mathcal{A}$ are the parameters as found
  in~\cite{LoeMePe2,LoeMePe3} and the numbers above them are recalculated with
  higher numerical accuracy as commented in~\cite{Ronniger}.
  The numbers in the column for model $\mathcal{C}$ represent an improved set
  with respect to the values quoted in~\cite{Ronniger} which are listed in
  brackets. Note that the Dirac structure for the confinement interaction
  kernel is different in models $\mathcal{A}$ and $\mathcal{C}$\,, see
  also~\cite{Ronniger}\,. 
  \label{tab:par}
}
\begin{tabular}{l@{\hspace*{5pt}}l@{\hspace*{5pt}}r@{\hspace*{5pt}}r}
\toprule
parameter                    &                                        & model $\mathcal{C}$      & model $\mathcal{A}$ \\
\midrule
masses                       & $m_n$ [MeV]                            &   350.0                  & \multirow{2}{*}{330.0} \\
                             &                                        &  [325.0]                 &                       \\
                             & $m_s$ [MeV]                            &   625.0                  & \multirow{2}{*}{670.0} \\
                             &                                        &  [600.0]                 &                       \\
\midrule
\multirow{2}{*}{confinement} & \multirow{2}{*}{$a$ [MeV]}             & -370.8                   &  -734.6 \\
                             &                                        &[-366.8]                 & [-700.0]\\
                             & \multirow{2}{*}{$b$ [MeV/fm]}          &  208.4                   &   453.6 \\
                             &                                        & [212.86]                 &  [440.0]\\
\midrule
\multirow{2}{*}{instanton}   & \multirow{2}{*}{$g_{nn}$ [MeV fm$^3$]} &  317.9                   &   130.3 \\
                             &                                        & [341.5]                 &  [136.0]\\
\multirow{2}{*}{induced}     & \multirow{2}{*}{$g_{ns}$ [MeV fm$^3$]} &  260.0                   &    81.8 \\
                             &                                        & [273.6]                 &   [96.0]\\
interaction                  & $\lambda$ [fm]                         &  0.4                     & 0.4 \\
\midrule
octet                        & \multirow{2}{*}{$\frac{g_{8}^2}{4\pi}$  [MeV fm$^3$]}   &  118.0                   & \multirow{2}{*}{--} \\
exchange                     &                                        & [100.86]                 & \\
singlet                      & \multirow{2}{*}{$\frac{g_{0}^2}{4\pi}$ [MeV fm$^3$]}    & 1715.5                   & \multirow{2}{*}{--} \\
exchange                     &                                        &[1897.4]                 & \\
                             & $\lambda_{8}=\lambda_0$ [fm]           &    0.25                  & -- \\
\bottomrule
\end{tabular}
\end{table}
Note that the two models $\mathcal{A}$ and $\mathcal{C}$, specified in
table~\ref{tab:par} employ different confinement Dirac-structures for the
constant part (offset) $\Gamma_0$ and the linear part (slope) $\Gamma_s$
(see~\cite{Ronniger,LoeMePe2} for more information). All calculations in the
present paper are based on the parameter values of table~\ref{tab:par}.

Furthermore we want to point out, that the calculation of helicity amplitudes
or transition form factors (such as that for the nucleon-$\Delta(1232)$
magnetic transition) in lowest order as considered here does not introduce any
additional parameter in the underlying models as discussed in~\cite{Ronniger}
before. Since in the new model $\mathcal{C}$ we can account for more baryon
excitations accurately we can now also offer predictions for some
$\Delta$-resonances which could not be covered before in model $\mathcal{A}$.
In particular these are: $\Delta_{1/2^+}(1750)$, $\Delta_{3/2^+}(1600)$,
$\Delta_{1/2^-}(1900)$, $\Delta_{3/2^-}(1940)$ and $\Delta_{5/2^-}(1930)$ as
reported in~\cite{Ronniger}. Additionally there now exists new data for photon
decay amplitudes from Anisovich \textit{et
al.}~\cite{Anisovich_1,Anisovich_2,Anisovich_3} and for helicity amplitudes
from Aznauryan \textit{et
al.}~\cite{Aznauryan05_1,Aznauryan05_2,Aznauryan2009,Aznauryan2012} as well as
the MAID analysis~\cite{Drechsel,Tiator}, in particular the analysis and
parametrisations in the recent overview~\cite{Tiator2011}, with information
also on longitudinal amplitudes which can serve as a test of the present model beyond the
comparison done previously in~\cite{Merten,Kreuzer,PhDMerten} on the basis of
the amplitudes determined in model $\mathcal{A}$ of~\cite{LoeMePe2}. For the
definition of the helicity amplitudes we use the conventions as in 
Tiator \textit{et al.}~\cite{Tiator2011} as mentioned in Eqs.~(\ref{TransFF_eq4a})
to~(\ref{TransFF_eq4c}) in the subsequent sec.~\ref{TransFF}.

The paper is organised as follows: After a brief recapitulation on some
improvements concerning model $\mathcal{C}$ in sec.~\ref{ImprovedModel} and
the determination of the helicity amplitudes for electro-excitation in the
Salpeter model, we shall present the results in sec.~\ref{TransFF}, which
contains three subsections: Sec.~\ref{NNHelAmpl}, covers the helicity
amplitudes for the electro-excitation of nucleon resonances,
sec.~\ref{NDeltaHelAmpl} contains the helicity amplitudes for the
electro-excitation of $\Delta$-resonances, while sec.~\ref{PhotonCoupl}
summarises the photon decay amplitudes. Sec.~\ref{MagnTransFF} contains a
short discussion of the magnetic and electric transition form factor of the
$\Delta(1232)$ resonance before we conclude with a summary in
sec.~\ref{Summary}.

\section{Improvements to model $\mathcal{C}$\label{ImprovedModel}}
In the course of the investigations within the novel model
$\mathcal{C}$\cite{Ronniger} a new parameter set was found which led to an
improved description in particular of the nucleon form factors. This new set
of parameters is listed in table~\ref{tab:par} of the introduction. The
corresponding baryon mass spectra are very similar to those published
in~\cite{Ronniger}; only for some higher excitation deviations up to
$30\,\textrm{MeV}$ with respect to the values presented in~\cite{Ronniger}
were found. We therefore refrain from displaying the mass spectra here. The
calculated masses of those baryons which enter the helicity amplitudes
calculated in this work can be found in tables~\ref{tab:PhotonCoupl1} and
\ref{tab:PhotonCoupl2}\,.  However for nucleon form factors which were also
given in~\cite{Ronniger} some small but partially significant modifications
were found and the results will be discussed subsequently.

\begin{figure}[!htb]
\centering
\psfrag{x-axis}[c][c]{$Q^2\,\,[\textrm{GeV}^2]$}
\psfrag{y-axis}[c][c]{$G_E^p(Q^2)/G_D(Q^2)$}
\psfrag{MMD}[r][r]{\scriptsize MMD~\cite{Mergell}}
\psfrag{Christy}[r][r]{\scriptsize Christy~\cite{Christy}}
\psfrag{Qattan}[r][r]{\scriptsize Qattan~\cite{Qattan}}
\psfrag{Gauss-old}[r][r]{\scriptsize model $\mathcal{C}$~\cite{Ronniger}}
\psfrag{Gauss-new}[r][r]{\scriptsize improved model $\mathcal{C}$}
\includegraphics[width=\linewidth]{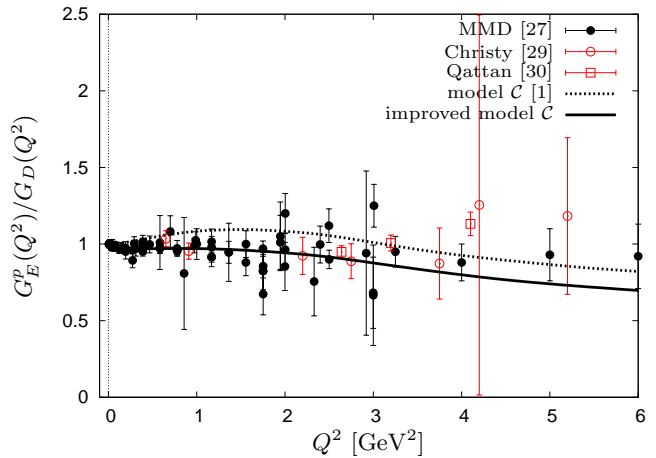}
\caption{
 The electric form factor of the proton divided by the dipole form $G_D(Q^2)$, Eq.~(\ref{eq:EFF}). 
 MMD-Data are taken from Mergell \textit{et al.}~\cite{Mergell}, supplemented by
 data from Christy \textit{et al.}~\cite{Christy} and Qattan \textit{et al.}~\cite{Qattan}\,.
 The solid line represents the results from model $\mathcal{C}$ with the new
 set of parameters, while the dashed line those from 
 model $\mathcal{C}$~\cite{Ronniger}. Red data points are taken
 from polarisation experiments and black ones are obtained by Rosenbluth separation.
 \label{EFF_Proton} }
\end{figure}

In Fig.~\ref{EFF_Proton} the calculated electric proton form factor (divided by
its dipole-shape) 
\begin{align}
   \label{eq:EFF}
   G_D(Q^2) = \frac{1}{(1+Q^2/M_V^2)^2}\,,
\end{align}
with $M_V^2=0.71\,\textrm{GeV}^2$ (see~\cite{Mergell,Bodek}) in both versions
of model $\mathcal C$ is compared to experimental data. 

\begin{figure}[!htb]
\centering
\psfrag{x-axis}[c][c]{$Q^2\,\,[\textrm{GeV}^2]$}
\psfrag{y-axis}[c][c]{$G_E^n(Q^2)$}
\psfrag{MMD}[r][r]{\scriptsize MMD \cite{Mergell}}
\psfrag{Rosenbluth}[r][r]{\scriptsize \cite{Eden,Herberg,Ostrick,Passchier,Schiavilla}}
\psfrag{Polarisation}[r][r]{\scriptsize \cite{Rohe,Golak,Zhu,Madey,Warren,Glazier,Alarcon}}
\psfrag{Gauss-old}[r][r]{\scriptsize model $\mathcal{C}$~\cite{Ronniger}}
\psfrag{Gauss-new}[r][r]{\scriptsize improved model $\mathcal{C}$}
\includegraphics[width=1.0\linewidth]{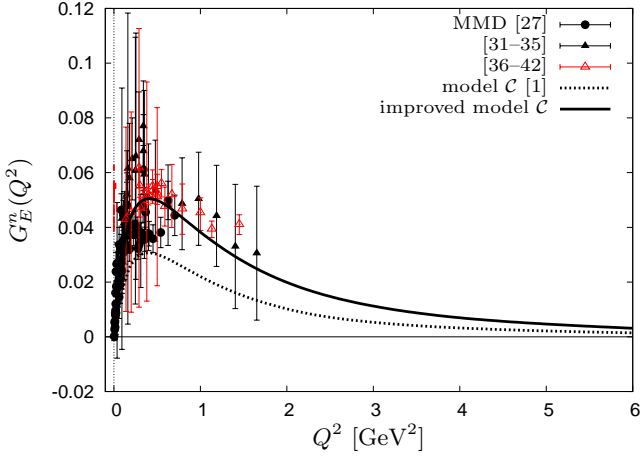}
\caption{
  The electric form factor of the neutron. MMD-Data are
  taken from the compilation of Mergell \emph{et al.}~\cite{Mergell}.
  The solid line represents the results from the improved model
  $\mathcal{C}$; the dashed line is the result from model
  $\mathcal{C}$~\cite{Ronniger}. Red data points are taken from
  polarisation experiments and black ones are obtained by Rosenbluth
  separation.\label{EFF_Neutron}
}
\end{figure}

It is found that with the new set of parameters model $\mathcal{C}$ describes
the data for momenta transfers $Q^2 \lesssim 3\,\textrm{GeV}^2$ slightly
better than with the older set of~\cite{Ronniger}\,, but the modification is
rather small.
Fig.~\ref{EFF_Neutron} shows the electric neutron form factor where the effect
of the new parameter set yields a significantly improved description,
reflecting the fact that this small quantity is thus very sensitive to
parameter changes. Although the deviation between both curves could be
interpreted as an estimate of the uncertainty in the model prediction the new
version demonstrates that within the model it is indeed possible to account
for the momentum transfer dependence and the position of the maximum rather
accurately.
The new results are very similar to the results from the Bhaduri-Cohler-Nogami
quark model quoted as BCN in~\cite{Melde07}\,, whereas the older version
from~\cite{Ronniger} is closer to the result quoted as GBE-model
in~\cite{Melde07}\,.
\begin{figure}[!htb]
\centering
\psfrag{x-axis}[c][c]{$Q^2\,\,[\textrm{GeV}^2]$}
\psfrag{y-axis}[c][c]{$G_M^p(Q^2)/G_D(Q^2)/\mu_p$}
\psfrag{MMD}[r][r]{\scriptsize MMD \cite{Mergell}}
\psfrag{Christy}[r][r]{\scriptsize Christy \cite{Christy}}
\psfrag{Qattan}[r][r]{\scriptsize Qattan \cite{Qattan}}
\psfrag{Bartel}[r][r]{\scriptsize Bartel \cite{Bartel}}
\psfrag{Gauss-old}[r][r]{\scriptsize model $\mathcal{C}$~\cite{Ronniger}}
\psfrag{Gauss-new}[r][r]{\scriptsize improved model $\mathcal{C}$}
\includegraphics[width=1.0\linewidth]{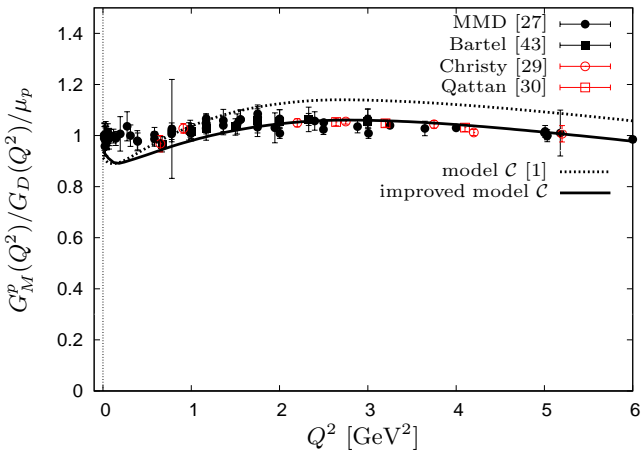}
\caption{
  The magnetic form factor of the proton divided by the dipole form
  $G_D(Q^2)$, Eq.~(\ref{eq:EFF}) and the magnetic moment of the
  proton $\mu_p=2.793\,\mu_N$\,. MMD-Data are taken from the compilation
  of Mergell \emph{et al.}~\cite{Mergell}. Additionally, polarisation
  experiments are marked in red. The black marked data points are obtained by
  Rosenbluth separation.
  \label{MFF_Proton}
}
\end{figure}

\begin{figure}[!htb]
\centering
\psfrag{x-axis}[c][c]{$Q^2\,\,[\textrm{GeV}^2]$}
\psfrag{y-axis}[c][c]{$G_M^n(Q^2)/G_D(Q^2)/\mu_n$}
\psfrag{MMD}[r][r]{\scriptsize MMD \cite{Mergell}}
\psfrag{Anklin}[r][r]{\scriptsize Anklin \cite{Anklin}}
\psfrag{Kubon}[r][r]{\scriptsize Kubon \cite{Kubon}}
\psfrag{Xu}[r][r]{\scriptsize Xu \cite{Xu}}
\psfrag{Madey}[r][r]{\scriptsize Madey \cite{Madey}}
\psfrag{Alarcon}[r][r]{\scriptsize Alarcon \cite{Alarcon}}
\psfrag{Gauss-old}[r][r]{\scriptsize model $\mathcal{C}$~\cite{Ronniger}}
\psfrag{Gauss-new}[r][r]{\scriptsize improved model $\mathcal{C}$}
\includegraphics[width=1.0\linewidth]{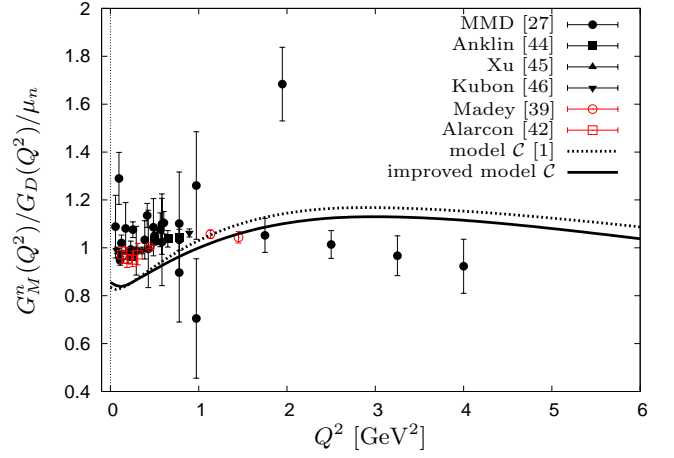}
\caption{
  The magnetic form factor of the neutron divided by the dipole form $G_D(Q^2)$,
  Eq.~(\ref{eq:EFF}) and the magnetic moment of the neutron $\mu_n=-1.913\,\mu_N$.
  MMD-Data are taken from the compilation by Mergell \emph{et al.}~\cite{Mergell} and
  from more recent results from MAMI~\cite{Anklin,Kubon}\,. Additionally, polarisation
  experiments are marked by red data points. The black marked ones are obtained by
  Rosenbluth separation.
  \label{MFF_Neutron}
}
\end{figure}

Also for the magnetic form factors displayed in Fig.~\ref{MFF_Proton}
and~\ref{MFF_Neutron} improvements are observed: this concerns in particular
the description of the magnetic proton form factor at higher momentum
transfers $Q^2 \gtrsim 1\,\textrm{GeV}^2$\,.

\begin{figure}[!htb]
\centering
\psfrag{x-axis}[c][c]{$Q^2\,\,[\textrm{GeV}^2]$}
\psfrag{y-axis}[c][c]{$G_A^3(Q^2)/G_D^A(Q^2)/g_A$}
\psfrag{Amaldi}[r][r]{\scriptsize Amaldi \cite{Amaldi}}
\psfrag{Brauel}[r][r]{\scriptsize Brauel \cite{Brauel}}
\psfrag{Bloom}[r][r]{\scriptsize Bloom \cite{Bloom}}
\psfrag{Del Guerra}[r][r]{\scriptsize D. Guerra \cite{DGuerra}}
\psfrag{Joos}[r][r]{\scriptsize Joos \cite{Joos}}
\psfrag{Baker}[r][r]{\scriptsize Baker \cite{Baker}}
\psfrag{Miller}[r][r]{\scriptsize Miller \cite{Miller}}
\psfrag{Kitagaki}[r][r]{\scriptsize Kita. \cite{Kitagaki83,Kitagaki90}}
\psfrag{Allasia}[r][r]{\scriptsize Allasia \cite{Allasia}}
\psfrag{Gauss-old}[r][r]{\scriptsize model $\mathcal{C}$ \cite{Ronniger}}
\psfrag{Gauss-new}[r][r]{\scriptsize impr. model $\mathcal{C}$}
\includegraphics[width=1.0\linewidth]{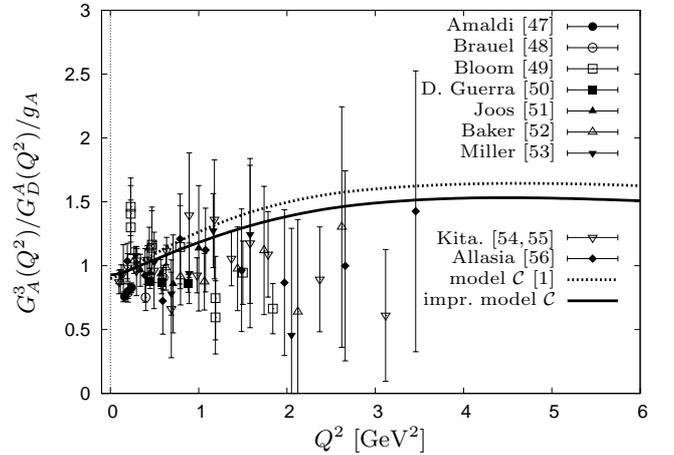}
\caption{The axial form factor of the nucleon divided by the axial dipole
 form in Eq.~(\ref{eq:AFF}) and the axial coupling $g_A=1.267$. The solid
 line is the improved result of model $\mathcal C$, the dashed line the
 result of model $\mathcal{C}$ in~\cite{Ronniger}. Experimental
 data are taken from the compilation by Bernard \emph{et al.}~\cite{Bernard}\,.
 \label{AFF_3}} 
\end{figure}
The axial form factor divided by its dipole-shape
\begin{align}
   \label{eq:AFF}
   G^A_D(Q^2)=\frac{g_A}{(1+Q^2/M_A^2)^2}\,,
\end{align}
with the parameters $M_A=1.014\pm0.014\,\textrm{GeV}$ and $g_A=1.267$ taken
from Bodek \textit{et al.}~\cite{Bodek} is shown in Fig.~\ref{AFF_3}. Here a
clear preference for either set of parameters cannot be inferred; in general
the description is satisfactory in both versions.  In subsequent sections all
calculations will use the parameters of the improved model $\mathcal C$ and
shall be quoted simply as model $\mathcal C$\,.

\section{Helicity amplitudes and transition form factors
  from the current matrix elements\label{TransFF}} 

Following the elaboration on the transition current matrix elements in Merten
\textit{et al.}~\cite{Merten} one finds in lowest order for a initial baryon
state with four-momentum $\bar{P}_i=M_i=(M_i,\vec 0)$ in its rest frame and a
final baryon state with four-momentum $\bar{P}_f$ the expression
\begin{eqnarray}
  \label{TransFF_eq1}
    &&\langle \bar P_f | j^\mu(0) | M_i \rangle
    =
    \!-3\!\!\int\!\!\frac{\textrm{d}^4p_\xi}{(2\pi)^4}\!\!
    \int\!\!\frac{\textrm{d}^4p_\eta}{(2\pi)^4}\!
    \textstyle
    {\bar{\Gamma}}^\Lambda_{\bar{P}_f}\!\left(p_\xi,p_\eta\!-\!\tfrac{2}{3}q\right)
  \nonumber
  \\
  &&\hspace*{1em}
  \textstyle
  S^1_F\!\left(\tfrac{1}{3}M_i+\!p_\xi\!+\!\tfrac{1}{2}p_\eta\right)\!
  \otimes\!
  S^2_F\!\left(\tfrac{1}{3}M_i-\!p_\xi\!+\!\tfrac{1}{2}p_\eta\right)
  \nonumber
  \\
  &&\hspace*{1em}
  \textstyle
  \otimes
  S^3_F\!\left(\tfrac{1}{3}M_i\!-\!p_\xi\!-\!p_\eta\!+\!q\right)
  \widehat q\gamma^\mu
  S^3_F\!\left(\tfrac{1}{3}M_i\!-\!p_\xi\!-\!p_\eta\right)
  \nonumber
  \\
  &&
  \hspace*{10em}
  \textstyle
  \Gamma^\Lambda_{M_i}\left(\vec p_\xi,\vec p_\eta\right)\,,
\end{eqnarray}
where the so-called vertex-function $\Gamma^\Lambda_{M_i}\left(\vec p_\xi,\vec
p_\eta\right)$ is given in the rest frame by
\begin{eqnarray}
  \label{TransFF_eq2}
  \lefteqn{
    \Gamma^\Lambda_{M_i}\left(\vec p_\xi,\vec p_\eta\right)
    :=
    -\textrm{i}
    \int\frac{\textrm{d}p'_\xi}{(2\pi)^4}
    \int\frac{\textrm{d}p'_\eta}{(2\pi)^4}
  }\nonumber
  \\
  &&
  \hspace*{1em}\left[
    V^{(3)}_\Lambda\left(\vec p_\xi,\vec p_\eta;\vec p'_\xi,\vec p'_\eta\right)
    +
    V^{\textrm{eff}}_\Lambda\left(\vec p_\xi,\vec p_\eta;\vec p'_\xi,\vec
      p'_\eta\right)
  \right]\nonumber
  \\
  &&
  \hspace*{10em}\Phi^\Lambda_{M_i}\left(\vec p'_\xi,\vec p'_\eta\right)\,,
\end{eqnarray}
and where the Salpeter-amplitude $\Phi^\Lambda_{M_i}$ is normalised to
$\sqrt{2M_i}$. For an arbitrary on-shell momentum $\bar{P}_f$ with
$\bar{P}_f^2 = M_f^2$ the vertex function
$\Gamma^\Lambda_{\bar{P}_f}\!\left(p_\xi,p_\eta\!-\!\tfrac{2}{3}q\right)$ is
obtained from $\Gamma^\Lambda_{M_f}\left(\vec p_\xi,\vec p_\eta\right)$ by
applying a boost.
Note that by this procedure we can determine the required vertex function only
on the mass shell, which precludes a calculation of transition amplitudes in
the time-like region.

The electromagnetic current operator is then defined as
\begin{align}
 j^E_{\mu}(x) =\,\,:\bar{\Psi}(x)\hat{q}\gamma_{\mu}\Psi(x):\label{TransFF_eq3}
\end{align}
in terms of the charge operator $\hat{q}$ and the quark field-operator
$\Psi(x)$\,.
With
\begin{align}
  j^E_{\pm}(x) = j^E_1(x) \pm \textrm{i} j^E_{2}(x)\,,\label{TransFF_eq4}
\end{align}
and with our normalisation of the Salpeter amplitudes, in accordance with the
definitions in Warns \textit{et al.}~\cite{Warns1990}, Tiator \textit{et
al.}~\cite{Tiator2011} and Aznauryan and Burkert~\cite{Aznauryan2012}\,, the
transverse and longitudinal helicity amplitudes $A^N_{\nicefrac{1}{2}}$\,,
$A^N_{\nicefrac{3}{2}}$ and $S^N_{\nicefrac{1}{2}}$, respectively, are related
to the transition current matrix elements in the rest frame of the baryon $B$
with rest mass $M_B$ via 
\begin{subequations}
\begin{align}
  A^N_{\frac{1}{2}}\!(Q^2)
  = 
  & 
  \frac{\zeta}{\sqrt{2}}\,K\,
  \Big\langle
  B, M_B,\tfrac{1}{2} \Big| j_+^E(0) \Big| N,\bar P_N,-\tfrac{1}{2}
  \Big\rangle,\label{TransFF_eq4a}
  \\ 
  A^N_{\frac{3}{2}}\!(Q^2)
  = 
  & 
  \frac{\zeta}{\sqrt{2}}\,K\,
  \Big\langle
  B, M_B,\tfrac{3}{2} \Big| j_+^E(0) \Big| N,\bar P_N,\tfrac{1}{2}
  \Big\rangle,\label{TransFF_eq4b}
  \\ 
  S^N_{\frac{1}{2}}\!(Q^2)
  = 
  & 
  \zeta\,K\,
  \Big\langle
  B, M_B,\tfrac{1}{2} \Big| j_0^E(0) \Big| N,\bar P_N,\tfrac{1}{2}
  \Big\rangle,\label{TransFF_eq4c}
\end{align}
\end{subequations}
where $K := \sqrt{(\pi\,\alpha)/(M_N(M_B^2-M_N^2))}$\,, $\alpha$ is the fine
structure constant\,, $N$ denotes the ground state 
nucleon ($N=(p,n)$) with four-momentum $\bar P_N=(\sqrt{M_N^2+\vec k^2},-\vec
k)$ related to the momentum transfer $Q^2$ by $\vec k^2 = (M_B^2-M_N^2-Q^2)^2
/ (4 M_B^2) + Q^2$\,.      

Note that the common phase $\zeta$ is not determined in the present
calculation. 
In most cases we shall fix $\zeta$ such as to reproduce the sign of the proton
decay amplitude reported in~\cite{PDG}\,.
Furthermore, note that 
$\langle p,\,\bar P_N,\tfrac{1}{2}|j_0^E(0)|p,\,M_N,\tfrac{1}{2}\rangle$ 
is normalised to +1 at $Q^2=0$\,.

\subsection{Helicity amplitudes for electro-excitation\label{ElecHelAmpl}}

In the last decade new experiments were performed at the Jefferson-Laboratory
in order to study helicity amplitudes up to $6\,\textrm{GeV}^2$. These new
experiments were designed to determine the helicity amplitudes for the
electro-excitation of the $P_{11}(1440)$, $S_{11}(1535)$ and $D_{13}(1520)$
resonances. The results can be found
in~\cite{Aznauryan05_1,Aznauryan05_2,Aznauryan2009} and~\cite{Drechsel}. In
addition novel data for the longitudinal $S_{1/2}^N$ amplitudes were obtained.

We calculated the corresponding helicity amplitudes of these and other states
on the basis of the Salpeter amplitudes obtained in the novel model $\mathcal
C$~\cite{Ronniger}\,. As mentioned above we are now able to solve the
eigenvalue problem with higher numerical accuracy by an expansion into a
larger basis which presently includes all three-particle harmonic oscillator
states up to an excitation quantum number $N_{\textrm{max}}=18$\,, whereas
previously~\cite{LoeMePe2,LoeMePe3,Merten} the results for baryon masses and
amplitudes in model $\mathcal{A}$ were obtained with
$N_{\textrm{max}}=12$\,. For comparison and to study the effects of the newly
introduced phenomenological flavour-dependent interaction of model
$\mathcal{C}$ we thus also recalculated the spectrum and the amplitudes for
model $\mathcal{A}$ within the same larger model space. The corresponding
changes in the determination of the interaction parameters are indicated in
table~\ref{tab:par}\,.

\subsubsection{Helicity amplitudes for nucleons\label{NNHelAmpl}}

We will now turn to the discussion of $N\to N^\ast$ helicity amplitudes for each angular
momentum $J$ and parity $\pi$\,.

\paragraph{The $J=1/2$ resonances:}

\begin{figure}[ht!]

\centering
\psfrag{x-axis}[c][c]{$Q^2\,[\textrm{GeV}^2]$}
\psfrag{y-axis}[c][c]{$A_{\tfrac{1}{2}}^N(Q^2)\,[10^{-3}\textrm{GeV}^{-\tfrac{1}{2}}]$}
\psfrag{Anisovich}[r][r]{\scriptsize Anisovich~\cite{Anisovich_3}, p}
\psfrag{PDG-p}[r][r]{\scriptsize PDG~\cite{PDG}, p}
\psfrag{PDG-n}[r][r]{\scriptsize PDG~\cite{PDG}, n}
\psfrag{Ku-Th}[r][r]{\scriptsize~\cite{Kummer,Beck,Alder,Breuker,Brasse78,Benmerrouche,Krusche,Armstrong,Thompson}, p}
\psfrag{Capstick}[r][r]{\scriptsize Capstick~\cite{Keister}, p}
\psfrag{Aznauryan}[r][r]{\scriptsize Aznauryan~\cite{Aznauryan05_1,Aznauryan05_2,Aznauryan2009}, p}
\psfrag{Aznauryan-fit}[r][r]{\scriptsize Fit: Aznau.~\cite{Aznauryan2012}, p}
\psfrag{Maid}[r][r]{\scriptsize MAID~\cite{Drechsel,Tiator}, p}
\psfrag{Tiator}[r][r]{\scriptsize Fit: Tiator~\cite{Tiator2011}, p}
\psfrag{Merten-p}[r][r]{\scriptsize model $\mathcal A$, p}
\psfrag{Merten-n}[r][r]{\scriptsize model $\mathcal A$, n}
\psfrag{Gauss-p}[r][r]{\scriptsize model $\mathcal C$, p}
\psfrag{Gauss-n}[r][r]{\scriptsize model $\mathcal C$, n}
\includegraphics[width=\linewidth]{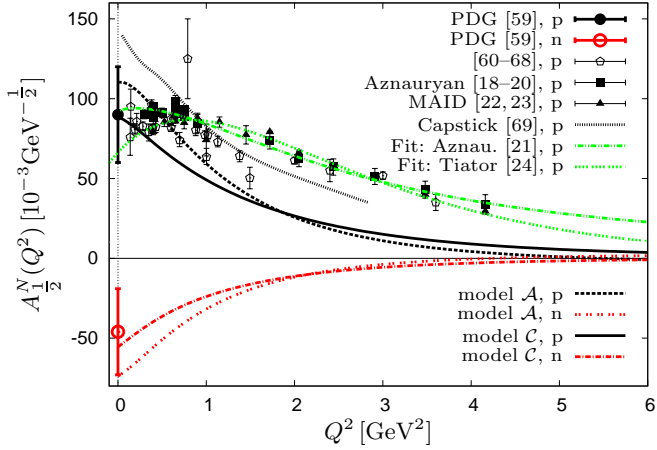}
\caption{%
  Comparison of the $S_{11}(1535)$ transverse helicity amplitude $A_{1/2}^N$
  for electro-excitation from the proton and the neutron calculated in the
  model $\mathcal C$ (solid and dashed-dotted line) and model $\mathcal A$ (dashed lines) to
  experimental
  data~\cite{PDG,Kummer,Beck,Alder,Breuker,Brasse78,%
    Benmerrouche,Krusche,Armstrong,Thompson,%
    Aznauryan05_1,Aznauryan05_2,Aznauryan2009,%
    Drechsel,Tiator}\,.The dotted
  line is the result obtained by Keister and Capstick~\cite{Keister}.
  Additionally 
  recent fits obtained by Tiator \textit{et al.}~\cite{Tiator2011}
  and by Aznauryan \textit{et al.}~\cite{Aznauryan2012}
  are displayed as green dotted and dashed-dotted lines, respectively. 
  Note that the results
  for model $\mathcal A$ were recalculated with higher numerical accuracy and
  thus deviate from the results published previously in~\cite{Merten}\,.%
  \label{TRANSFF_A12_S11_1535}
}
\end{figure}
\begin{figure}[ht!]
\centering
\psfrag{x-axis}[c][c]{$Q^2\,[\textrm{GeV}^2]$}
\psfrag{y-axis}[c][c]{$S_{\tfrac{1}{2}}^N(Q^2)\,[10^{-3}\textrm{GeV}^{-\tfrac{1}{2}}]$}
\psfrag{Aznauryan}[r][r]{\scriptsize Aznauryan~\cite{Aznauryan05_1,Aznauryan05_2,Aznauryan2009}, p}
\psfrag{Maid}[r][r]{\scriptsize MAID~\cite{Drechsel,Tiator}, p}
\psfrag{Tiator}[r][r]{\scriptsize Fit: Tiator~\cite{Tiator2011}, p}
\psfrag{Kreuzer-p}[r][r]{\scriptsize model $\mathcal A$, p}
\psfrag{Kreuzer-n}[r][r]{\scriptsize model $\mathcal A$, n}
\psfrag{Gauss-p}[r][r]{\scriptsize model $\mathcal C$, p}
\psfrag{Gauss-n}[r][r]{\scriptsize model $\mathcal C$, n}
\includegraphics[width=\linewidth]{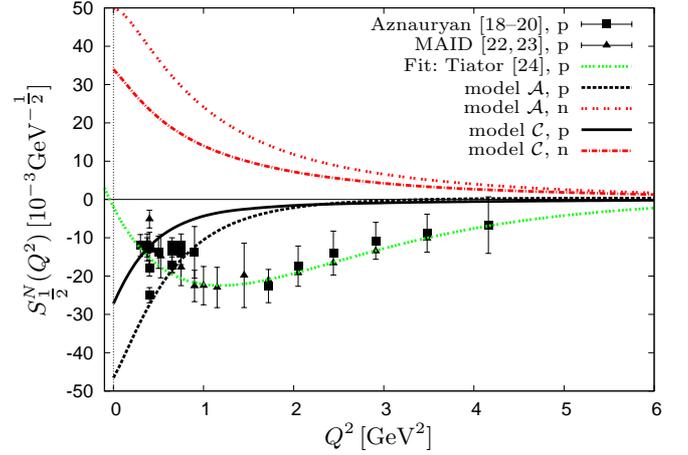}
\caption{%
  Comparison of the $S_{11}(1535)$ longitudinal electro-excitation helicity
  amplitude $S_{1/2}^N$ of proton and neutron calculated in model $\mathcal C$
  (solid and dashed-dotted line) and model $\mathcal A$ (dashed lines) with experimental
  data~\cite{Aznauryan05_1,Aznauryan05_2,Aznauryan2009,%
    Drechsel,Tiator}. Note that for the data points of the MAID-analysis by
  Tiator \textit{et al.}~\cite{Tiator} no errors are quoted. See also caption to
  Fig.~\ref{TRANSFF_A12_S11_1535}.
  \label{TRANSFF_S_S11_1535}} 

\end{figure}

A comparison of calculated transverse and longitudinal helicity amplitudes
with experimental data for the electro-excitation of the $S_{11}(1535)$
resonance is given in Figs.~\ref{TRANSFF_A12_S11_1535}
and~\ref{TRANSFF_S_S11_1535}, respectively.  Whereas the value of the
transverse amplitudes at the photon point ($Q^2=0$) both for the proton and
the neutron are accurately reproduced in particular by the new model
$\mathcal{C}$\,, in general the calculated transverse amplitudes are too small
by a factor of two; in comparison to the results from model $\mathcal{A}$ the
amplitudes of model $\mathcal{C}$ decrease more slowly with increasing
momentum transfer, in better agreement with the experimental data. But, in
particular the near constancy of the proton data for $0<Q^2 < 1\,\textnormal{GeV}^2$
is not reflected by any of the calculated results. For
comparison we also plotted the results from the quark model calculation of the
transverse $A_{1/2}^p$-amplitude by Keister and Capstick~\cite{Keister} for
$Q^2\lesssim3\,\textrm{GeV}^2$ and the fits obtained by Aznauryan \textit{et
al.}~\cite{Aznauryan2012} and Tiator \textit{et al.}~\cite{Tiator2011}.

Contrary to this, the momentum transfer dependence of the calculated longitudinal
helicity amplitudes hardly bear any resemblance to what has been determined
experimentally, in particular the minimum found for the proton at $Q^2 \approx
1.5 \textnormal{ GeV}^2$ is not reproduced. Only the non-relativistic calculation
of Capstick and Keister~\cite{Capstick1995} shows a pronounced minimum for the
longitudinal $S_{11}(1535)$ amplitude, however this minimum is predicted at the
wrong position.

\begin{figure}[ht!]
\centering
\psfrag{x-axis}[c][c]{$Q^2\,[\textrm{GeV}^2]$}
\psfrag{y-axis}[c][c]{$A_{\tfrac{1}{2}}^N(Q^2)\,[10^{-3}\textrm{GeV}^{-\tfrac{1}{2}}]$}
\psfrag{Anisovich}[r][r]{\scriptsize Anisovich~\cite{Anisovich_3}, p}
\psfrag{PDG-p}[r][r]{\scriptsize PDG~\cite{PDG}, p}
\psfrag{PDG-n}[r][r]{\scriptsize PDG~\cite{PDG}, n}
\psfrag{Burkert}[r][r]{\scriptsize Burkert~\cite{Burkert}, p}
\psfrag{Aznauryan}[r][r]{\scriptsize Aznauryan~\cite{Aznauryan05_2}, p}
\psfrag{Maid}[r][r]{\scriptsize MAID~\cite{Drechsel,Tiator}, p}
\psfrag{Tiator}[r][r]{\scriptsize Fit: Tiator~\cite{Tiator2011}, p}
\psfrag{Merten-p}[r][r]{\scriptsize model $\mathcal A$, p}
\psfrag{Merten-n}[r][r]{\scriptsize model $\mathcal A$, n}
\psfrag{Gauss-p}[r][r]{\scriptsize model $\mathcal C$, p}
\psfrag{Gauss-n}[r][r]{\scriptsize model $\mathcal C$, n}
\includegraphics[width=\linewidth]{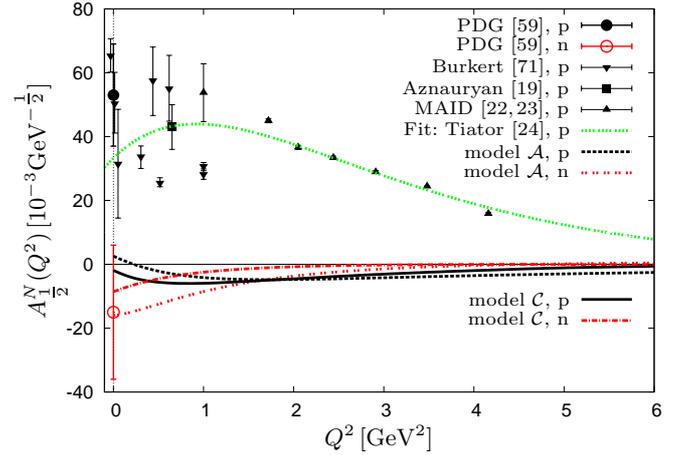}
\caption{%
  Comparison of the $S_{11}(1650)$ transverse electro-excitation helicity
  amplitude $A_{1/2}^N$ of proton and neutron calculated in model
  $\mathcal C$ (solid and dashed-dotted line) and model $\mathcal A$ (dashed lines) to experimental
  data from~\cite{PDG,Burkert,Aznauryan05_2,Drechsel,Tiator}\,. See also caption to
  Fig.~\ref{TRANSFF_A12_S11_1535}.
  \label{TRANSFF_A12_S11_1650}
}
\end{figure}
\begin{figure}[ht!]
\centering
\psfrag{x-axis}[c][c]{$Q^2\,[\textrm{GeV}^2]$}
\psfrag{y-axis}[c][c]{$S_{\tfrac{1}{2}}^N(Q^2)\,[10^{-3}\textrm{GeV}^{-\tfrac{1}{2}}]$}
\psfrag{Aznauryan}[r][r]{\scriptsize Aznauryan~\cite{Aznauryan05_2}}
\psfrag{Maid}[r][r]{\scriptsize MAID~\cite{Drechsel,Tiator}, p}
\psfrag{Tiator}[r][r]{\scriptsize Fit: Tiator~\cite{Tiator2011}, p}
\psfrag{Merten-p}[r][r]{\scriptsize model $\mathcal A$, p}
\psfrag{Merten-n}[r][r]{\scriptsize model $\mathcal A$, n}
\psfrag{Gauss-p}[r][r]{\scriptsize model $\mathcal C$, p}
\psfrag{Gauss-n}[r][r]{\scriptsize model $\mathcal C$, n}
\includegraphics[width=\linewidth]{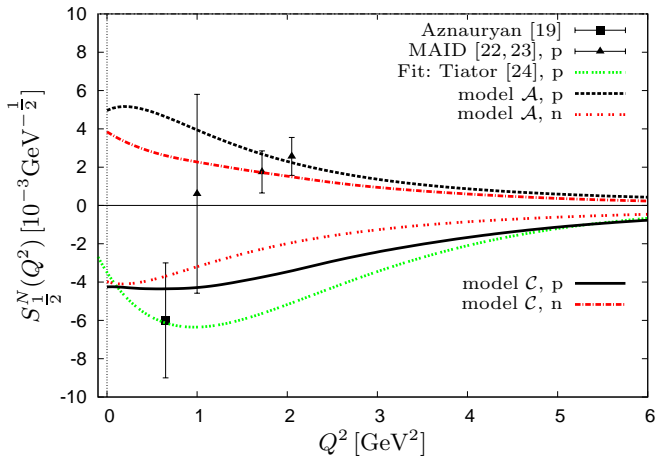}
\caption{
  Comparison of the $S_{11}(1650)$ longitudinal electro-excitation helicity
  amplitude $S_{1/2}^N$ of proton and neutron calculated in model $\mathcal C$
  (solid and dashed-dotted line) and model $\mathcal A$ (dashed lines).
  See also caption to Fig.~\ref{TRANSFF_A12_S11_1535}.\label{TRANSFF_S_S11_1650}
}
\end{figure}
Also the calculated transverse proton helicity amplitude $A_{1/2}^p$ for the
next $S_{11}(1650)$ resonance shows a large disagreement
with experimental data as shown in Fig.~\ref{TRANSFF_A12_S11_1650}.
This discrepancy was already found in the previous calculation of Merten
\textit{et al.}~\cite{Merten} and obviously is not resolved within model
$\mathcal C$\,. Note, however that the neutron amplitude $A_{1/2}^n$ calculated
at the photon point does correspond to the data from PDG~\cite{PDG}, as
illustrated in Fig~\ref{TRANSFF_A12_S11_1650}. 
The rather small longitudinal $S_{11}(1650)$ amplitude $S_{1/2}^N$ seems to
agree with the scarce medium $Q^2$ data from the MAID-analysis
of~\cite{Drechsel,Tiator}\,, however for lower $Q^2$ the single data point of
Aznauryan \textit{et al.}~\cite{Aznauryan05_2} seems to indicate a zero
crossing of this amplitude not reproduced by either of the model calculations of
the $S_{1/2}^p$-amplitude for the $S_{11}(1650)$-resonance (see
Fig.~\ref{TRANSFF_S_S11_1650}).

\begin{figure}[ht!]
\centering
\psfrag{x-axis}[c][c]{$Q^2\,[\textrm{GeV}^2]$}
\psfrag{y-axis}[c][c]{$A_{\tfrac{1}{2}}^N(Q^2)\,[10^{-3}\textrm{GeV}^{-\tfrac{1}{2}}]$}
\psfrag{Anisovich}[r][r]{\scriptsize Anisovich~\cite{Anisovich_3}, p}
\psfrag{Merten-p-A}[r][r]{\scriptsize model $\mathcal A$, p 3rd}
\psfrag{Merten-n-A}[r][r]{\scriptsize model $\mathcal A$, n 3rd}
\psfrag{Merten-p-B}[r][r]{\scriptsize model $\mathcal A$, p 4th}
\psfrag{Merten-n-B}[r][r]{\scriptsize model $\mathcal A$, n 4th}
\psfrag{Gauss-p-A}[r][r]{\scriptsize model $\mathcal C$, p 3rd}
\psfrag{Gauss-n-A}[r][r]{\scriptsize model $\mathcal C$, n 3rd}
\psfrag{Gauss-p-B}[r][r]{\scriptsize model $\mathcal C$, p 4th}
\psfrag{Gauss-n-B}[r][r]{\scriptsize model $\mathcal C$, n 4th}
\includegraphics[width=\linewidth]{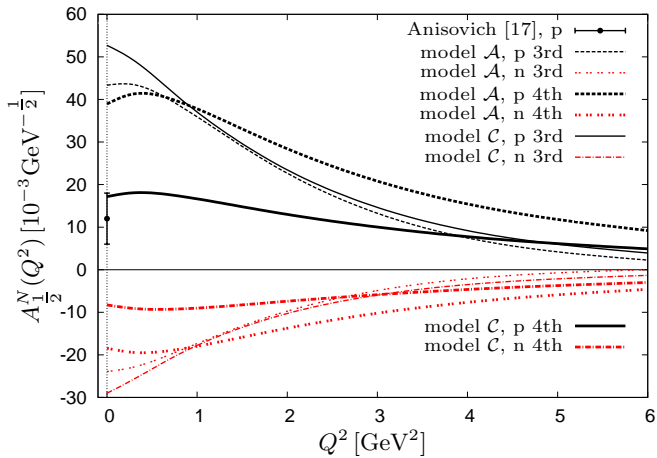}
\caption{%
  Comparison of the $S_{11}(1895)$ transverse helicity amplitude $A_{1/2}^N$
  for proton and neutron calculated in model $\mathcal C$ (solid and dashed-dotted line) and
  model $\mathcal A$ (dashed lines) with the single photon point value from
  Anisovich \textit{et al.}~\cite{Anisovich_1}. See also caption to
  Fig.~\ref{TRANSFF_A12_S11_1535}.
  \label{TRANSFF_A12_S11_1895}
}
\end{figure}

The third and fourth $J^\pi=1/2^-$ nucleon resonances are predicted in model $\mathcal A$
at $1872\,\textrm{MeV}$ and $1886\,\textrm{MeV}$ and in model $\mathcal C$ at  
$1839\,\textrm{MeV}$ and $1882\,\textrm{MeV}$\,, respectively. Indeed within the
Bonn-Gatchina Analysis of the CB-ELSA collaboration data~\cite{Anisovich_3}
evidence for a $J^\pi=1/2^-$ nucleon resonance at $1895\,\textrm{MeV}$ was
found. As can be seen from Fig.~\ref{TRANSFF_A12_S11_1895} the predicted
transverse amplitudes for the third resonance both models are rather large and the
calculated value at the photon point ($Q^2=0$) is much larger than the
experimental value quoted in~\cite{Anisovich_1,Anisovich_3}, but the value of the
fourth resonance matches the PDG photon decay amplitude.  

\begin{figure}[ht!]
\centering
\psfrag{x-axis}[c][c]{$Q^2\,[\textrm{GeV}^2]$}
\psfrag{y-axis}[c][c]{$A_{\tfrac{1}{2}}^N(Q^2)\,[10^{-3}\textrm{GeV}^{-\tfrac{1}{2}}]$}
\psfrag{Anisovich}[r][r]{\scriptsize Aniso.~\cite{Anisovich_3}, p}
\psfrag{PDG-p}[r][r]{\scriptsize PDG~\cite{PDG}, p}
\psfrag{PDG-n}[r][r]{\scriptsize PDG~\cite{PDG}, n}
\psfrag{BurkertGerhardt}[r][r]{\scriptsize~\cite{Burkert,Gerhardt}, p}
\psfrag{Tiator}[r][r]{\scriptsize Fit: Tiator~\cite{Tiator2011}, p}
\psfrag{Aznauryan}[r][r]{\scriptsize Aznau.~\cite{Aznauryan05_1,Aznauryan05_2,Aznauryan2009}, p}
\psfrag{Maid}[r][r]{\scriptsize MAID~\cite{Drechsel,Tiator}, p}
\psfrag{Merten-p}[r][r]{\scriptsize model $\mathcal A$, p}
\psfrag{Merten-n}[r][r]{\scriptsize model $\mathcal A$, n}
\psfrag{Gauss-p}[r][r]{\scriptsize model $\mathcal C$, p}
\psfrag{Gauss-n}[r][r]{\scriptsize model $\mathcal C$, n}
\includegraphics[width=\linewidth]{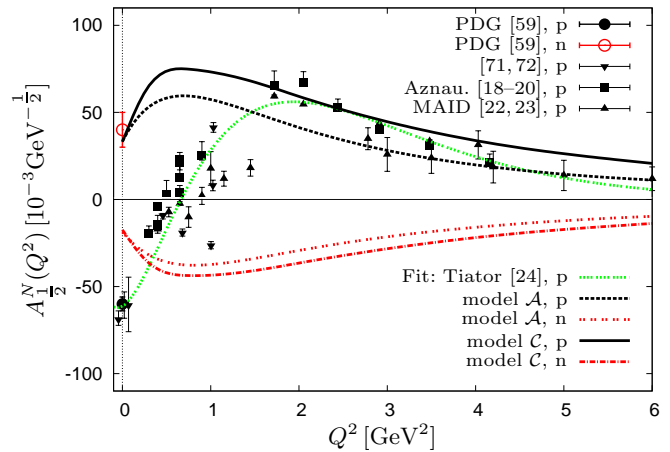}
\caption{
  Comparison of the $P_{11}(1440)$ transverse helicity amplitude $A_{1/2}^N$
  for proton and neutron calculated in model $\mathcal C$ (solid and dashed-dotted line) and
  model $\mathcal A$ (dashed lines). See also caption to
  Fig.~\ref{TRANSFF_A12_S11_1535}\,.
  \label{TRANSFF_P11_1440} 
}
\end{figure}
\begin{figure}[ht!]
\centering
\psfrag{x-axis}[c][c]{$Q^2\,[\textrm{GeV}^2]$}
\psfrag{y-axis}[c][c]{$S_{\tfrac{1}{2}}^N(Q^2)\,[10^{-3}\textrm{GeV}^{-\tfrac{1}{2}}]$}
\psfrag{PDG}[r][r]{\scriptsize $C^p$ PDG~\cite{PDG}}
\psfrag{Aznauryan}[r][r]{\scriptsize Aznauryan~\cite{Aznauryan05_1,Aznauryan05_2,Aznauryan2009}, p}
\psfrag{Maid}[r][r]{\scriptsize MAID~\cite{Drechsel,Tiator}, p}
\psfrag{Tiator}[r][r]{\scriptsize Fit: Tiator~\cite{Tiator2011}, p}
\psfrag{Kreuzer-p}[r][r]{\scriptsize model $\mathcal A$, p}
\psfrag{Kreuzer-n}[r][r]{\scriptsize model $\mathcal A$, n}
\psfrag{Gauss-p}[r][r]{\scriptsize model $\mathcal C$, p}
\psfrag{Gauss-n}[r][r]{\scriptsize model $\mathcal C$, n}
\includegraphics[width=\linewidth]{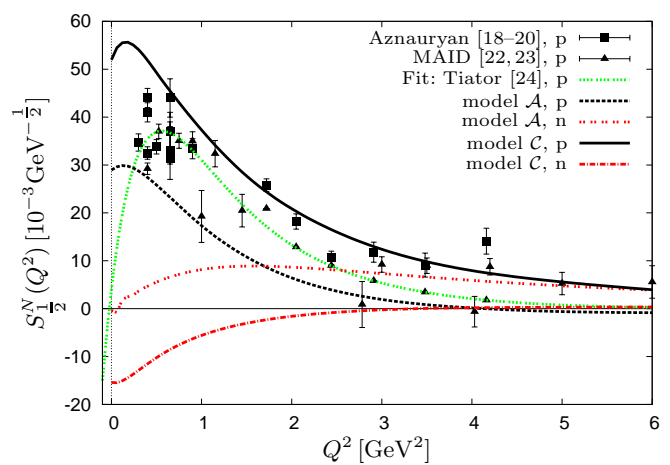}
\caption{
  Comparison of the $P_{11}(1440)$ longitudinal helicity amplitude $S_{1/2}^N$
  for proton and neutron calculated in model $\mathcal C$ (solid and dashed-dotted line) and
  model $\mathcal A$ (dashed lines). Note that for the data points of the
  MAID-analysis by Tiator \textit{et al.}~\cite{Tiator} no errors are quoted.
  See also caption to Fig.~\ref{TRANSFF_A12_S11_1535}.
  \label{TRANSFF_S_P11_1440}
}
\end{figure}
The transverse and longitudinal helicity amplitudes of the Roper resonance
$P_{11}(1440)$ are displayed in Figs.~\ref{TRANSFF_P11_1440} and~\ref{TRANSFF_S_P11_1440},
respectively. It is obvious, that the zero crossing
found in the data at $Q^2\approx 0.5\,\textnormal{GeV}^2$, see
Fig.~\ref{TRANSFF_P11_1440}, is not reproduced in the calculated curves, although
the $Q^2$ dependence of the positive values at higher momentum transfers can
be accounted for in both models after changing the sign of the old
prediction~\cite{Merten}\,. On the other hand we do find a satisfactory description
of the longitudinal $C^p_{1/2}$-amplitude displayed in
Fig.~\ref{TRANSFF_S_P11_1440} in particular in the new model $\mathcal C$\,.

\begin{figure}[ht!]
\centering
\psfrag{x-axis}[c][c]{$Q^2\,[\textrm{GeV}^2]$}
\psfrag{y-axis}[c][c]{$S_{\tfrac{1}{2}}^N(Q^2)\,[10^{-3}\textrm{GeV}^{-\tfrac{1}{2}}]$}
\psfrag{PDG-p}[r][r]{\scriptsize PDG~\cite{PDG}, p}
\psfrag{PDG-n}[r][r]{\scriptsize PDG~\cite{PDG}, n}
\psfrag{Merten-p}[r][r]{\scriptsize model $\mathcal A$, p}
\psfrag{Merten-n}[r][r]{\scriptsize model $\mathcal A$, n}
\psfrag{Gauss-p}[r][r]{\scriptsize model $\mathcal C$, p}
\psfrag{Gauss-n}[r][r]{\scriptsize model $\mathcal C$, n}
\includegraphics[width=\linewidth]{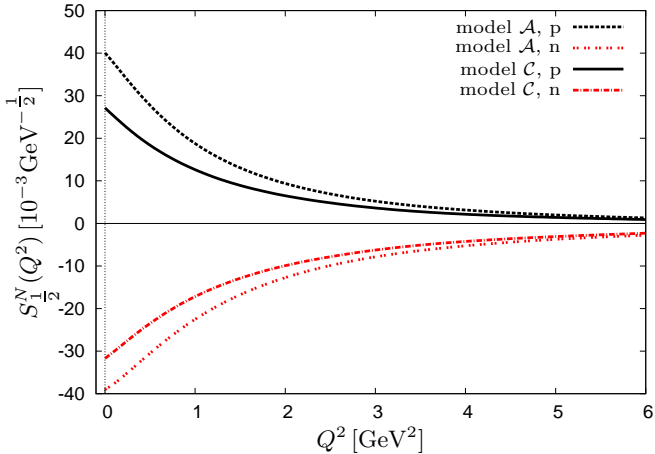}
\caption{
  Prediction of the $P_{11}(1710)$ longitudinal helicity amplitude $S_{1/2}^N$
  for proton and neutron calculated in model $\mathcal C$ (solid and dashed-dotted line) and
  model $\mathcal A$ (dashed lines). See also caption to Fig.~\ref{TRANSFF_A12_S11_1535}.
  \label{TRANSFF_S_P11_1710}
}

\end{figure}

Helicity amplitudes of higher lying resonances in the $P_{11}$ channel are
only poorly studied in experiments. Nevertheless we shall discuss briefly the
$P_{11}(1710)$ helicity amplitude before treating the higher excitations
$P_{11}(1880)$ and $P_{11}(2100)$\,.
\begin{figure}[ht!]
\centering
\psfrag{x-axis}[c][c]{$Q^2\,[\textrm{GeV}^2]$}
\psfrag{y-axis}[c][c]{$A_{\tfrac{1}{2}}^N(Q^2)\,[10^{-3}\textrm{GeV}^{-\tfrac{1}{2}}]$}
\psfrag{Anisovich}[r][r]{\scriptsize Anisovich~\cite{Anisovich_3}, p}
\psfrag{PDG-p}[r][r]{\scriptsize PDG~\cite{PDG}, p}
\psfrag{PDG-n}[r][r]{\scriptsize PDG~\cite{PDG}, n}
\psfrag{Merten-p}[r][r]{\scriptsize model $\mathcal A$, p}
\psfrag{Merten-n}[r][r]{\scriptsize model $\mathcal A$, n}
\psfrag{Gauss-p}[r][r]{\scriptsize model $\mathcal C$, p}
\psfrag{Gauss-n}[r][r]{\scriptsize model $\mathcal C$, n}
\includegraphics[width=\linewidth]{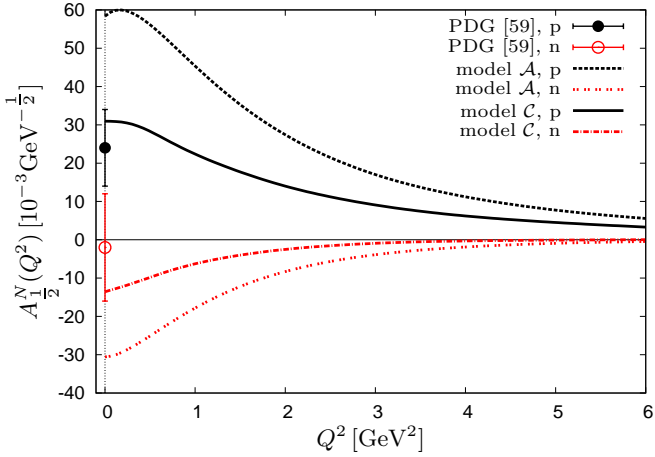}
\caption{
  Comparison of the $P_{11}(1710)$ transverse helicity
  amplitude $A_{1/2}^N$ for proton and neutron calculated in model $\mathcal C$
  (solid and dashed-dotted line) and model $\mathcal A$ (dashed lines). See also caption to
  Fig.~\ref{TRANSFF_A12_S11_1535}.
  \label{TRANSFF_P11_1710}
}

\end{figure}
For the $P_{11}(1710)$ resonance only the photon decay amplitude is reported~\cite{PDG}.
In Figs.~\ref{TRANSFF_P11_1710} and~\ref{TRANSFF_S_P11_1710} we display
our predictions for these amplitudes.
The transverse $A_{1/2}^N$-amplitude of model $\mathcal C$ matches the PDG data
at the photon point in contrast to model $\mathcal A$, which overestimates the
proton- and neutron amplitudes by a factor of two. On the other hand this would
be in accordance with the larger value obtained by Anisovich \textit{et al.}~\cite{Anisovich_3},
 $A^p_{1/2}=(52\pm15)\times10^{-3}\textrm{ GeV}^2$\,. The prediction of the longitudinal
$S_{1/2}^N$-amplitudes is given in Fig.~\ref{TRANSFF_S_P11_1710}.

\begin{figure}[ht!]
\centering
\psfrag{x-axis}[c][c]{$Q^2\,[\textrm{GeV}^2]$}
\psfrag{y-axis}[c][c]{$A_{\tfrac{1}{2}}^N(Q^2)\,[10^{-3}\textrm{GeV}^{-\tfrac{1}{2}}]$}
\psfrag{Anisovich}[r][r]{\scriptsize Anisovich $N_{1/2^+}(1880)$~\cite{Anisovich_1}}
\psfrag{Anisovich-2}[r][r]{\scriptsize Anisovich $N_{1/2^+}(1880)$~\cite{Anisovich_3}}
\psfrag{01}[c][c]{\scriptsize 01}
\psfrag{02}[c][c]{\scriptsize 02}
\psfrag{Merten-p-3rd}[r][r]{\scriptsize model $\mathcal A$, p 4th}
\psfrag{Merten-n-3rd}[r][r]{\scriptsize model $\mathcal A$, n 4th}
\psfrag{Merten-p-4th}[r][r]{\scriptsize model $\mathcal A$, p 5th}
\psfrag{Merten-n-4th}[r][r]{\scriptsize model $\mathcal A$, n 5th}
\psfrag{Gauss-p-3rd}[r][r]{\scriptsize model $\mathcal C$, p 4th}
\psfrag{Gauss-n-3rd}[r][r]{\scriptsize model $\mathcal C$, n 4th}
\psfrag{Gauss-p-4th}[r][r]{\scriptsize model $\mathcal C$, p 5th}
\psfrag{Gauss-n-4th}[r][r]{\scriptsize model $\mathcal C$, n 5th}
\includegraphics[width=\linewidth]{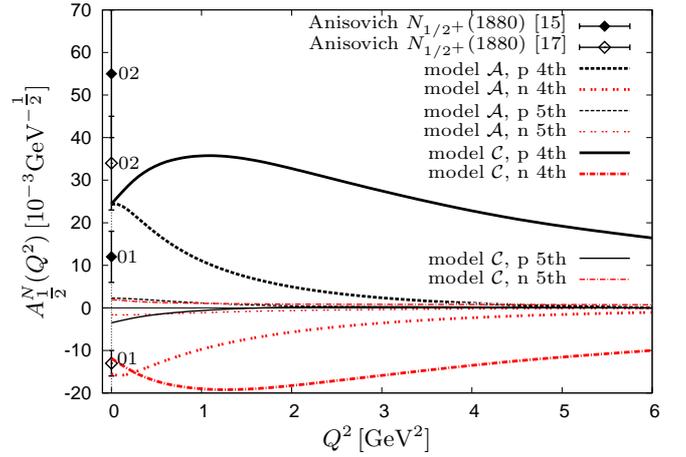}
\caption{
  Prediction of the $P_{11}$ transverse helicity amplitudes $A_{1/2}^N$ for
  the third and fourth excitation of the proton and the neutron calculated
  within model $\mathcal C$ (solid and dashed-dotted lines) and model $\mathcal A$ (dashed lines).
  The data at the photon point marked $''01''$ and $''02''$ were reported
  in~\cite{Anisovich_1,Anisovich_3} as alternatives for the $N_{1/2}^+(1880)$
  resonance. See also caption to Fig.~\ref{TRANSFF_A12_S11_1535}.
  \label{TRANSFF_P11_3rd-4th_Res}}
\end{figure}

Finally we present the results for the fourth and fifth
$J^\pi=\tfrac{1}{2}^+$-nucleon state in Fig.~\ref{TRANSFF_P11_3rd-4th_Res}\,,
where we show the transverse helicity amplitudes only. The corresponding\linebreak
masses predicted by model $\mathcal A$ are $1905\,\textrm{MeV}$ for the
fourth and $1953\,\textrm{MeV}$ for the fifth state; for model
$\mathcal C$ the predicted masses are $1872\,\textrm{MeV}$ and
$1968\,\textrm{MeV}$\,, respectively. The two data at the photon point marked
$''01''$ and $''02''$ were obtained by the CB-ELSA collaboration within the
Bonn-Gatchina Analysis as reported in~\cite{Anisovich_1,Anisovich_3} for the
$N_{1/2}^+(1880)$ resonance.
They correspond to two different partial wave solutions in order to extract the corresponding
baryon mass and helicity amplitudes. The prediction for the fourth state lies between these
values, the values found for the fifth state are much smaller. This also applies for higher
$J^\pi=\tfrac{1}{2}^+$ excitations not displayed here.

\paragraph{The $J=3/2$ resonances:}

\begin{figure}[ht!]
\centering
\psfrag{x-axis}[c][c]{$Q^2\,[\textrm{GeV}^2]$}
\psfrag{y-axis}[c][c]{$A_{\tfrac{1}{2}}^N(Q^2)\,[10^{-3}\textrm{GeV}^{-\tfrac{1}{2}}]$}
\psfrag{PDG-p}[r][r]{\scriptsize PDG~\cite{PDG}, p}
\psfrag{PDG-n}[r][r]{\scriptsize PDG~\cite{PDG}, n}
\psfrag{Aznauryan}[r][r]{\scriptsize Aznauryan~\cite{Aznauryan05_2}, p}
\psfrag{Maid}[r][r]{\scriptsize MAID~\cite{Drechsel,Tiator}, p}
\psfrag{Tiator}[r][r]{\scriptsize Fit: Tiator~\cite{Tiator2011}, p}
\psfrag{Merten-p}[r][r]{\scriptsize model $\mathcal A$, p}
\psfrag{Merten-n}[r][r]{\scriptsize model $\mathcal A$, n}
\psfrag{Gauss-p}[r][r]{\scriptsize model $\mathcal C$, p}
\psfrag{Gauss-n}[r][r]{\scriptsize model $\mathcal C$, n}
\includegraphics[width=\linewidth]{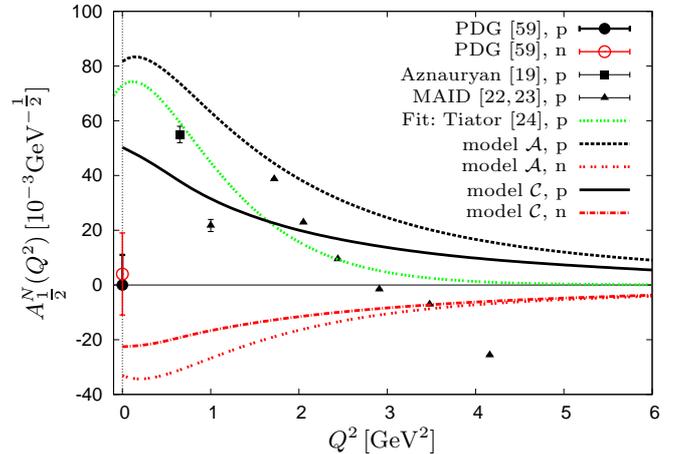}
\caption{
  Comparison of the $P_{13}(1720)$ transverse helicity amplitude $A_{1/2}^N$
  of proton and neutron calculated in model $\mathcal C$ (solid and dashed-dotted line)
  and model $\mathcal A$ (dashed lines). Note that for the data points
  of the MAID-analysis by Tiator \textit{et al.}~\cite{Tiator} no errors
  are quoted. See also caption to Fig.~\ref{TRANSFF_A12_S11_1535}.
\label{TRANSFF_A12_P13_1720}}
\end{figure}

\begin{figure}[ht!]
\centering
\psfrag{x-axis}[c][c]{$Q^2\,[\textrm{GeV}^2]$}
\psfrag{y-axis}[c][c]{$A_{\tfrac{3}{2}}^N(Q^2)\,[10^{-3}\textrm{GeV}^{-\tfrac{1}{2}}]$}
\psfrag{PDG-p}[r][r]{\scriptsize PDG~\cite{PDG}, p}
\psfrag{PDG-n}[r][r]{\scriptsize PDG~\cite{PDG}, n}
\psfrag{Aznauryan}[r][r]{\scriptsize Aznauryan~\cite{Aznauryan05_2}, p}
\psfrag{Maid}[r][r]{\scriptsize MAID~\cite{Drechsel,Tiator}, p}
\psfrag{Tiator}[r][r]{\scriptsize Fit: Tiator~\cite{Tiator2011}, p}
\psfrag{Merten-p}[r][r]{\scriptsize model $\mathcal A$, p}
\psfrag{Merten-n}[r][r]{\scriptsize model $\mathcal A$, n}
\psfrag{Gauss-p}[r][r]{\scriptsize model $\mathcal C$, p}
\psfrag{Gauss-n}[r][r]{\scriptsize model $\mathcal C$, n}
\includegraphics[width=\linewidth]{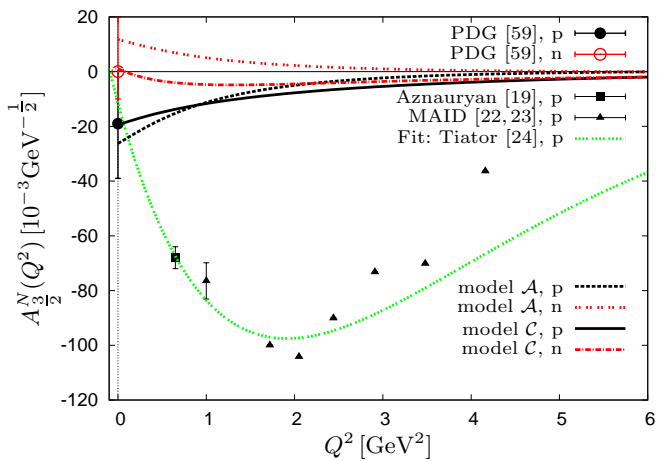}
\caption{
  Comparison of the $P_{13}(1720)$ transverse helicity amplitude $A_{3/2}^N$
  of proton and neutron calculated in model $\mathcal C$ (solid and dashed-dotted line) and
  model $\mathcal A$ (dashed lines). Note that for the data
  points of the MAID-analysis by Tiator \textit{et al.}~\cite{Tiator} no
  errors are quoted. See also caption to Fig.~\ref{TRANSFF_A12_S11_1535}.
  \label{TRANSFF_A32_P13_1720}}
\end{figure}

In Figs.~\ref{TRANSFF_A12_P13_1720} and~\ref{TRANSFF_A32_P13_1720} the transverse
helicity amplitudes of the $P_{13}(1720)$ resonance are displayed.
Although a reasonable agreement with the data of Aznauryan \textit{et
al.}~\cite{Aznauryan05_2} and with the photon decay amplitude is found for both
models, the data from the MAID analysis~\cite{Drechsel,Tiator} indicate a
sign change for the $A^p_{\nicefrac{1}{2}}$ amplitude at 
$Q^2 \approx 3\,\textnormal{GeV}^2$ not reproduced by either model.
\begin{figure}[ht!]
\centering
\psfrag{x-axis}[c][c]{$Q^2\,[\textrm{GeV}^2]$}
\psfrag{y-axis}[c][c]{$S_{\tfrac{1}{2}}^N(Q^2)\,[10^{-3}\textrm{GeV}^{-\tfrac{1}{2}}]$}
\psfrag{Maid}[r][r]{\scriptsize MAID~\cite{Drechsel,Tiator}, p}
\psfrag{Tiator}[r][r]{\scriptsize Fit: Tiator~\cite{Tiator2011}, p}
\psfrag{Merten-p}[r][r]{\scriptsize model $\mathcal A$, p}
\psfrag{Merten-n}[r][r]{\scriptsize model $\mathcal A$, n}
\psfrag{Gauss-p}[r][r]{\scriptsize model $\mathcal C$, p}
\psfrag{Gauss-n}[r][r]{\scriptsize model $\mathcal C$, n}
\includegraphics[width=\linewidth]{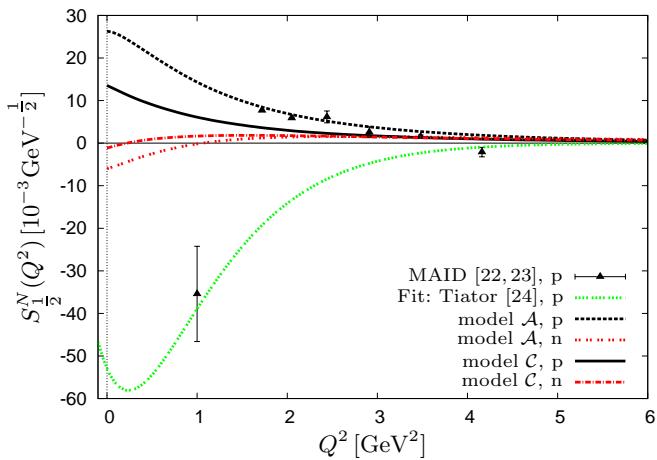}
\caption{
  Comparison of the $P_{13}(1720)$ longitudinal electro-excitation helicity amplitude $S_{1/2}^N$
  of proton and neutron calculated in model $\mathcal C$ (solid and dashed-dotted line)
  and model $\mathcal A$ (dashed lines). Note that for the data points of the
  MAID-analysis by Tiator \textit{et al.}~\cite{Tiator} no errors are quoted. 
  See also caption to Fig.~\ref{TRANSFF_A12_S11_1535}.
  \label{TRANSFF_S12_P13_1720}}
\end{figure}
In spite of not being able to account at all for the large $A^p_{\nicefrac{3}{2}}$
amplitude found experimentally, the longitudinal helicity amplitude as reported
in the MAID analysis with exception of the value at $Q^2 \approx 1\,
\textnormal{GeV}^2$ is reproduced by both models rather well, as shown in
Fig.~\ref{TRANSFF_S12_P13_1720}. 

\begin{figure}[ht!]
\centering
\psfrag{x-axis}[c][c]{$Q^2\,[\textrm{GeV}^2]$}
\psfrag{y-axis}[c][c]{$A_{\tfrac{1}{2}}^N(Q^2)\,[10^{-3}\textrm{GeV}^{-\tfrac{1}{2}}]$}
\psfrag{PDG-p}[r][r]{\scriptsize PDG~\cite{PDG}, p}
\psfrag{PDG-n}[r][r]{\scriptsize PDG~\cite{PDG}, n}
\psfrag{Burkert}[r][r]{\scriptsize Burkert~\cite{Burkert}, p}
\psfrag{Gerhardt}[r][r]{\scriptsize Gerhardt~\cite{Gerhardt}, p}
\psfrag{Aznauryan}[r][r]{\scriptsize Aznauryan~\cite{Aznauryan05_1,Aznauryan05_2,Aznauryan2009}, p}
\psfrag{Aznauryan-fit}[r][r]{\scriptsize Fit: Aznauryan~\cite{Aznauryan2012}, p}
\psfrag{Ahrens}[r][r]{\scriptsize Ahrens~\cite{Ahrens}, p}
\psfrag{Maid}[r][r]{\scriptsize MAID~\cite{Drechsel,Tiator}, p}
\psfrag{Tiator}[r][r]{\scriptsize Fit: Tiator~\cite{Tiator2011}, p}
\psfrag{Merten-p}[r][r]{\scriptsize model $\mathcal A$, p}
\psfrag{Merten-n}[r][r]{\scriptsize model $\mathcal A$, n}
\psfrag{Gauss-p}[r][r]{\scriptsize model $\mathcal C$, p}
\psfrag{Gauss-n}[r][r]{\scriptsize model $\mathcal C$, n}
\includegraphics[width=\linewidth]{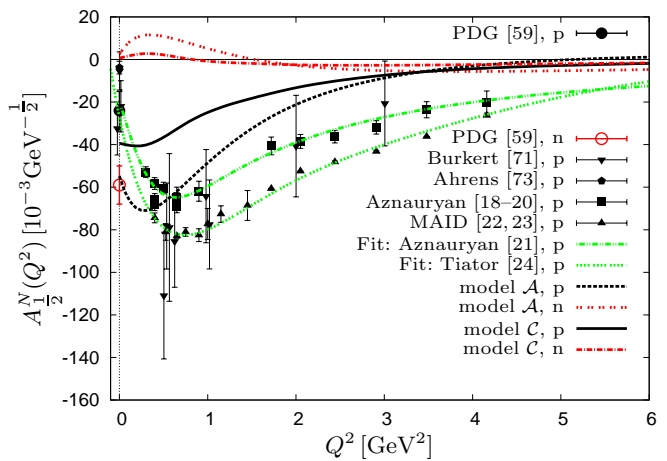}
\caption{
  Comparison of the $D_{13}(1520)$ transverse helicity amplitude $A_{1/2}^N$
  of proton and neutron calculated in model $\mathcal C$ (solid and dashed-dotted line) and
  model $\mathcal A$ (dashed lines). See also caption to Fig.~\ref{TRANSFF_A12_S11_1535}.
  \label{TRANSFF_A12_D13_1520}}
\end{figure}

\begin{figure}[ht!]
\centering
\psfrag{x-axis}[c][c]{$Q^2\,[\textrm{GeV}^2]$}
\psfrag{y-axis}[c][c]{$A_{\tfrac{3}{2}}^N(Q^2)\,[10^{-3}\textrm{GeV}^{-\tfrac{1}{2}}]$}
\psfrag{Anisovich}[r][r]{\scriptsize Anisovich~\cite{Anisovich_3}, p}
\psfrag{PDG-p}[r][r]{\scriptsize PDG~\cite{PDG}, p}
\psfrag{PDG-n}[r][r]{\scriptsize PDG~\cite{PDG}, n}
\psfrag{Burkert}[r][r]{\scriptsize Burkert~\cite{Burkert}, p}
\psfrag{Gerhardt}[r][r]{\scriptsize Gerhardt~\cite{Gerhardt}, p}
\psfrag{Aznauryan}[r][r]{\scriptsize Aznauryan~\cite{Aznauryan05_1,Aznauryan05_2,Aznauryan2009}, p}
\psfrag{Aznauryan-fit}[r][r]{\scriptsize Fit: Aznauryan~\cite{Aznauryan2012}, p}
\psfrag{Tiator}[r][r]{\scriptsize Fit: Tiator~\cite{Tiator2011}, p}
\psfrag{Ahrens}[r][r]{\scriptsize Ahrens~\cite{Ahrens}, p}
\psfrag{Maid}[r][r]{\scriptsize MAID~\cite{Drechsel,Tiator}, p}
\psfrag{Merten-p}[r][r]{\scriptsize model $\mathcal A$, p}
\psfrag{Merten-n}[r][r]{\scriptsize model $\mathcal A$, n}
\psfrag{Gauss-p}[r][r]{\scriptsize model $\mathcal C$, p}
\psfrag{Gauss-n}[r][r]{\scriptsize model $\mathcal C$, n}
\includegraphics[width=\linewidth]{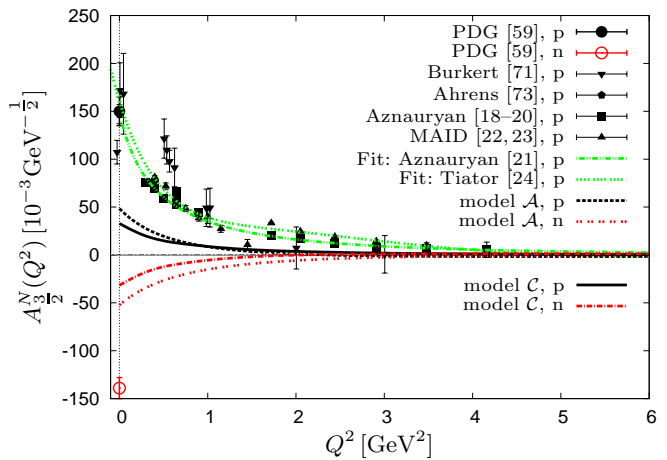}
\caption{
  Comparison of the $D_{13}(1520)$ transverse helicity amplitude $A_{3/2}^N$
  for proton and neutron calculated in model $\mathcal C$ (solid and dashed-dotted line) and
  model $\mathcal A$ (dashed lines). See also caption to Fig.~\ref{TRANSFF_A12_S11_1535}.
  \label{TRANSFF_A32_D13_1520}
}
\end{figure}

For the transverse helicity amplitude $A_{1/2}^p$ (see
Fig.~\ref{TRANSFF_A12_D13_1520}) of the $D_{13}(1520)$-resonance we find a
reasonable quantitative agreement with experimental data for low momentum
transfers, while apart from the fact that in model $\mathcal{C}$ the amplitude is too small by about
a factor of two the $Q^2$ dependence is reproduced up to $Q^2 \approx 6\,
\textnormal{GeV}^2$\,. The minimum at $Q^2 \approx 1\, \textnormal{GeV}^2$ is
clearly visible for model $\mathcal{A}$ whereas this feature is not so
pronounced in model $\mathcal{C}$\,.
The $A_{3/2}^p$-amplitudes are displayed in Fig.~\ref{TRANSFF_A32_D13_1520}; here both model
underestimates the data by more than a factor of three.
\begin{figure}[ht!]
\centering
\psfrag{x-axis}[c][c]{$Q^2\,[\textrm{GeV}^2]$}
\psfrag{y-axis}[c][c]{$S_{\tfrac{1}{2}}^N(Q^2)\,[10^{-3}\textrm{GeV}^{-\tfrac{1}{2}}]$}
\psfrag{Aznauryan}[r][r]{\scriptsize Aznauryan~\cite{Aznauryan05_1,Aznauryan05_2,Aznauryan2009}, p}
\psfrag{Maid}[r][r]{\scriptsize MAID~\cite{Drechsel,Tiator}, p}
\psfrag{Tiator}[r][r]{\scriptsize Fit: Tiator~\cite{Tiator2011}, p}
\psfrag{Kreuzer-p}[r][r]{\scriptsize model $\mathcal A$, p}
\psfrag{Kreuzer-n}[r][r]{\scriptsize model $\mathcal A$, n}
\psfrag{Gauss-p-S}[r][r]{\scriptsize model $\mathcal C$, p}
\psfrag{Gauss-n-S}[r][r]{\scriptsize model $\mathcal C$, n}
\includegraphics[width=\linewidth]{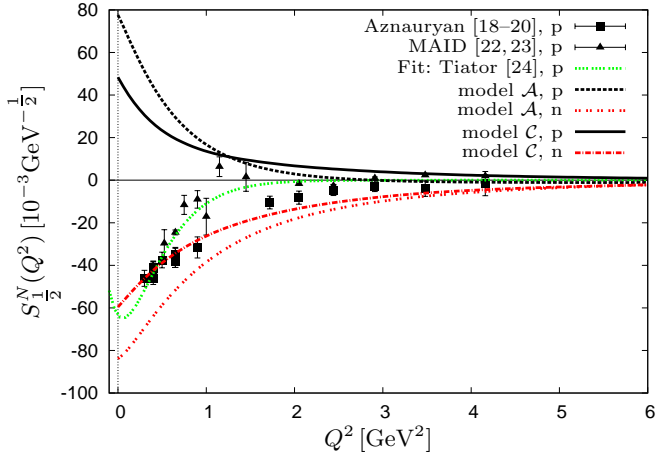}
\caption{
  Comparison of the $D_{13}(1520)$ longitudinal helicity amplitude $S_{1/2}^N$
  for proton and neutron calculated in model $\mathcal C$ (solid and dashed-dotted line) and
  model $\mathcal A$ (dashed lines). See also caption to Fig.~\ref{TRANSFF_A12_S11_1535}.
  \label{TRANSFF_S_D13_1520}
}
\end{figure}
Likewise the calculated neutron $A_{1/2}^n$- and $A_{3/2}^n$-amplitudes at the
photon point are too small. In particular for the $A_{1/2}^n$-amplitude the
predicted value close to zero is in contradiction to the 
experimental value $-59\pm9 \times 10^{-3}\textrm{ GeV}^{-1/2}$\, from PDG~\cite{PDG}.
Unfortunately, although the $Q^2$ dependence of the magnitude of the
longitudinal amplitude $S_{1/2}^p$\,, see Fig.~\ref{TRANSFF_S_D13_1520}\,,
would describe the experimental data of Aznauryan \textit{et
al.}~\cite{Aznauryan05_1,Aznauryan05_2,Aznauryan2009} and 
MAID~\cite{Drechsel} very well, the amplitude has the wrong sign. Note that
although, as mentioned above, the common phase $\zeta$ in the definition of
the helicity amplitudes is not determined in our framework, relative signs
beween the three helicity amplitudes are fixed.

\begin{figure}[ht!]
\centering
\psfrag{x-axis}[c][c]{$Q^2\,[\textrm{GeV}^2]$}
\psfrag{y-axis}[c][c]{$A_{\tfrac{1}{2}}^N(Q^2)\,[10^{-3}\textrm{GeV}^{-\tfrac{1}{2}}]$}
\psfrag{Anisovich}[r][r]{\scriptsize Anisovich~\cite{Anisovich_3}, p}
\psfrag{PDG-p}[r][r]{\scriptsize PDG~\cite{PDG}, p}
\psfrag{PDG-n}[r][r]{\scriptsize PDG~\cite{PDG}, n}
\psfrag{Aznauryan}[r][r]{\scriptsize Aznauryan~\cite{Aznauryan05_2}, p}
\psfrag{Merten-p}[r][r]{\scriptsize model $\mathcal A$, p}
\psfrag{Merten-n}[r][r]{\scriptsize model $\mathcal A$, n}
\psfrag{Gauss-p}[r][r]{\scriptsize model $\mathcal C$, p}
\psfrag{Gauss-n}[r][r]{\scriptsize model $\mathcal C$, n}
\includegraphics[width=\linewidth]{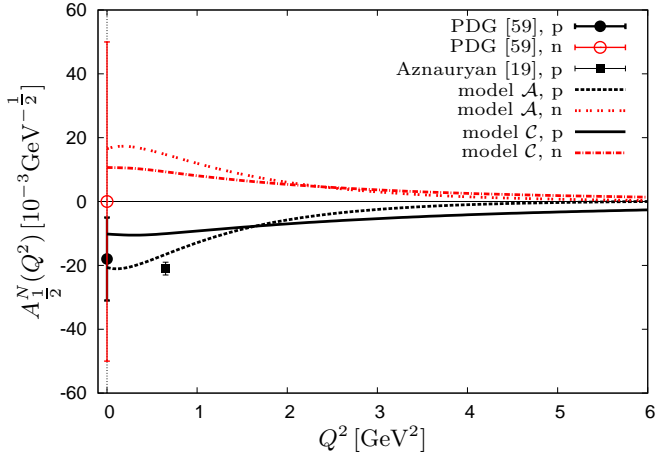}
\caption{
  Comparison of the $D_{13}(1700)$ transverse helicity amplitude $A_{1/2}^N$
  for proton and neutron calculated in model $\mathcal C$ (solid and dashed-dotted line) and
  model $\mathcal A$ (dashed lines). See also caption to Fig.~\ref{TRANSFF_A12_S11_1535}.
  \label{TRANSFF_A12_D13_1700}
}
\end{figure}
\begin{figure}[ht!]
\centering
\psfrag{x-axis}[c][c]{$Q^2\,[\textrm{GeV}^2]$}
\psfrag{y-axis}[c][c]{$A_{\tfrac{3}{2}}^N(Q^2)\,[10^{-3}\textrm{GeV}^{-\tfrac{1}{2}}]$}
\psfrag{Anisovich}[r][r]{\scriptsize Anisovich~\cite{Anisovich_3}, p}
\psfrag{PDG-p}[r][r]{\scriptsize PDG~\cite{PDG}, p}
\psfrag{PDG-n}[r][r]{\scriptsize PDG~\cite{PDG}, n}
\psfrag{Aznauryan}[r][r]{\scriptsize Aznauryan~\cite{Aznauryan05_2}, p}
\psfrag{Merten-p}[r][r]{\scriptsize model $\mathcal A$, p}
\psfrag{Merten-n}[r][r]{\scriptsize model $\mathcal A$, n}
\psfrag{Gauss-p}[r][r]{\scriptsize model $\mathcal C$, p}
\psfrag{Gauss-n}[r][r]{\scriptsize model $\mathcal C$, n}
\includegraphics[width=\linewidth]{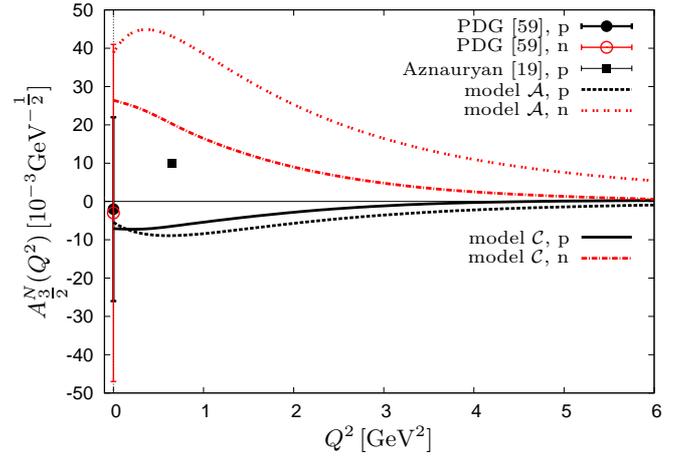}
\caption{
  Comparison of the $D_{13}(1700)$ transverse helicity amplitude $A_{3/2}^N$
  for proton and neutron calculated in model $\mathcal C$ (solid and dashed-dotted line) and
  model $\mathcal A$ (dashed lines). See also caption to Fig.~\ref{TRANSFF_A12_S11_1535}.
  \label{TRANSFF_A32_D13_1700}}
\end{figure}

The transverse amplitudes for the next ${3/2}^-$ nucleon resonance,
\textit{i.e.} $D_{13}(1700)$\,, are displayed in Figs.~\ref{TRANSFF_A12_D13_1700} 
and~\ref{TRANSFF_A32_D13_1700}.
In contrast to the situation for the $D_{13}(1520)$-resonance described above,
here both models are in accordance with the PDG-data~\cite{PDG} as well as with
the data from Aznauryan \textit{et al.}~\cite{Aznauryan05_2} for the
$A_{1/2}$-amplitude, whereas the $A_{3/2}$-amplitude only reproduces the PDG-data~\cite{PDG}
and not the data point from Aznauryan \textit{et al.}~\cite{Aznauryan05_2} at
finite momentum transfer.
\begin{figure}[ht!]
\centering
\psfrag{x-axis}[c][c]{$Q^2\,[\textrm{GeV}^2]$}
\psfrag{y-axis}[c][c]{$S_{\tfrac{1}{2}}^N(Q^2)\,[10^{-3}\textrm{GeV}^{-\tfrac{1}{2}}]$}
\psfrag{Merten-p}[r][r]{\scriptsize model $\mathcal A$, p}
\psfrag{Merten-n}[r][r]{\scriptsize model $\mathcal A$, n}
\psfrag{Gauss-p}[r][r]{\scriptsize model $\mathcal C$, p}
\psfrag{Gauss-n}[r][r]{\scriptsize model $\mathcal C$, n}
\includegraphics[width=\linewidth]{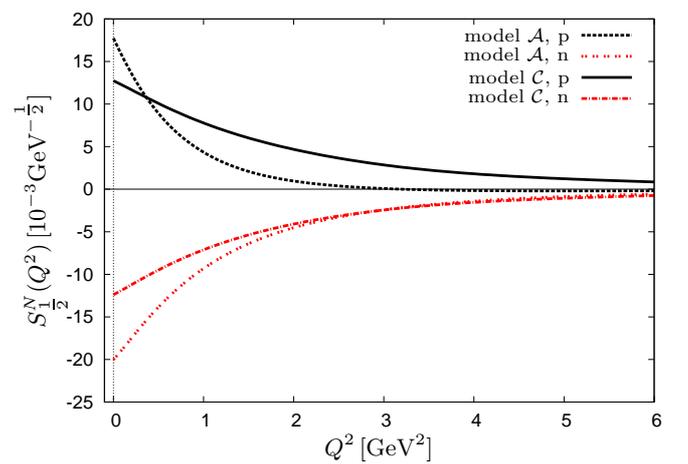}
\caption{
  Prediction of the $D_{13}(1700)$ longitudinal helicity amplitude $S_{1/2}^N$
  for proton and neutron calculated in model $\mathcal C$ (solid and dashed-dotted line) and model
  $\mathcal A$ (dashed lines). See also caption to Fig.~\ref{TRANSFF_A12_S11_1535}.
  \label{TRANSFF_S_D13_1700}}
\end{figure}
The prediction for the longitudinal $D_{13}(1700)$ amplitudes is given in
Fig.~\ref{TRANSFF_S_D13_1700}. The calculated amplitudes turn out to be rather small.

\paragraph{The $J=5/2$ resonances:}

\begin{figure}[ht!]
\centering
\psfrag{x-axis}[c][c]{$Q^2\,[\textrm{GeV}^2]$}
\psfrag{y-axis}[c][c]{$S_{\tfrac{1}{2}}^N(Q^2)\,[10^{-3}\textrm{GeV}^{-\tfrac{1}{2}}]$}
\psfrag{PDG-p}[r][r]{\scriptsize PDG~\cite{PDG}, p}
\psfrag{PDG-n}[r][r]{\scriptsize PDG~\cite{PDG}, n}
\psfrag{Burkert}[r][r]{\scriptsize Burkert~\cite{Burkert}, p}
\psfrag{Aznauryan}[r][r]{\scriptsize Aznauryan~\cite{Aznauryan05_1,Aznauryan05_2}, p}
\psfrag{Maid}[r][r]{\scriptsize MAID~\cite{Drechsel,Tiator}, p}
\psfrag{Tiator}[r][r]{\scriptsize Fit: Tiator~\cite{Tiator2011}, p}
\psfrag{Kreuzer-p}[r][r]{\scriptsize model $\mathcal A$, p}
\psfrag{Kreuzer-n}[r][r]{\scriptsize model $\mathcal A$, n}
\psfrag{Gauss-p}[r][r]{\scriptsize model $\mathcal C$, p}
\psfrag{Gauss-n}[r][r]{\scriptsize model $\mathcal C$, n}
\includegraphics[width=\linewidth]{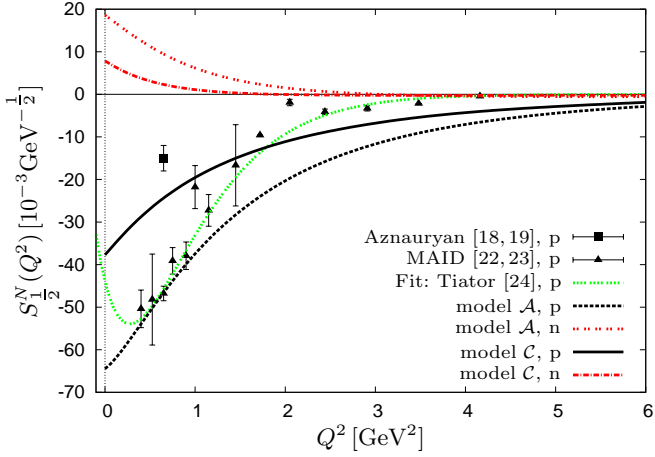}
\caption{
  Comparison of the $F_{15}(1680)$ longitudinal helicity amplitude $S_{1/2}^N$
  for proton and neutron calculated in model $\mathcal C$ (solid and dashed-dotted line) and
  model $\mathcal A$ (dashed lines). See also caption to Fig.~\ref{TRANSFF_A12_S11_1535}.
  \label{TRANSFF_S_F15_1680}
}
\end{figure}

Although the transverse\linebreak $D_{15}(1675)$ helicity amplitudes at the photon point
\begin{figure}[ht!]
\centering
\psfrag{x-axis}[c][c]{$Q^2\,[\textrm{GeV}^2]$}
\psfrag{y-axis}[c][c]{$A_{\tfrac{1}{2}}^N(Q^2)\,[10^{-3}\textrm{GeV}^{-\tfrac{1}{2}}]$}
\psfrag{Anisovich}[r][r]{\scriptsize Anisovich~\cite{Anisovich_3}, p}
\psfrag{PDG-p}[r][r]{\scriptsize PDG~\cite{PDG}, p}
\psfrag{PDG-n}[r][r]{\scriptsize PDG~\cite{PDG}, n}
\psfrag{Park}[r][r]{\scriptsize MAID~\cite{Tiator}, p}
\psfrag{Tiator}[r][r]{\scriptsize Fit: Tiator~\cite{Tiator2011}, p}
\psfrag{Merten-p}[r][r]{\scriptsize model $\mathcal A$, p}
\psfrag{Merten-n}[r][r]{\scriptsize model $\mathcal A$, n}
\psfrag{Gauss-p}[r][r]{\scriptsize model $\mathcal C$, p}
\psfrag{Gauss-n}[r][r]{\scriptsize model $\mathcal C$, n}
\includegraphics[width=\linewidth]{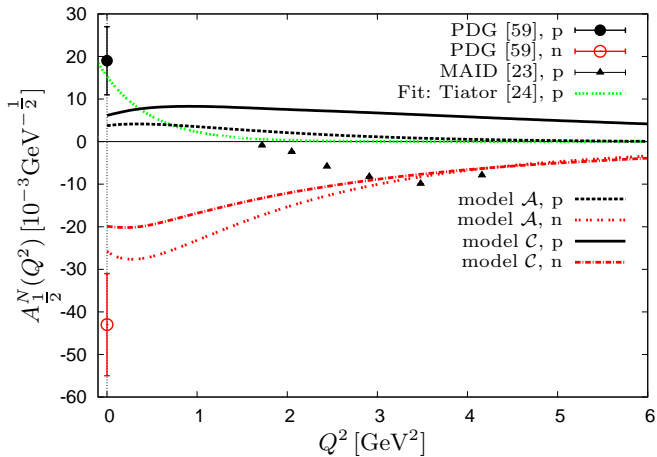}
\caption{
  Comparison of the $D_{15}(1675)$ transverse helicity amplitude $A_{1/2}^N$
  for proton and neutron calculated in model $\mathcal C$ (solid and dashed-dotted line) and
  model $\mathcal A$ (dashed lines). See also caption to Fig.~\ref{TRANSFF_A12_S11_1535}.
  \label{TRANSFF_A12_D15_1675}
}
\end{figure}
\begin{figure}[ht!]
\centering
\psfrag{x-axis}[c][c]{$Q^2\,[\textrm{GeV}^2]$}
\psfrag{y-axis}[c][c]{$A_{\tfrac{3}{2}}^N(Q^2)\,[10^{-3}\textrm{GeV}^{-\tfrac{1}{2}}]$}
\psfrag{Anisovich}[r][r]{\scriptsize Anisovich~\cite{Anisovich_3}, p}
\psfrag{PDG-p}[r][r]{\scriptsize PDG~\cite{PDG}, p}
\psfrag{PDG-n}[r][r]{\scriptsize PDG~\cite{PDG}, n}
\psfrag{Park}[r][r]{\scriptsize MAID~\cite{Tiator}, p}
\psfrag{Tiator}[r][r]{\scriptsize Fit: Tiator~\cite{Tiator2011}, p}
\psfrag{Merten-p}[r][r]{\scriptsize model $\mathcal A$, p}
\psfrag{Merten-n}[r][r]{\scriptsize model $\mathcal A$, n}
\psfrag{Gauss-p}[r][r]{\scriptsize model $\mathcal C$, p}
\psfrag{Gauss-n}[r][r]{\scriptsize model $\mathcal C$, n}
\includegraphics[width=\linewidth]{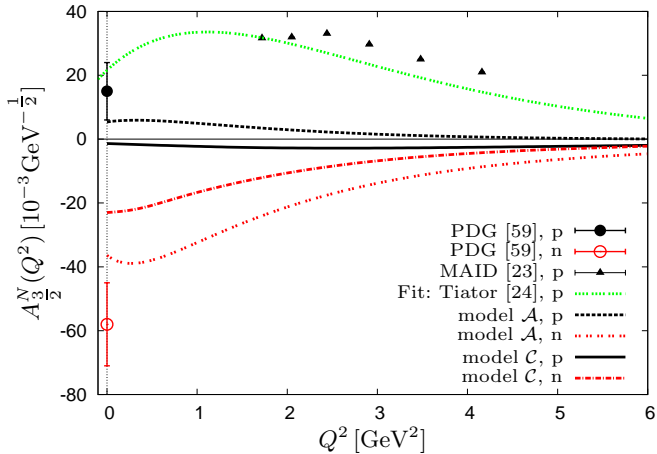}
\caption{Comparison of the $D_{15}(1675)$ transverse electro-excitation helicity
  amplitude $A_{3/2}^N$ for proton and neutron calculated in model $\mathcal C$
  (solid and dashed-dotted line) and model $\mathcal A$ (dashed lines). See also caption to
  Fig.~\ref{TRANSFF_A12_S11_1535}.\label{TRANSFF_A32_D15_1675}}
\end{figure}
reproduce the experimental data from MAID~\cite{Drechsel,Tiator} and the
PDG~\cite{PDG} rather well, as displayed in Figs.~\ref{TRANSFF_A12_D15_1675}
and~\ref{TRANSFF_A32_D15_1675}, both calculations cannot account for the
apparent zero of the experimental $A_{1/2}^p$-amplitude at $Q^2 \approx 1.5\,
\textnormal{GeV}^2$\,. Furthermore the $A_{3/2}^p$-amplitude, displayed in
Fig.~\ref{TRANSFF_A32_D15_1675} is severely underestimated in magnitude by
both models and model $\mathcal C$ even yields the wrong sign. The
transverse amplitudes for the neutron are predicted to be negative, here the
calculated value at the photon point for model $\mathcal A$ is closer to the
experimental value than for model $\mathcal C$\,.
\begin{figure}[ht!]
\centering
\psfrag{x-axis}[c][c]{$Q^2\,[\textrm{GeV}^2]$}
\psfrag{y-axis}[c][c]{$S_{\tfrac{1}{2}}^N(Q^2)\,[10^{-3}\textrm{GeV}^{-\tfrac{1}{2}}]$}
\psfrag{PDG-p}[r][r]{\scriptsize PDG~\cite{PDG}, p}
\psfrag{PDG-n}[r][r]{\scriptsize PDG~\cite{PDG}, n}
\psfrag{Park}[r][r]{\scriptsize MAID~\cite{Tiator}, p}
\psfrag{Tiator}[r][r]{\scriptsize Fit: Tiator~\cite{Tiator2011}, p}
\psfrag{Merten-p}[r][r]{\scriptsize model $\mathcal A$, p}
\psfrag{Merten-n}[r][r]{\scriptsize model $\mathcal A$, n}
\psfrag{Gauss-p}[r][r]{\scriptsize model $\mathcal C$, p}
\psfrag{Gauss-n}[r][r]{\scriptsize model $\mathcal C$, n}
\includegraphics[width=\linewidth]{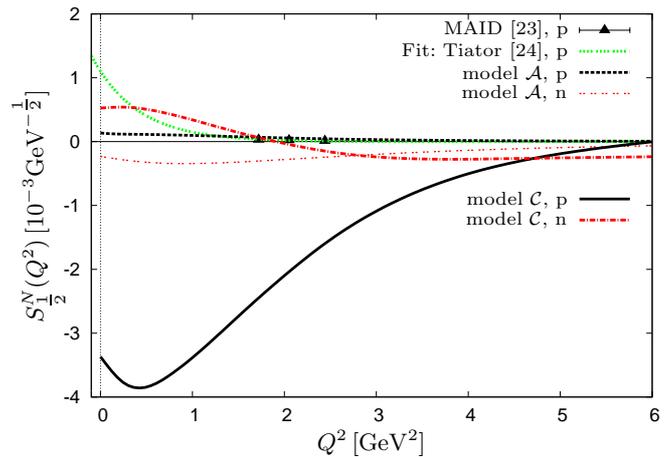}
\caption{
  Comparison of the $D_{15}(1675)$ longitudinal helicity amplitude $S_{1/2}^N$
  for proton and neutron calculated in model $\mathcal C$ (solid and dashed-dotted line) and
  model $\mathcal A$ (dashed lines). See also caption to Fig.~\ref{TRANSFF_A12_S11_1535}.
  \label{TRANSFF_S_D15_1675}
}
\end{figure}
The longitudinal amplitudes are calculated to be very small for both models. 
There exists only experimental data from the MAID-analysis~\cite{Tiator},
indicating that the experimental values are consistent with zero (see
Fig.~\ref{TRANSFF_S_F15_1680}).

\begin{figure}[ht!]
\centering
\psfrag{x-axis}[c][c]{$Q^2\,[\textrm{GeV}^2]$}
\psfrag{y-axis}[c][c]{$A_{\tfrac{1}{2}}^N(Q^2)\,[10^{-3}\textrm{GeV}^{-\tfrac{1}{2}}]$}
\psfrag{Anisovich}[r][r]{\scriptsize Anisovich~\cite{Anisovich_3}, p}
\psfrag{PDG-p}[r][r]{\scriptsize PDG~\cite{PDG}, p}
\psfrag{PDG-n}[r][r]{\scriptsize PDG~\cite{PDG}, n}
\psfrag{Burkert}[r][r]{\scriptsize Burkert~\cite{Burkert}, p}
\psfrag{Aznauryan}[r][r]{\scriptsize Aznauryan~\cite{Aznauryan05_1,Aznauryan05_2}, p}
\psfrag{Maid}[r][r]{\scriptsize MAID~\cite{Drechsel,Tiator}, p}
\psfrag{Tiator}[r][r]{\scriptsize Fit: Tiator~\cite{Tiator2011}, p}
\psfrag{Merten-p}[r][r]{\scriptsize model $\mathcal A$, p}
\psfrag{Merten-n}[r][r]{\scriptsize model $\mathcal A$, n}
\psfrag{Gauss-p}[r][r]{\scriptsize model $\mathcal C$, p}
\psfrag{Gauss-n}[r][r]{\scriptsize model $\mathcal C$, n}
\includegraphics[width=\linewidth]{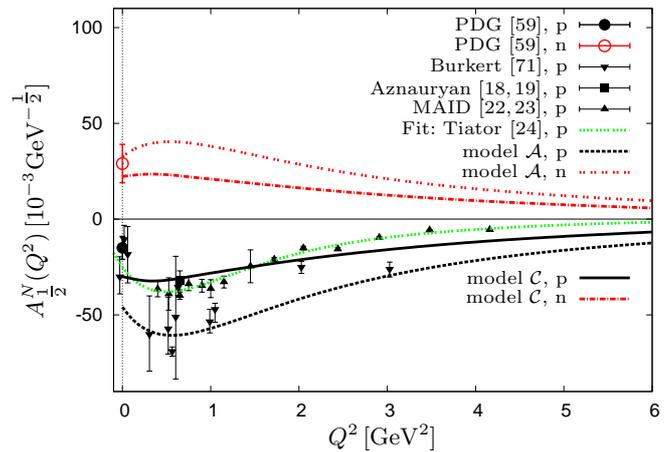}
\caption{
  Comparison of the $F_{15}(1680)$ transverse helicity amplitude $A_{1/2}^N$
  for proton and neutron calculated in model $\mathcal C$ (solid and dashed-dotted line) and
  model $\mathcal A$ (dashed lines). See also caption to Fig.~\ref{TRANSFF_A12_S11_1535}.
  \label{TRANSFF_A12_F15_1680}}
\end{figure}

\begin{figure}[ht!]
\centering
\psfrag{x-axis}[c][c]{$Q^2\,[\textrm{GeV}^2]$}
\psfrag{y-axis}[c][c]{$A_{\tfrac{3}{2}}^N(Q^2)\,[10^{-3}\textrm{GeV}^{-\tfrac{1}{2}}]$}
\psfrag{Anisovich}[r][r]{\scriptsize Anisovich~\cite{Anisovich_3}, p}
\psfrag{PDG-p}[r][r]{\scriptsize PDG~\cite{PDG}, p}
\psfrag{PDG-n}[r][r]{\scriptsize PDG~\cite{PDG}, n}
\psfrag{Burkert}[r][r]{\scriptsize Burkert~\cite{Burkert}, p}
\psfrag{Aznauryan}[r][r]{\scriptsize Aznauryan~\cite{Aznauryan05_1,Aznauryan05_2}, p}
\psfrag{Maid}[r][r]{\scriptsize MAID~\cite{Drechsel,Tiator}, p}
\psfrag{Tiator}[r][r]{\scriptsize Fit: Tiator~\cite{Tiator2011}, p}
\psfrag{Merten-p}[r][r]{\scriptsize model $\mathcal A$, p}
\psfrag{Merten-n}[r][r]{\scriptsize model $\mathcal A$, n}
\psfrag{Kreuzer-p}[r][r]{\scriptsize model $\mathcal A$, p}
\psfrag{Kreuzer-n}[r][r]{\scriptsize model $\mathcal A$, n}
\psfrag{Gauss-p}[r][r]{\scriptsize model $\mathcal C$, p}
\psfrag{Gauss-n}[r][r]{\scriptsize model $\mathcal C$, n}
\includegraphics[width=\linewidth]{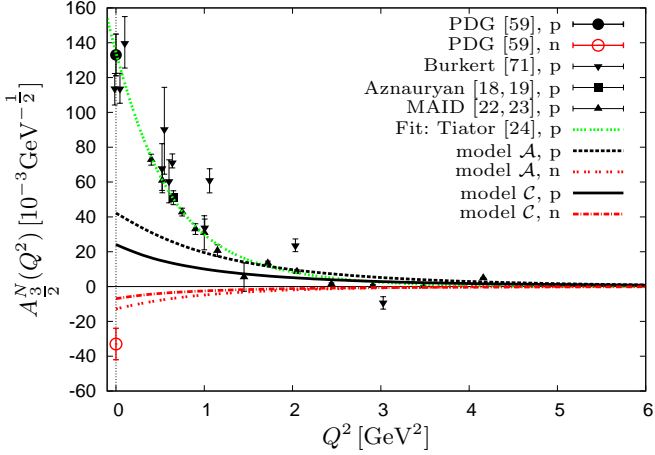}
\caption{
  Comparison of the $F_{15}(1680)$ transverse helicity amplitude $A_{3/2}^N$
  for proton and neutron calculated in model $\mathcal C$ (solid and dashed-dotted line) and
  model $\mathcal A$ (dashed line). See also caption to Fig.~\ref{TRANSFF_A12_S11_1535}.
  \label{TRANSFF_A32_F15_1680}}
\end{figure}

There also exist data for the helicity amplitudes of the
$F_{15}(1680)$-resonance. The comparison with the calculated values is given in 
Figs.~\ref{TRANSFF_A12_F15_1680} and~\ref{TRANSFF_A32_F15_1680}.
In particular in model $\mathcal{C}$ a reasonable description of the
$A_{1/2}^p$ amplitudes is found for the newer data from Aznauryan \textit{et
  al.}~\cite{Aznauryan05_1,Aznauryan05_2} and MAID~\cite{Drechsel,Tiator} 
both at the photon point and for the values at higher momentum transfers. 
The calculated values in model $\mathcal{A}$ are in better accordance with the
the older data from Burkert \textit{et al.}~\cite{Burkert} which are larger in
magnitude.
In contrast to this the $A_{3/2}^p$-amplitudes are again severely underestimated in
magnitude, see Fig.~\ref{TRANSFF_A32_F15_1680}\,.
In contrast, for the longitudinal $S_{1/2}^p$-amplitude we observe a rather good agreement
with the data as displayed in Fig.~\ref{TRANSFF_S_F15_1680}, the values
obtained in  model $\mathcal C$ being too small at lower momentum transfers.

\begin{figure}[ht!]
\centering
\psfrag{x-axis}[c][c]{$Q^2\,[\textrm{GeV}^2]$}
\psfrag{y-axis}[c][c]{$A_{\tfrac{1}{2}}^N/A_{\tfrac{3}{2}}^N(Q^2)\,[10^{-3}\textrm{GeV}^{-\tfrac{1}{2}}]$}
\psfrag{Anisovich}[r][r]{\scriptsize Anisovich~\cite{Anisovich_3}, p}
\psfrag{Merten-M2-p}[r][r]{\scriptsize model $\mathcal A$, p $A_{1/2}$}
\psfrag{Merten-M2-n}[r][r]{\scriptsize model $\mathcal A$, n $A_{1/2}$}
\psfrag{Merten-M4-p}[r][r]{\scriptsize model $\mathcal A$, p $A_{3/2}$}
\psfrag{Merten-M4-n}[r][r]{\scriptsize model $\mathcal A$, n $A_{3/2}$}
\psfrag{Gauss-M2-p}[r][r]{\scriptsize model $\mathcal C$, p $A_{1/2}$}
\psfrag{Gauss-M2-n}[r][r]{\scriptsize model $\mathcal C$, n $A_{1/2}$}
\psfrag{Gauss-M4-p}[r][r]{\scriptsize model $\mathcal C$, p $A_{3/2}$}
\psfrag{Gauss-M4-n}[r][r]{\scriptsize model $\mathcal C$, n $A_{3/2}$}
\includegraphics[width=\linewidth]{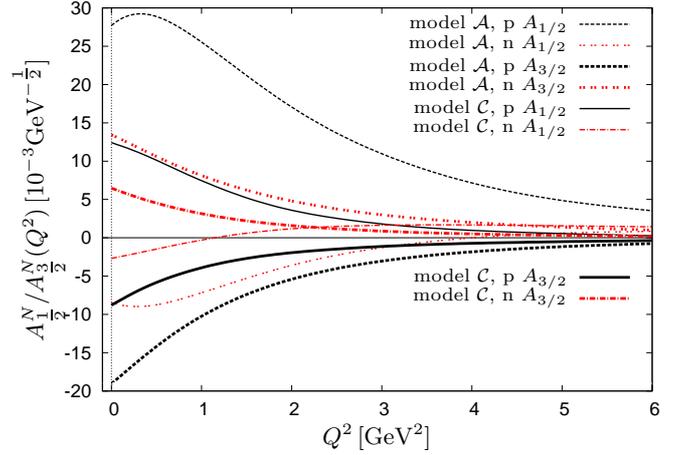}
\caption{
  Prediction of the $G_{17}(2190)$ transverse helicity amplitudes $A_{1/2}^N$ and
  $A_{3/2}^N$ for proton and neutron calculated in model $\mathcal C$ (solid and dashed-dotted line)
  and model $\mathcal A$ (dashed lines). See also caption to Fig.~\ref{TRANSFF_A12_S11_1535}.
  \label{TRANSFF_A12_G17_2190}
}
\end{figure}

\begin{figure}[ht!]
\centering
\psfrag{x-axis}[c][c]{$Q^2\,[\textrm{GeV}^2]$}
\psfrag{y-axis}[c][c]{$S_{\tfrac{1}{2}}^N(Q^2)\,[10^{-3}\textrm{GeV}^{-\tfrac{1}{2}}]$}
\psfrag{PDG-p}[r][r]{\scriptsize PDG~\cite{PDG}, p}
\psfrag{PDG-n}[r][r]{\scriptsize PDG~\cite{PDG}, n}
\psfrag{Burkert}[r][r]{\scriptsize Burkert~\cite{Burkert}, p}
\psfrag{Aznauryan}[r][r]{\scriptsize Aznauryan~\cite{Aznauryan05_1,Aznauryan05_2}, p}
\psfrag{Maid}[r][r]{\scriptsize MAID~\cite{Drechsel,Tiator}, p}
\psfrag{Merten-p}[r][r]{\scriptsize model $\mathcal A$, p}
\psfrag{Merten-n}[r][r]{\scriptsize model $\mathcal A$, n}
\psfrag{Gauss-p}[r][r]{\scriptsize model $\mathcal C$, p}
\psfrag{Gauss-n}[r][r]{\scriptsize model $\mathcal C$, n}
\includegraphics[width=\linewidth]{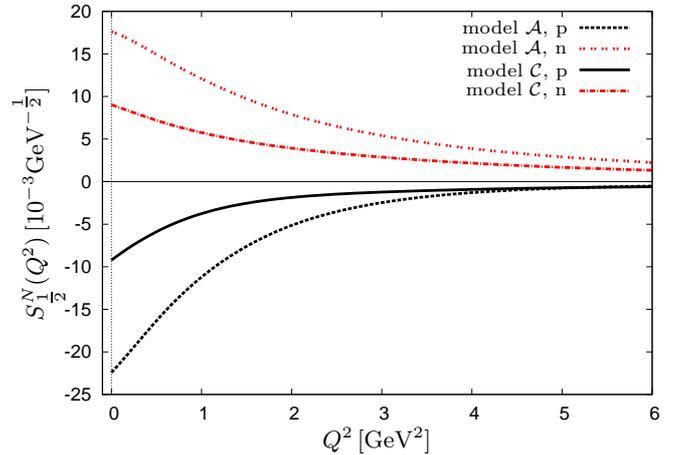}
\caption{
  Prediction of the $G_{17}(2190)$ longitudinal helicity amplitude $S_{1/2}^N$
  for proton and neutron calculated in model $\mathcal C$ (solid and dashed-dotted line) and
  model $\mathcal A$ (dashed lines). See also caption to Fig.~\ref{TRANSFF_A12_S11_1535}.
  \label{TRANSFF_S_G17_2190}
}
\end{figure}

\paragraph{The $J=7/2$ resonances:}
For positive parity PDG~\cite{PDG} lists the $F_{17}(1990)$ resonance rated
with two stars. Both in model $\mathcal{A}$ and in model $\mathcal{C}$ we
can relate this to states with a calculated mass of 1954 MeV and 1997 MeV,
respectively. The corresponding photon amplitudes are very small, see
Table~\ref{tab:PhotonCoupl1}. Otherwise, concerning the
$J=7/2$ resonances there exists only one negative parity resonance with more than at
least a three star rating, the $G_{17}(2190)$. The corresponding predictions for
transverse and longitudinal helicity amplitudes are shown in Figs.~\ref{TRANSFF_A12_G17_2190}
and~\ref{TRANSFF_S_G17_2190}.

\paragraph{The $J=9/2$ resonances:}

\begin{figure}[ht!]
\centering
\psfrag{x-axis}[c][c]{$Q^2\,[\textrm{GeV}^2]$}
\psfrag{y-axis}[c][c]{$A_{\tfrac{1}{2}}^N/A_{\tfrac{3}{2}}^N(Q^2)\,[10^{-3}\textrm{GeV}^{-\tfrac{1}{2}}]$}
\psfrag{Anisovich}[r][r]{\scriptsize Anisovich~\cite{Anisovich_3}, p}
\psfrag{Merten-M2-p}[r][r]{\scriptsize model $\mathcal A$, p $A_{1/2}$}
\psfrag{Merten-M2-n}[r][r]{\scriptsize model $\mathcal A$, n $A_{1/2}$}
\psfrag{Merten-M4-p}[r][r]{\scriptsize model $\mathcal A$, p $A_{3/2}$}
\psfrag{Merten-M4-n}[r][r]{\scriptsize model $\mathcal A$, n $A_{3/2}$}
\psfrag{Gauss-M2-p}[r][r]{\scriptsize model $\mathcal C$, p $A_{1/2}$}
\psfrag{Gauss-M2-n}[r][r]{\scriptsize model $\mathcal C$, n $A_{1/2}$}
\psfrag{Gauss-M4-p}[r][r]{\scriptsize model $\mathcal C$, p $A_{3/2}$}
\psfrag{Gauss-M4-n}[r][r]{\scriptsize model $\mathcal C$, n $A_{3/2}$}
\includegraphics[width=\linewidth]{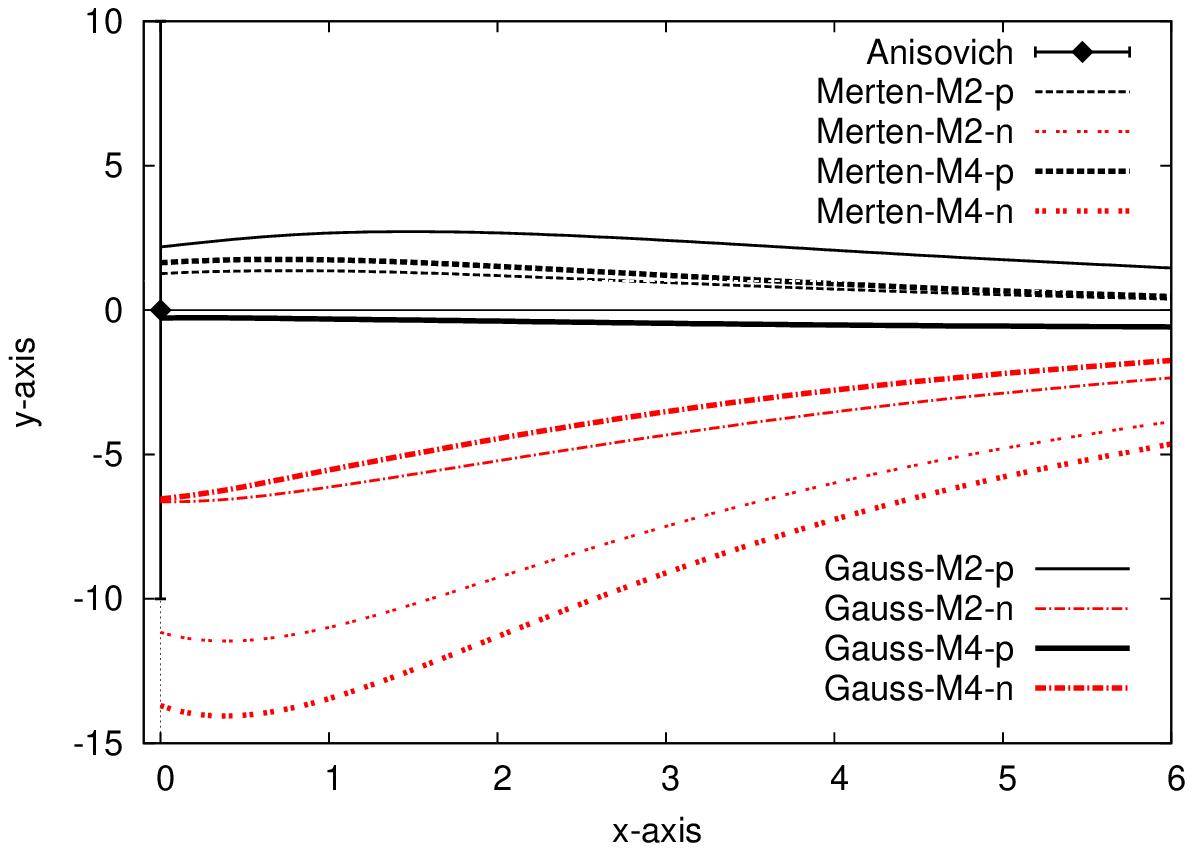}
\caption{
  Comparison of the $G_{19}(2250)$ transverse helicity amplitudes
  $A_{1/2}^N$ and $A_{3/2}^N$ for proton and neutron calculated in model $\mathcal C$
  (solid and dashed-dotted line) and model $\mathcal A$ (dashed lines). The error bar at the photon point
  corresponds to an estimate by Anisovich \textit{et al.}~\cite{Anisovich_3}
  for$A_{1/2}^N$ and $A_{3/2}^N$ within $|A^p|<10\times10^{-3}\,\textrm{GeV}^{-\nicefrac{1}{2}}$\,.
  See also caption to Fig.~\ref{TRANSFF_A12_S11_1535}.\label{TRANSFF_A12_G19_2250}
}
\end{figure}

\begin{figure}[ht!]
\centering
\psfrag{x-axis}[c][c]{$Q^2\,[\textrm{GeV}^2]$}
\psfrag{y-axis}[c][c]{$S_{\tfrac{1}{2}}^N(Q^2)\,[10^{-3}\textrm{GeV}^{-\tfrac{1}{2}}]$}
\psfrag{Merten-p}[r][r]{\scriptsize model $\mathcal A$, p}
\psfrag{Merten-n}[r][r]{\scriptsize model $\mathcal A$, n}
\psfrag{Gauss-p}[r][r]{\scriptsize model $\mathcal C$, p}
\psfrag{Gauss-n}[r][r]{\scriptsize model $\mathcal C$, n}
\includegraphics[width=\linewidth]{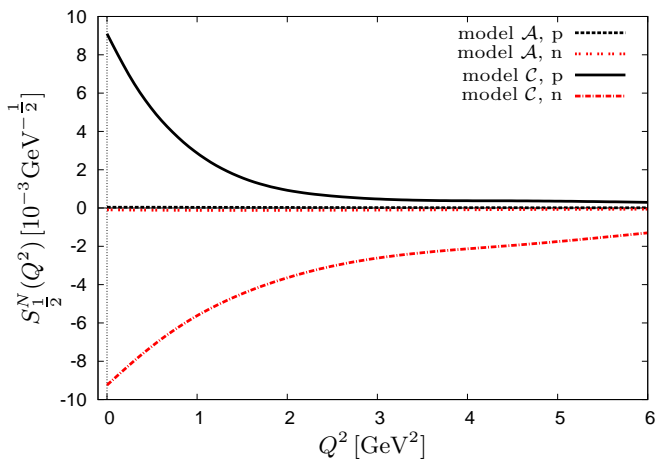}
\caption{
  Prediction of the $G_{19}(2250)$ longitudinal helicity amplitude $S_{1/2}^N$
  for proton and neutron calculated in model $\mathcal C$ (solid and dashed-dotted line) and
  model $\mathcal A$ (dashed lines). See also caption to Fig.~\ref{TRANSFF_A12_S11_1535}.
  \label{TRANSFF_S_G19_2250}
}
\end{figure}

The transverse and longitudinal helicity amplitudes of the $J^\pi=9/2^+$ resonance
$G_{19}(2250)$ are predicted to be very small as shown in Figs.~\ref{TRANSFF_A12_G19_2250}
and~\ref{TRANSFF_S_G19_2250} and coincide with the estimate by Anisovich
\textit{et al.}~\cite{Anisovich_3} for the transverse amplitudes. Obviously,
the $A^p_{\nicefrac{3}{2}}$ amplitude of model $\mathcal C$ and the longitudinal
amplitudes of model $\mathcal A$ are effectively zero.

\begin{figure}[ht!]
\centering
\psfrag{x-axis}[c][c]{$Q^2\,[\textrm{GeV}^2]$}
\psfrag{y-axis}[c][c]{$A_{\tfrac{1}{2}}^N/A_{\tfrac{3}{2}}^N(Q^2)\,[10^{-3}\textrm{GeV}^{-\tfrac{1}{2}}]$}
\psfrag{Anisovich}[r][r]{\scriptsize Anisovich~\cite{Anisovich_3}, p}
\psfrag{Merten-M2-p}[r][r]{\scriptsize model $\mathcal A$, p $A_{1/2}$}
\psfrag{Merten-M2-n}[r][r]{\scriptsize model $\mathcal A$, n $A_{1/2}$}
\psfrag{Merten-M4-p}[r][r]{\scriptsize model $\mathcal A$, p $A_{3/2}$}
\psfrag{Merten-M4-n}[r][r]{\scriptsize model $\mathcal A$, n $A_{3/2}$}
\psfrag{Gauss-M2-p}[r][r]{\scriptsize model $\mathcal C$, p $A_{1/2}$}
\psfrag{Gauss-M2-n}[r][r]{\scriptsize model $\mathcal C$, n $A_{1/2}$}
\psfrag{Gauss-M4-p}[r][r]{\scriptsize model $\mathcal C$, p $A_{3/2}$}
\psfrag{Gauss-M4-n}[r][r]{\scriptsize model $\mathcal C$, n $A_{3/2}$}
\includegraphics[width=\linewidth]{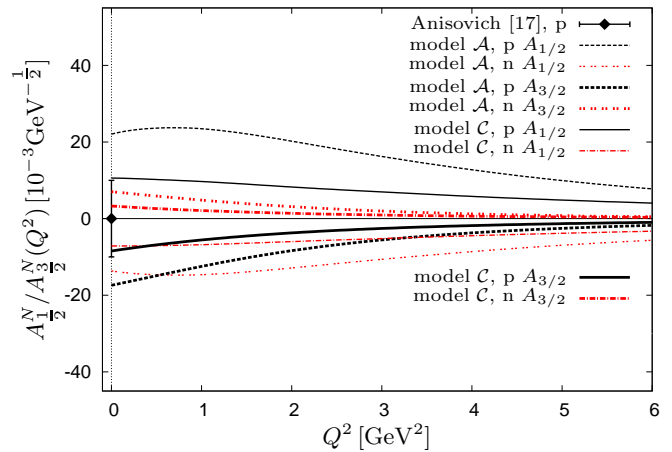}
\caption{
  Comparison of the $H_{19}(2220)$ transverse helicity amplitudes
  $A_{1/2}^N$ and $A_{3/2}^N$ for proton and neutron calculated in model $\mathcal C$
  (solid and dashed-dotted lines) and model $\mathcal A$ (dashed lines). The error bar at the photon point
  corresponds to an estimate by Anisovich \textit{et al.}~\cite{Anisovich_3}
  for$A_{1/2}^N$ and $A_{3/2}^N$ within $|A^p|<10\times10^{-3}\,\textrm{GeV}^{-\nicefrac{1}{2}}$\,.
  See also caption to Fig.~\ref{TRANSFF_A12_S11_1535}.\label{TRANSFF_A12_H19_2220}
}
\end{figure}

\begin{figure}[ht!]
\centering
\psfrag{x-axis}[c][c]{$Q^2\,[\textrm{GeV}^2]$}
\psfrag{y-axis}[c][c]{$S_{\tfrac{1}{2}}^N(Q^2)\,[10^{-3}\textrm{GeV}^{-\tfrac{1}{2}}]$}
\psfrag{Merten-p}[r][r]{\scriptsize model $\mathcal A$, p}
\psfrag{Merten-n}[r][r]{\scriptsize model $\mathcal A$, n}
\psfrag{Gauss-p}[r][r]{\scriptsize model $\mathcal C$, p}
\psfrag{Gauss-n}[r][r]{\scriptsize model $\mathcal C$, n}
\includegraphics[width=\linewidth]{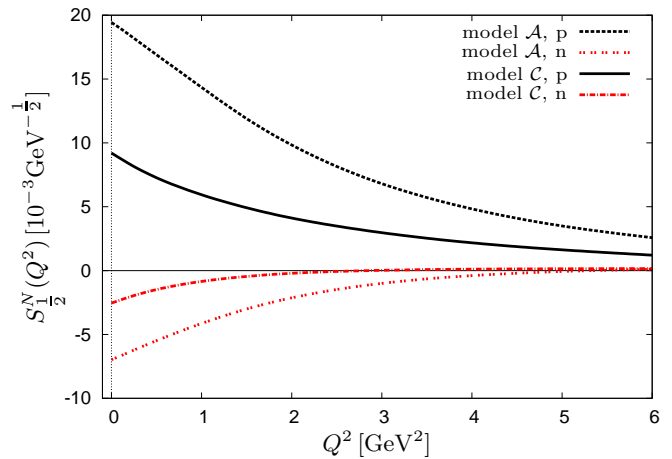}
\caption{
  Prediction of the $H_{19}(2220)$ longitudinal helicity amplitude $S_{1/2}^N$
  for proton and neutron calculated in model $\mathcal C$ (solid and dashed-dotted line) and
  model $\mathcal A$ (dashed lines). See also caption to Fig.~\ref{TRANSFF_A12_S11_1535}.
  \label{TRANSFF_S_H19_2220}
}
\end{figure}

Although the resonance with $J^\pi=9/2^-$\,, $H_{19}(2220)$ has a four star
rating by the PDG, only the proton photon decay amplitude has been estimated
in~\cite{Anisovich_3}. The calculated values are displayed in Fig.~\ref{TRANSFF_A12_H19_2220}
and Fig.~\ref{TRANSFF_S_H19_2220}\,; 
the amplitudes turn out to be smaller in model $\mathcal{C}$ than in
model $\mathcal{A}$ in better agreement with the estimate of~\cite{Anisovich_3}.

\begin{figure}[ht!]
\centering
\psfrag{x-axis}[c][c]{$Q^2\,[\textrm{GeV}^2]$}
\psfrag{y-axis}[c][c]{$A_{\tfrac{1}{2}}^N(Q^2)/A_{\tfrac{3}{2}}^N(Q^2)\,[10^{-3}\textrm{GeV}^{-\tfrac{1}{2}}]$}
\psfrag{Merten-M2-p}[r][r]{\scriptsize model $\mathcal A$, p $A_{1/2}$}
\psfrag{Merten-M2-n}[r][r]{\scriptsize model $\mathcal A$, n $A_{1/2}$}
\psfrag{Merten-M4-p}[r][r]{\scriptsize model $\mathcal A$, p $A_{3/2}$}
\psfrag{Merten-M4-n}[r][r]{\scriptsize model $\mathcal A$, n $A_{3/2}$}
\psfrag{Gauss-M2-p}[r][r]{\scriptsize model $\mathcal C$, p $A_{1/2}$}
\psfrag{Gauss-M2-n}[r][r]{\scriptsize model $\mathcal C$, n $A_{1/2}$}
\psfrag{Gauss-M4-p}[r][r]{\scriptsize model $\mathcal C$, p $A_{3/2}$}
\psfrag{Gauss-M4-n}[r][r]{\scriptsize model $\mathcal C$, n $A_{3/2}$}
\includegraphics[width=\linewidth]{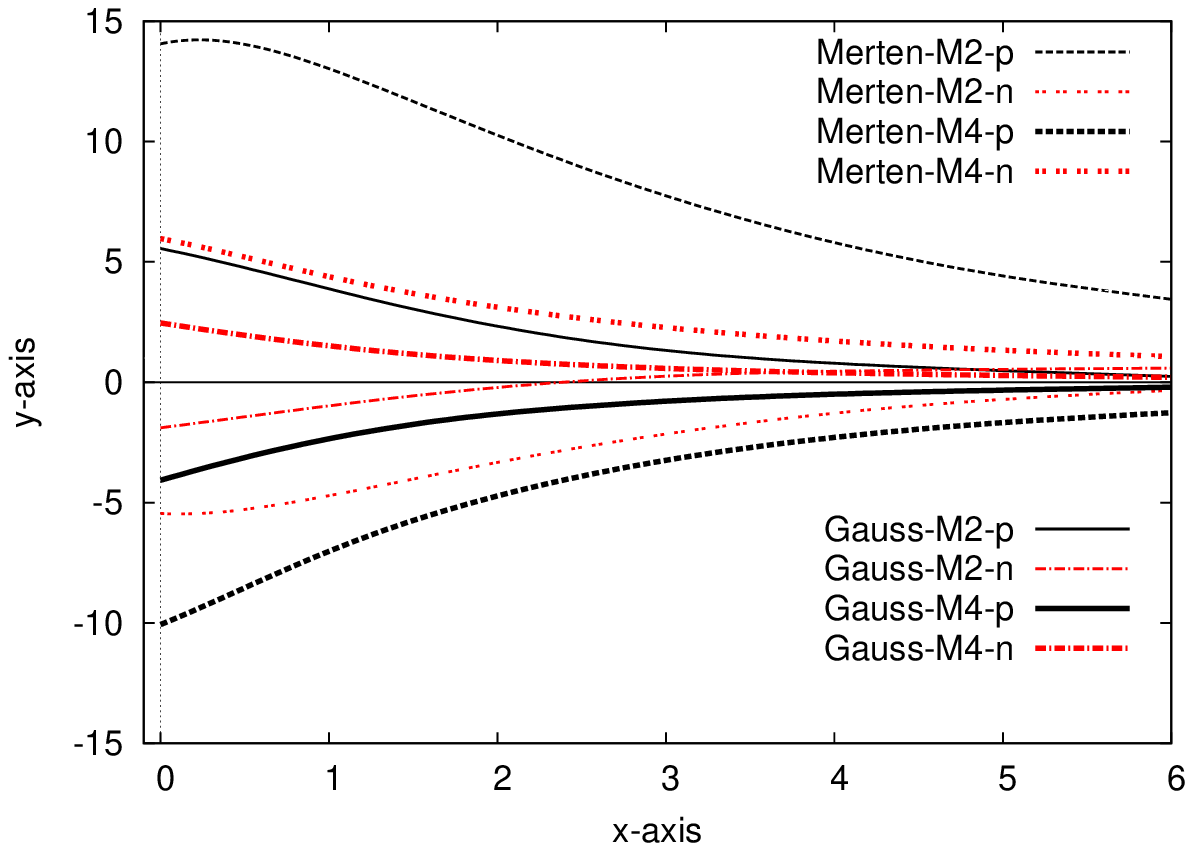}
\caption{
  Prediction of the $I_{1\,11}(2600)$ transverse helicity amplitudes $A_{1/2}^N$ and
  $A_{3/2}^N$ for proton and neutron calculated in model $\mathcal C$ (solid and dashed-dotted line)
  and model $\mathcal A$ (dashed lines). See also caption to Fig.~\ref{TRANSFF_A12_S11_1535}.
  \label{TRANSFF_A12_I111_2600}
}
\end{figure}

\begin{figure}[ht!]
\centering
\psfrag{x-axis}[c][c]{$Q^2\,[\textrm{GeV}^2]$}
\psfrag{y-axis}[c][c]{$S_{\tfrac{1}{2}}^N(Q^2)\,[10^{-3}\textrm{GeV}^{-\tfrac{1}{2}}]$}
\psfrag{Merten-p}[r][r]{\scriptsize model $\mathcal A$, p}
\psfrag{Merten-n}[r][r]{\scriptsize model $\mathcal A$, n}
\psfrag{Gauss-p}[r][r]{\scriptsize model $\mathcal C$, p}
\psfrag{Gauss-n}[r][r]{\scriptsize model $\mathcal C$, n}
\includegraphics[width=\linewidth]{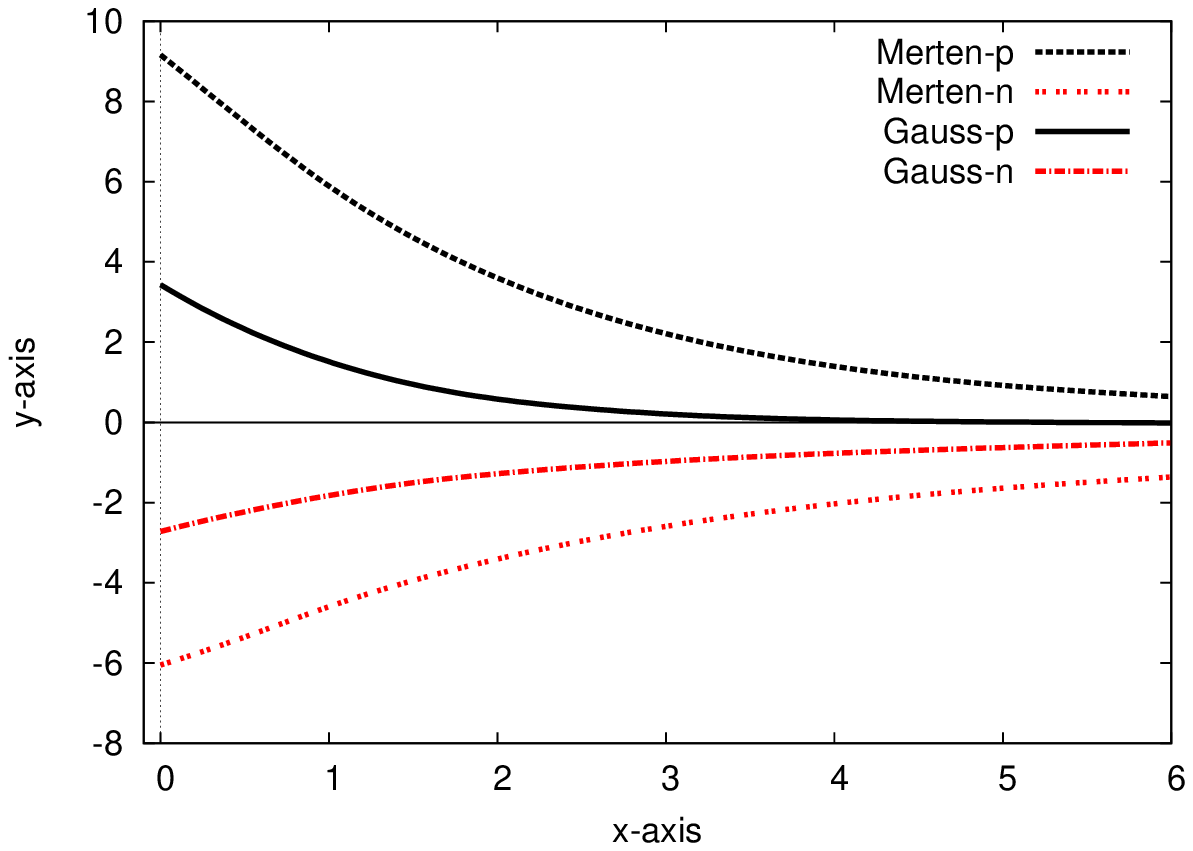}
\caption{
  Prediction of the $I_{1\,11}(2600)$ longitudinal helicity amplitude $S_{1/2}^N$
  for proton and neutron calculated in model $\mathcal C$ (solid and dashed-dotted line) and
  model $\mathcal A$ (dashed lines). See also caption to Fig.~\ref{TRANSFF_A12_S11_1535}.
  \label{TRANSFF_S_I111_2600}
}

\end{figure}

\paragraph{The $J=11/2$ resonances:}
Figs.~\ref{TRANSFF_A12_I111_2600} and~\ref{TRANSFF_S_I111_2600} shows predictions
of the transverse and longitudinal helicity amplitudes for the
$J^{\pi}=1/2^-$ $I_{1\,11}(2600)$-resonances. So far no data available.

\subsubsection{Nucleon$\to \Delta$ helicity amplitudes\label{NDeltaHelAmpl}}

We now turn to a discussion of the results for $N \to \Delta$ electro-excitation.

\paragraph{The $J=1/2$ resonances:}
We start the discussion with the negative parity $S_{31}(1620)$-resonance. 
\begin{figure}[ht!]
\centering
\psfrag{x-axis}[c][c]{$Q^2\,[\textrm{GeV}^2]$}
\psfrag{y-axis}[c][c]{$A_{\tfrac{1}{2}}^N/S_{\tfrac{1}{2}}^N(Q^2)\,[10^{-3}\textrm{GeV}^{-\tfrac{1}{2}}]$}
\psfrag{Anisovich}[r][r]{\scriptsize Anisovich~\cite{Anisovich_3}, $A_{1/2}$}
\psfrag{PDG}[r][r]{\scriptsize PDG~\cite{PDG}, $A_{1/2}$}
\psfrag{Burkert}[r][r]{\scriptsize Burkert~\cite{Burkert}, $A_{1/2}$}
\psfrag{Aznauryan}[r][r]{\scriptsize Aznauryan~\cite{Aznauryan05_2}, $A_{1/2}$}
\psfrag{Aznauryan-S}[r][r]{\scriptsize Aznauryan~\cite{Aznauryan05_2}, $S_{1/2}$}
\psfrag{Maid}[r][r]{\scriptsize MAID~\cite{Drechsel,Tiator}, $A_{1/2}$}
\psfrag{Maid-S}[r][r]{\scriptsize MAID~\cite{Drechsel,Tiator}, $S_{1/2}$}
\psfrag{Tiator-A12}[r][r]{\scriptsize Fit: Tiator~\cite{Tiator2011}, $A_{1/2}$}
\psfrag{Tiator-S12}[r][r]{\scriptsize Fit: Tiator~\cite{Tiator2011}, $S_{1/2}$}
\psfrag{Merten}[r][r]{\scriptsize model $\mathcal A$, $A_{1/2}$}
\psfrag{Merten-S}[r][r]{\scriptsize model $\mathcal A$, $S_{1/2}$}
\psfrag{Gauss}[r][r]{\scriptsize model $\mathcal C$, $A_{1/2}$}
\psfrag{Gauss-S}[r][r]{\scriptsize model $\mathcal C$, $S_{1/2}$}
\includegraphics[width=\linewidth]{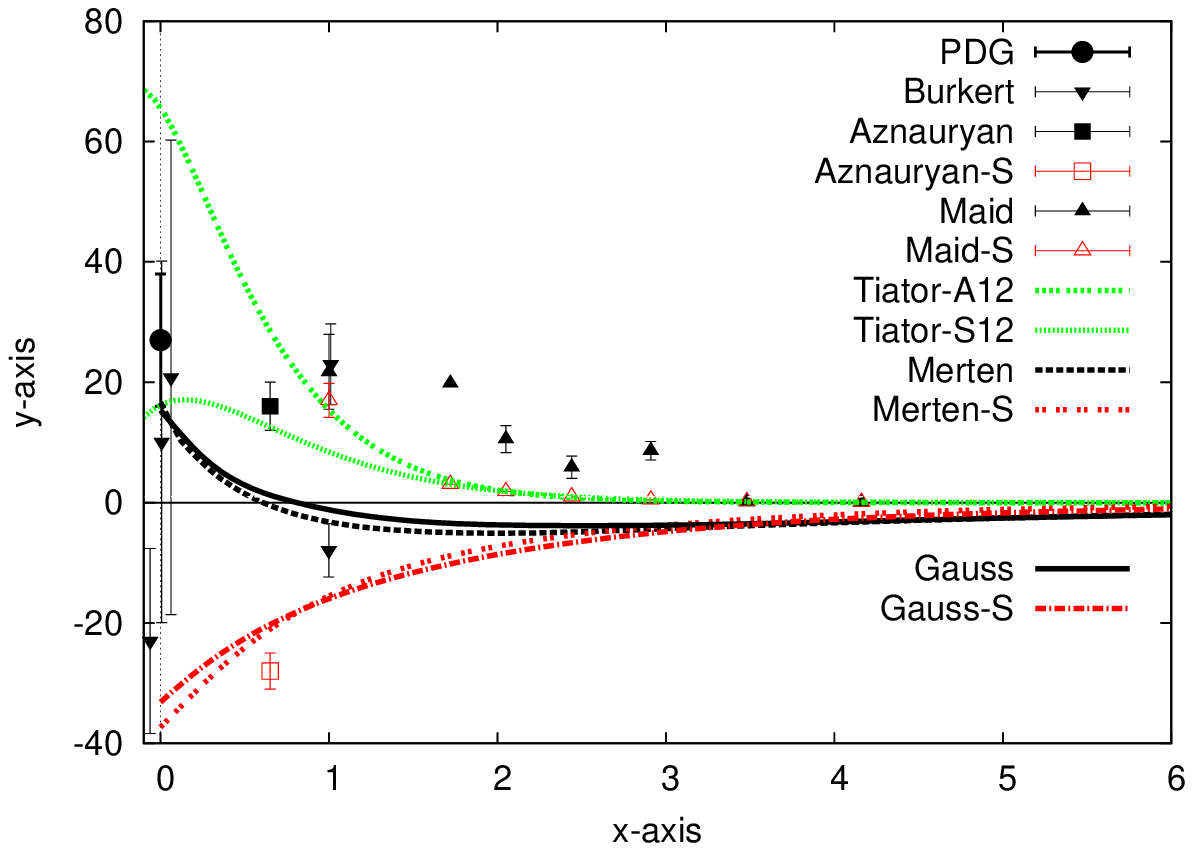}
\caption{
  Comparison of the $S_{31}(1620)$ transverse and the longitudinal helicity
  amplitudes  $A_{1/2}^N$ and $S_{1/2}^N$ calculated in model $\mathcal C$
  (solid and dashed-dotted line) and model $\mathcal A$ (dashed lines) with experimental data
  from~\cite{PDG,Burkert,Aznauryan05_2,Drechsel,Tiator}\,. Note that for
  the data points of the MAID-analysis by Tiator \textit{et al.}~\cite{Tiator}
  no errors are quoted. See also caption to Fig.~\ref{TRANSFF_A12_S11_1535}.
  \label{TRANSFF_AS12_S31_1620}
}
\end{figure}
For the $S_{31}(1620)$ transverse and longitudinal helicity amplitudes,
depicted in Fig.~\ref{TRANSFF_AS12_S31_1620}, a wide variety of experimental
data at and near the photon point exists. The calculated values lie well within
the region of experimental data obtained due to the spread in partially contradictory
experimental data but an assessment of the quality is hardly
possible. The positive longitudinal amplitude $S_{1/2}^N$ in
Fig.~\ref{TRANSFF_AS12_S31_1620} as determined in~\cite{Drechsel,Tiator}
together with the single data point from~\cite{Aznauryan05_2} suggest a sign
change in the region $Q^2 \approx 0.7-1.0\, \textnormal{GeV}^2$ not reproduced by
both calculations, this clearly needs more experimental clarification.
\begin{figure}[ht!]
\centering
\psfrag{x-axis}[c][c]{$Q^2\,[\textrm{GeV}^2]$}
\psfrag{y-axis}[c][c]{$A_{\tfrac{1}{2}}^N/S_{\tfrac{1}{2}}^N(Q^2)\,[10^{-3}\textrm{GeV}^{-\tfrac{1}{2}}]$}
\psfrag{Anisovich-M2}[r][r]{\scriptsize Anisovich~\cite{Anisovich_3}, $A_{1/2}$}
\psfrag{Awaji}[r][r]{\scriptsize Awaji~\cite{Awaji}, $A_{1/2}$}
\psfrag{Crawford}[r][r]{\scriptsize Crawford~\cite{Crawford}, $A_{1/2}$}
\psfrag{Gauss-A}[r][r]{\scriptsize model $\mathcal C$, $A_{1/2}$}
\psfrag{Gauss-S}[r][r]{\scriptsize model $\mathcal C$, $S_{1/2}$}
\includegraphics[width=\linewidth]{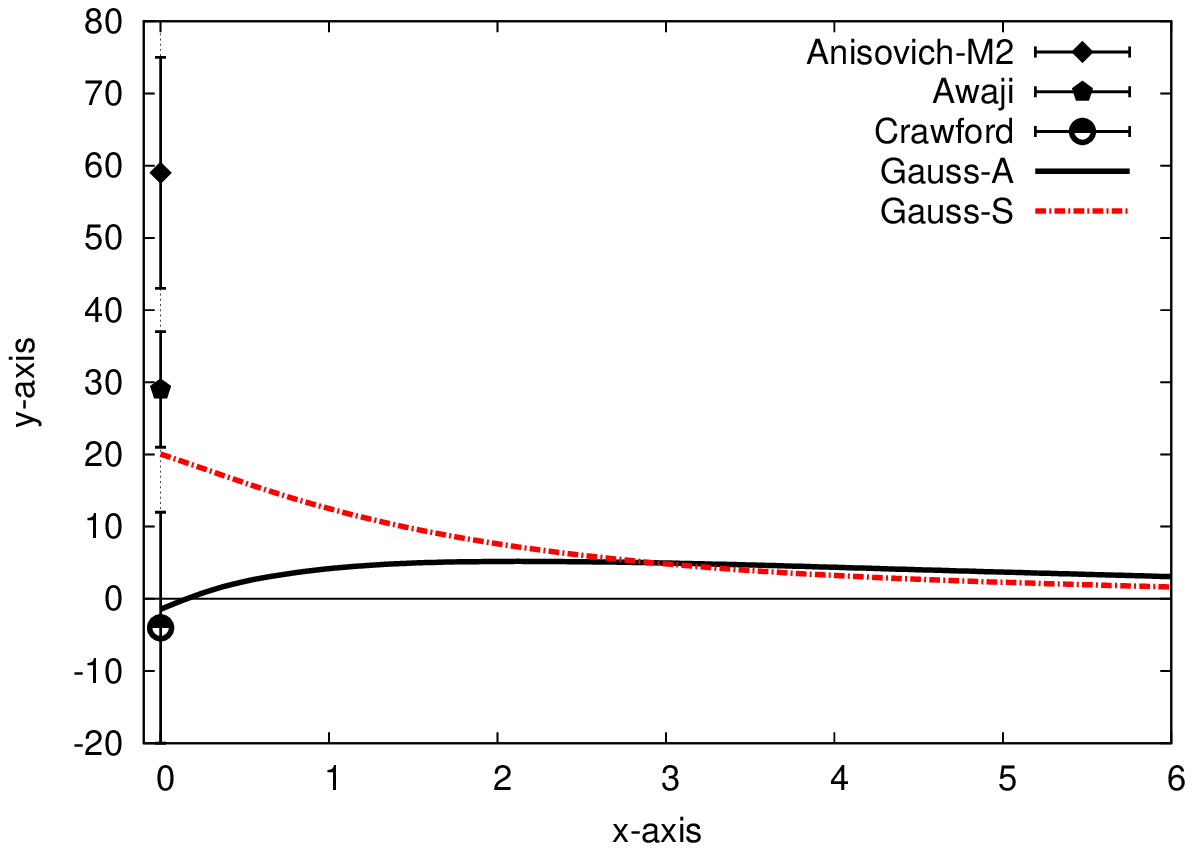}
\caption{
  Comparison of the $S_{31}(1900)$ transverse $A_{1/2}^N$ (solid and dashed-dotted line) and
  longitudinal helicity amplitude $S_{1/2}^N$ (dashed-dotted line) in model
  $\mathcal C$\,. See also caption to Fig.~\ref{TRANSFF_A12_S11_1535}.
  \label{TRANSFF_S31_1900}
}
\end{figure}

The next excitation in this channel is the $S_{31}(1900)$ resonance; the
corresponding transverse and longitudinal helicity amplitudes are displayed in
Fig.~\ref{TRANSFF_S31_1900}\,. Here we only give the results for model
$\mathcal{C}$ since the original model $\mathcal{A}$ does not describe a
resonance in this region. The values at the photon point seems to be in better
agreement with the data from Crawford \textit{et al.}~\cite{Crawford} than
with the data from Awaji \textit{et al.}~\cite{Awaji} and Anisovich
\textit{et al.}~\cite{Anisovich_3}.
\begin{figure}[ht!]
\centering
\psfrag{x-axis}[c][c]{$Q^2\,[\textrm{GeV}^2]$}
\psfrag{y-axis}[c][c]{$A_{\tfrac{1}{2}}^N(Q^2)/S_{\tfrac{1}{2}}^N(Q^2)\,[10^{-3}\textrm{GeV}^{-\tfrac{1}{2}}]$}
\psfrag{Penner}[r][r]{\scriptsize Penner~\cite{Penner}}
\psfrag{Gauss-A}[r][r]{\scriptsize model $\mathcal C$, $A_{1/2}$}
\psfrag{Gauss-S}[r][r]{\scriptsize model $\mathcal C$, $S_{1/2}$}
\includegraphics[width=\linewidth]{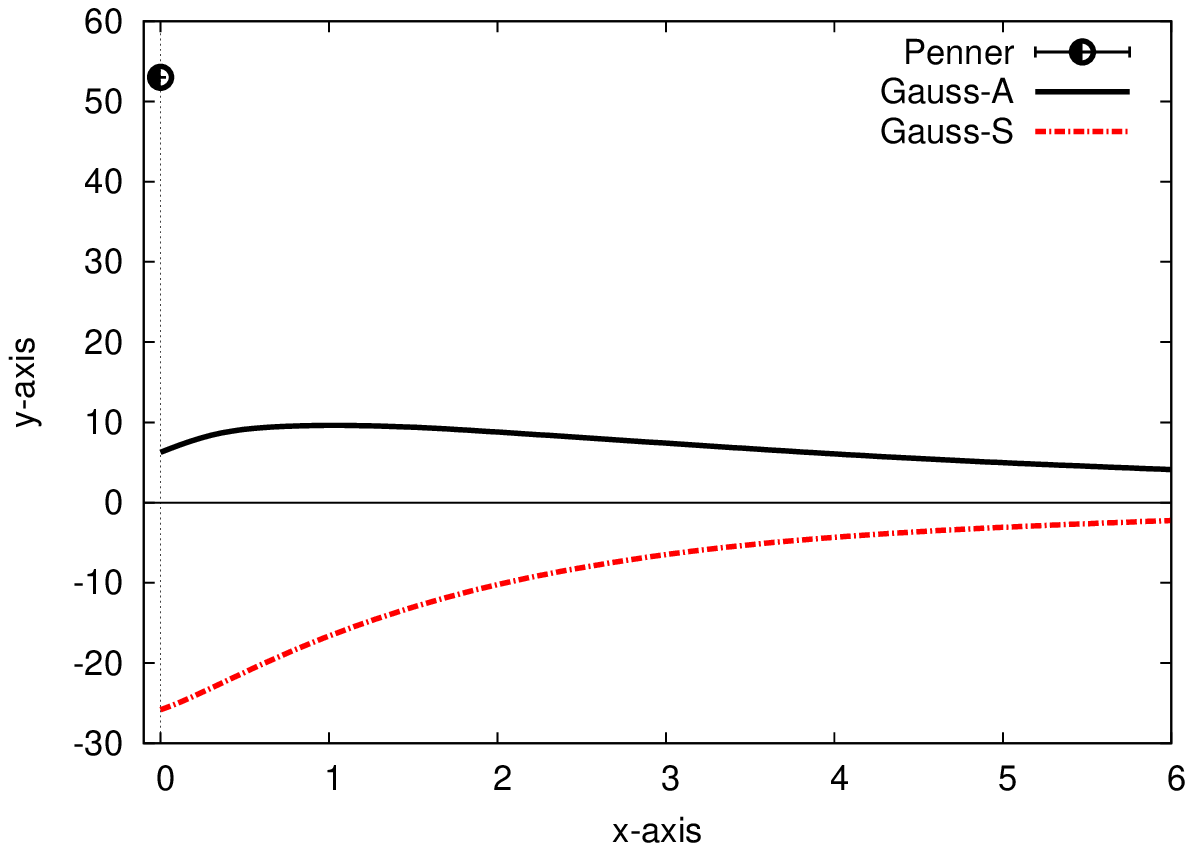}
\caption{
  Comparison of the $P_{31}(1750)$ transverse $A_{1/2}^N$ and longitudinal
  $S_{1/2}^N$ helicity amplitude calculated in model $\mathcal C$ (solid line)
  and the data from Penner \textit{et al.}~\cite{Penner}. See also caption
  to Fig.~\ref{TRANSFF_A12_S11_1535}.\label{TRANSFF_A12_P31_1750}}
\end{figure}
Note that for both $S_{31}$-resonances we judiciously fixed the
phase $\zeta$ in order to reproduce the sign of the PDG value at the photon
point, as has been mentioned above. 
Reversing the sign of $\zeta$ would in case of the $S_{31}(1620)$-resonance
in fact better reproduce the data at larger momentum transfers.

Also the lowest positive parity $\Delta$-resonance $P_{31}(1750)$ is only
reproduced in model $\mathcal C$ as shown in~\cite{Ronniger}. The calculation
does not account for the large value found by Penner \textit{et al.}~\cite{Penner}
at the photon point, see Fig.~\ref{TRANSFF_A12_P31_1750}\,. The longitudinal
amplitude is predicted to be negative for this resonance.

\begin{figure}[ht!]
\centering
\psfrag{x-axis}[c][c]{$Q^2\,[\textrm{GeV}^2]$}
\psfrag{y-axis}[c][c]{$A_{\tfrac{1}{2}}^N\,/\,S_{\tfrac{1}{2}}^N(Q^2)\,[10^{-3}\textrm{GeV}^{-\tfrac{1}{2}}]$}
\psfrag{PDG}[r][r]{\scriptsize PDG~\cite{PDG}}
\psfrag{Anisovich}[r][r]{\scriptsize Anisovich~\cite{Anisovich_3}}
\psfrag{Merten-A-A}[r][r]{\scriptsize model $\mathcal A$, $A_{1/2}$ 1st}
\psfrag{Merten-B-A}[r][r]{\scriptsize model $\mathcal A$, $A_{1/2}$ 2nd}
\psfrag{Merten-A-S}[r][r]{\scriptsize model $\mathcal A$, $S_{1/2}$ 1st}
\psfrag{Merten-B-S}[r][r]{\scriptsize model $\mathcal A$, $S_{1/2}$ 2nd}
\psfrag{Gauss-A}[r][r]{\scriptsize model $\mathcal C$, $A_{1/2}$}
\psfrag{Gauss-S}[r][r]{\scriptsize model $\mathcal C$, $S_{1/2}$}
\includegraphics[width=\linewidth]{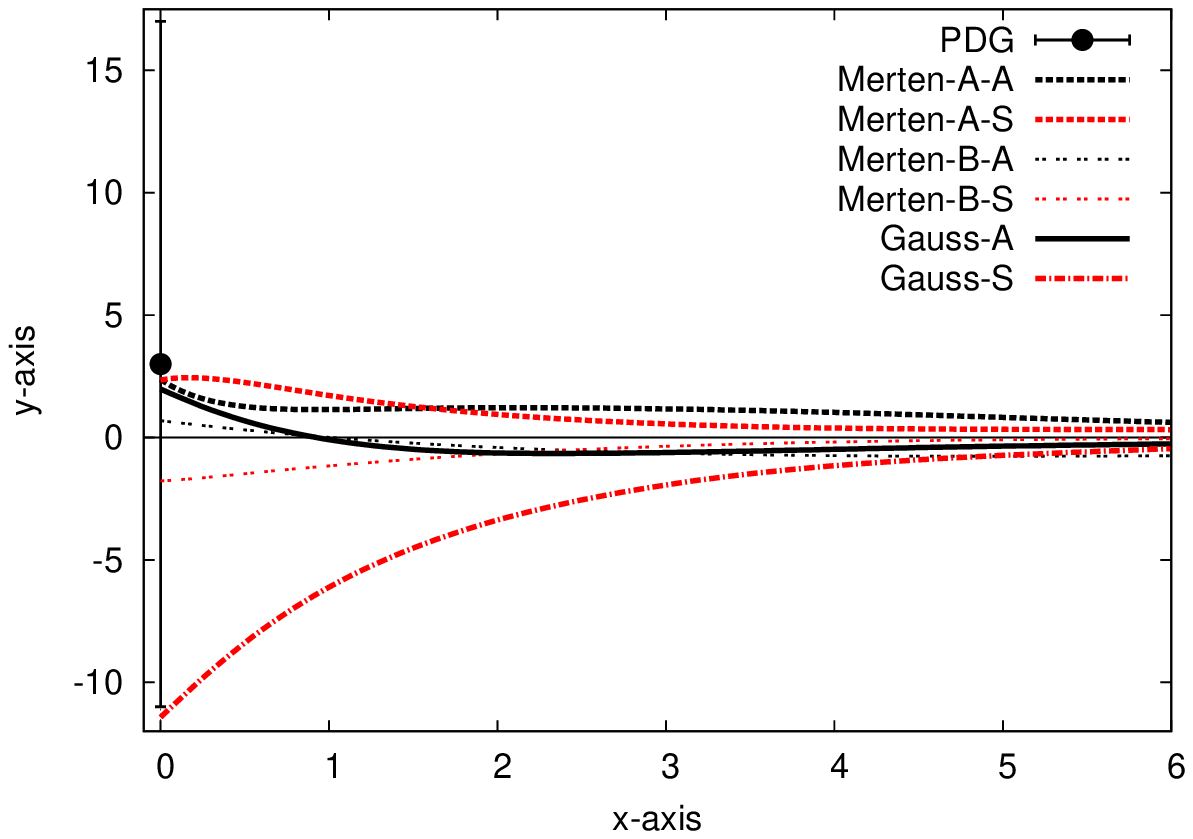}
\caption{
  Comparison of the $P_{31}(1910)$ transverse $A_{1/2}^N$ and longitudinal
  $S_{1/2}^N$ helicity amplitude calculated in model $\mathcal C$ (solid and dashed-dotted line)
  and model $\mathcal A$ (dashed lines). See also caption to Fig.~\ref{TRANSFF_A12_S11_1535}.
  \label{TRANSFF_AS12_P31_1910}
}
\end{figure}

The helicity amplitudes for the next excited state,\linebreak $P_{31}(1910)$ are shown
in Fig.~\ref{TRANSFF_AS12_P31_1910}.
Note that model $\mathcal A$ does produce two nearby resonances at the position of
the\linebreak $P_{31}(1910)$ resonance, see ~\cite{LoeMePe2,Ronniger}. The calculated
amplitudes for both resonances as well as the calculated amplitude in model
$\mathcal{C}$ are very small and in rough agreement with the experimental
value found at the photon point which has a large error. Again the
assessment cannot be conclusive. Also shown are the predictions for the
rather small longitudinal amplitudes.

\paragraph{The $J=3/2$ resonances:}

\begin{figure}[ht!]
\centering
\psfrag{x-axis}[c][c]{$Q^2\,[\textrm{GeV}^2]$}
\psfrag{y-axis}[c][c]{$A_{\tfrac{1}{2}}^N(Q^2)/A_{\tfrac{3}{2}}^N(Q^2)\,[10^{-3}\textrm{GeV}^{-\tfrac{1}{2}}]$}
\psfrag{Anisovich-A12}[r][r]{\scriptsize Anisovich~\cite{Anisovich_3}, $A_{1/2}$}
\psfrag{Anisovich-A32}[r][r]{\scriptsize Anisovich~\cite{Anisovich_3}, $A_{3/2}$}
\psfrag{PDG-A12}[r][r]{\scriptsize PDG~\cite{PDG}, $A_{1/2}$}
\psfrag{PDG-A32}[r][r]{\scriptsize PDG~\cite{PDG}, $A_{3/2}$}
\psfrag{Maid-A12}[r][r]{\scriptsize MAID~\cite{Drechsel,Tiator}, $A_{1/2}$}
\psfrag{Maid-A32}[r][r]{\scriptsize MAID~\cite{Drechsel,Tiator}, $A_{3/2}$}
\psfrag{Merten-M2}[r][r]{\scriptsize model $\mathcal A$, $A_{1/2}$}
\psfrag{Merten-M4}[r][r]{\scriptsize model $\mathcal A$, $A_{3/2}$}
\psfrag{Gauss-M2}[r][r]{\scriptsize model $\mathcal C$, $A_{1/2}$}
\psfrag{Gauss-M4}[r][r]{\scriptsize model $\mathcal C$, $A_{3/2}$}
\includegraphics[width=\linewidth]{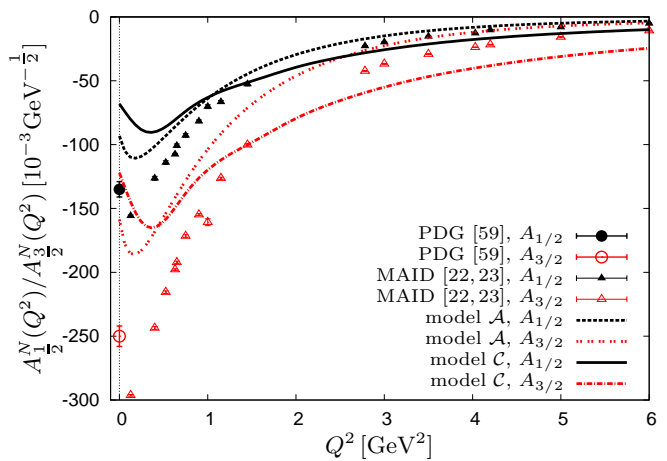}
\caption{
  Comparison of the $P_{33}(1232)$ transverse helicity amplitudes
  $A_{1/2}^N$ and $A_{3/2}^N$ as calculated in model $\mathcal C$
  (solid and dashed-dotted line) and model $\mathcal A$ (dashed lines).
  See also caption to Fig.~\ref{TRANSFF_A12_S11_1535} and the magnetic
  and electric transition form factor in Figs.~\ref{MagneticTRANSFF}
  and~\ref{ElectricTRANSFF}, respectively.\label{TRANSFF_A132_P33_1232}}
\end{figure}

\begin{figure}[ht!]
\centering
\psfrag{x-axis}[c][c]{$Q^2\,[\textrm{GeV}^2]$}
\psfrag{y-axis}[c][c]{$S_{\tfrac{1}{2}}^N(Q^2)\,[10^{-3}\textrm{GeV}^{-\tfrac{1}{2}}]$}
\psfrag{PDG}[r][r]{\scriptsize PDG~\cite{PDG}}
\psfrag{Maid}[r][r]{\scriptsize MAID~\cite{Drechsel,Tiator}}
\psfrag{Merten}[r][r]{\scriptsize model $\mathcal A$}
\psfrag{Gauss}[r][r]{\scriptsize model $\mathcal C$}
\includegraphics[width=\linewidth]{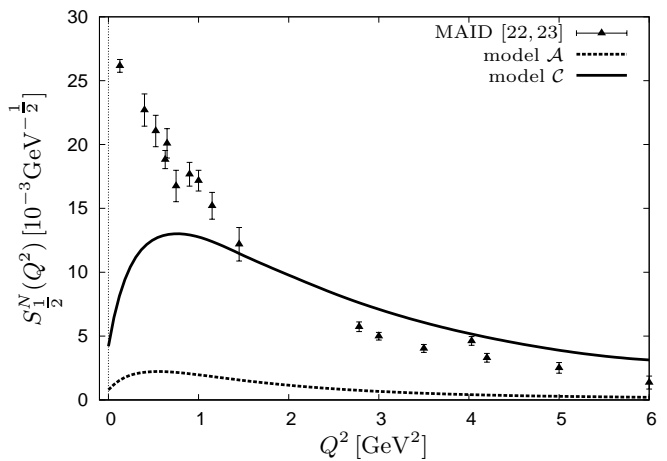}
\caption{
  Comparison of the $P_{33}(1232)$ longitudinal helicity amplitude $S_{1/2}^N$
  calculated in model $\mathcal C$ (solid line) and model $\mathcal A$ (dashed line)
  to experimental data from~\cite{PDG,Drechsel,Tiator}\,. See also caption to
  Fig.~\ref{TRANSFF_A12_S11_1535} and the Coulomb transition form factor in
  Fig.~\ref{CoulombTRANSFF}.\label{TRANSFF_S_P33_1232}
}
\end{figure}

We shall discuss the electro-excitation of the ground-state $\Delta$ resonance,
$P_{33}(1232)$ in some more detail below; the 
transverse amplitudes are shown in Fig.~\ref{TRANSFF_A132_P33_1232} while  
Fig.~\ref{TRANSFF_S_P33_1232} displays the results for the longitudinal amplitude.
With the exception of the low momentum transfer region $Q^2<0.5\,\textrm{GeV}^2$
we observe a fair agreement with experimental data for the transverse
amplitude $A_{1/2}^N$ for both models, which, however, both show a minimum
in the amplitudes for $Q^2 \lesssim 0.5\, \textnormal{GeV}^2$ (which, in
contradiction to data, also persists in the magnetic form factor, see
Fig.~\ref{MagneticTRANSFF})\,, whereas the data show a minimum of some
kinematical origin at much smaller momentum transfers $Q^2 \lesssim 0.1\,
\textnormal{GeV}^2$. 
Also the experimental data for the $S_{1/2}^N$ helicity amplitude 
can be accounted for by the calculated curve for model $\mathcal{C}$ at the
highest momentum transfers only, while the amplitude calculated in $\mathcal{A}$
is much smaller. Note that more data is available for the magnetic transition
form factor, which is a linear combination of the $A_{1/2}^N$ and $A_{3/2}^N$
amplitudes, see sec.~\ref{MagnTransFF}.

\begin{figure}[ht!]
\centering
\psfrag{x-axis}[c][c]{$Q^2\,[\textrm{GeV}^2]$}
\psfrag{y-axis}[c][c]{$A^N_{\tfrac{1}{2}}/A^N_{\tfrac{3}{2}}/S_{\tfrac{1}{2}}^N(Q^2)\,[10^{-3}\textrm{GeV}^{-\tfrac{1}{2}}]$}
\psfrag{Anisovich-M2}[r][r]{\scriptsize Anisovich~\cite{Anisovich_3}, $A_{1/2}$}
\psfrag{Anisovich-M4}[r][r]{\scriptsize Anisovich~\cite{Anisovich_3}, $A_{3/2}$}
\psfrag{PDG-M2}[r][r]{\scriptsize PDG~\cite{PDG}, $A_{1/2}$}
\psfrag{PDG-M4}[r][r]{\scriptsize PDG~\cite{PDG}, $A_{3/2}$}
\psfrag{Gauss-M2}[r][r]{\scriptsize model $\mathcal C$, $A_{1/2}$}
\psfrag{Gauss-M4}[r][r]{\scriptsize model $\mathcal C$, $A_{3/2}$}
\psfrag{Gauss-S}[r][r]{\scriptsize model $\mathcal C$, $S_{1/2}$}
\includegraphics[width=\linewidth]{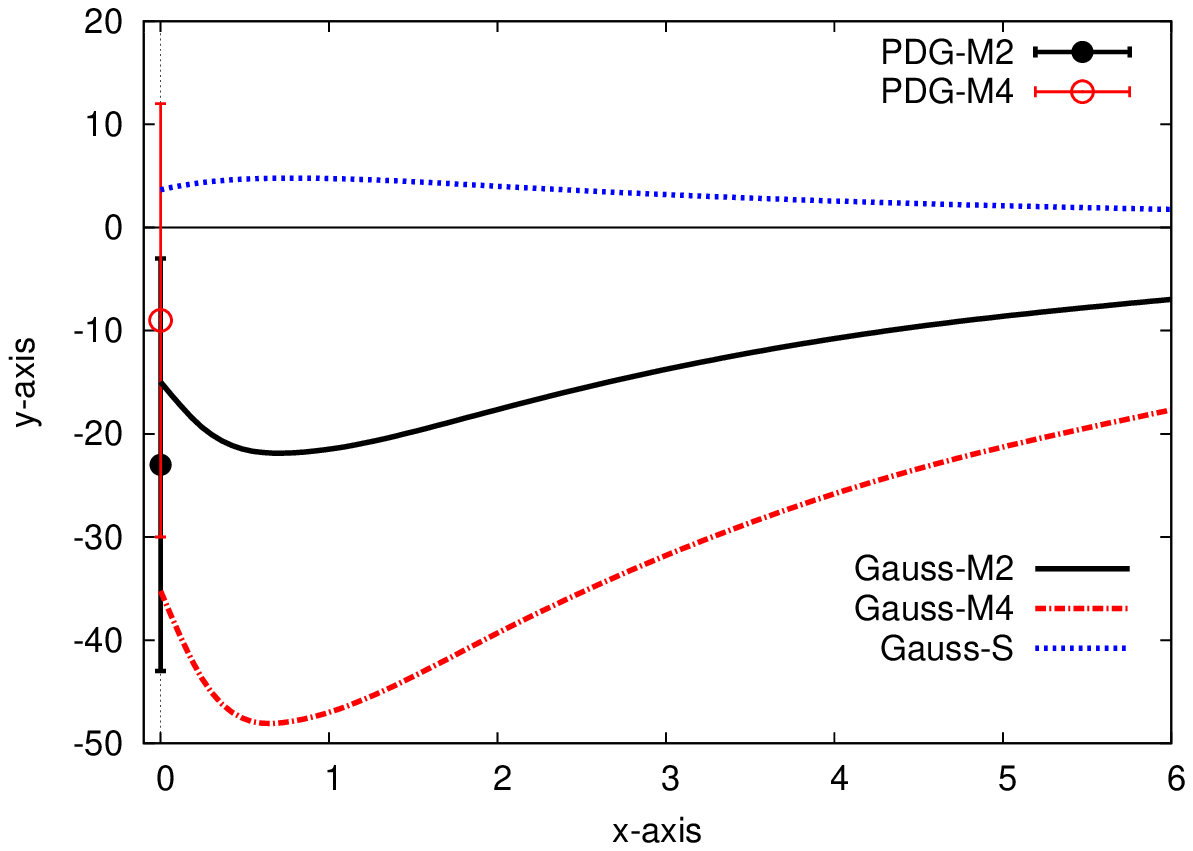}
\caption{
  Comparison of the $P_{33}(1600)$ transverse and longitudinal helicity
  amplitudes $A_{1/2}^N$, $A_{3/2}^N$ and $S_{1/2}^N$ calculated in model
  $\mathcal C$ with the PDG-data~\cite{PDG} and~\cite{Anisovich_3}. 
  See also caption to Fig.~\ref{TRANSFF_A12_S11_1535}.
  \label{TRANSFF_AS132_P33_1600}
}
\end{figure}

The Roper-like excitation of the ground state $\Delta$ resonance,
$P_{33}(1600)$ , is only described adequately in model $\mathcal{C}$\,.
The corresponding helicity amplitudes $A^N_{1/2}$\,, $A^N_{3/2}$
and $S^N_{1/2}$ are displayed in Fig.~\ref{TRANSFF_AS132_P33_1600}.
The $A^N_{\nicefrac{1}{2}}$ amplitude is calculated to be smaller than the decay
amplitude quoted by the PDG~\cite{PDG}. Contrary to this we find
a rather large $A^N_{\nicefrac{3}{2}}$ amplitude with a pronounced minimum around $Q^2
\approx 0.75\,\textrm{GeV}^2$. However in this case the value at the photon point is
overestimated.

\begin{figure}[ht!]
\centering
\psfrag{x-axis}[c][c]{$Q^2\,[\textrm{GeV}^2]$}
\psfrag{y-axis}[c][c]{$A^N_{\tfrac{1}{2}}/A^N_{\tfrac{3}{2}}\,[10^{-3}\textrm{GeV}^{-\tfrac{1}{2}}]$}
\psfrag{APHA-M2}[r][r]{\scriptsize~\cite{Awaji,Horn,Penner}, $A_{1/2}$}
\psfrag{APHA-M4}[r][r]{\scriptsize~\cite{Awaji,Horn,Penner}, $A_{3/2}$}
\psfrag{Merten-M2-2nd}[r][r]{\scriptsize model $\mathcal A$, $A_{1/2}$ 2nd}
\psfrag{Merten-M2-3rd}[r][r]{\scriptsize model $\mathcal A$, $A_{1/2}$ 3rd}
\psfrag{Merten-M4-2nd}[r][r]{\scriptsize model $\mathcal A$, $A_{3/2}$ 2nd}
\psfrag{Merten-M4-3rd}[r][r]{\scriptsize model $\mathcal A$, $A_{3/2}$ 3rd}
\psfrag{Gauss-M2-2nd}[r][r]{\scriptsize model $\mathcal C$, $A_{1/2}$ 2nd}
\psfrag{Gauss-M2-3rd}[r][r]{\scriptsize model $\mathcal C$, $A_{1/2}$ 3rd}
\psfrag{Gauss-M4-2nd}[r][r]{\scriptsize model $\mathcal C$, $A_{3/2}$ 2nd}
\psfrag{Gauss-M4-3rd}[r][r]{\scriptsize model $\mathcal C$, $A_{3/2}$ 3rd}
\includegraphics[width=\linewidth]{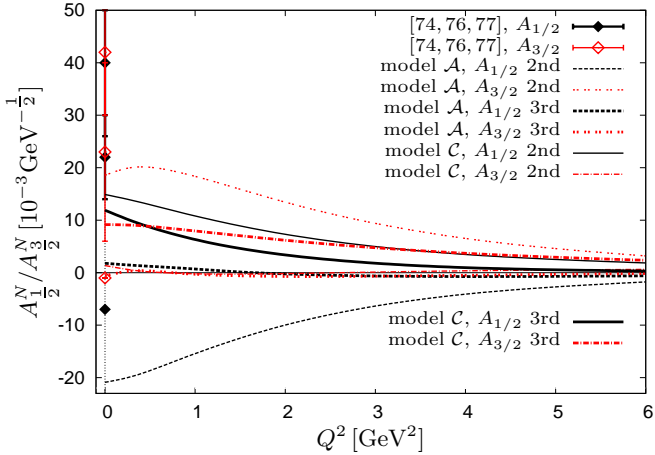}
\caption{
  Comparison of the $P_{33}(1920)$ transverse helicity amplitudes
  $A_{1/2}^N$ and $A_{3/2}^N$ as calculated in model $\mathcal C$ (solid and dashed-dotted line)
  and model $\mathcal A$ (dashed lines) with the data from~\cite{Awaji,Horn,Penner}. Note
  that the values at $Q^2=0$ of Anisovich \textit{et al.}~\cite{Anisovich_3},
  $A^p_{1/2}=130^{+30}_{-60}\times10^{-3}\textrm{GeV}^{-1/2}$ and
  $A^p_{3/2}=-150^{+25}_{-50}\times10^{-3}\textrm{GeV}^{-1/2}$, are beyond the range displayed.
  See also caption to Fig.~\ref{TRANSFF_A12_S11_1535}.
  \label{TRANSFF_AS132_P33_1920}
}
\end{figure}

\begin{figure}[ht!]
\centering
\psfrag{x-axis}[c][c]{$Q^2\,[\textrm{GeV}^2]$}
\psfrag{y-axis}[c][c]{$S_{\tfrac{1}{2}}^N(Q^2)\,[10^{-3}\textrm{GeV}^{-\tfrac{1}{2}}]$}
\psfrag{Merten-S-2nd}[r][r]{\scriptsize model $\mathcal A$, 2nd}
\psfrag{Merten-S-3rd}[r][r]{\scriptsize model $\mathcal A$, 3rd}
\psfrag{Gauss-S-2nd}[r][r]{\scriptsize model $\mathcal C$, 2nd}
\psfrag{Gauss-S-3rd}[r][r]{\scriptsize model $\mathcal C$, 3rd}
\includegraphics[width=\linewidth]{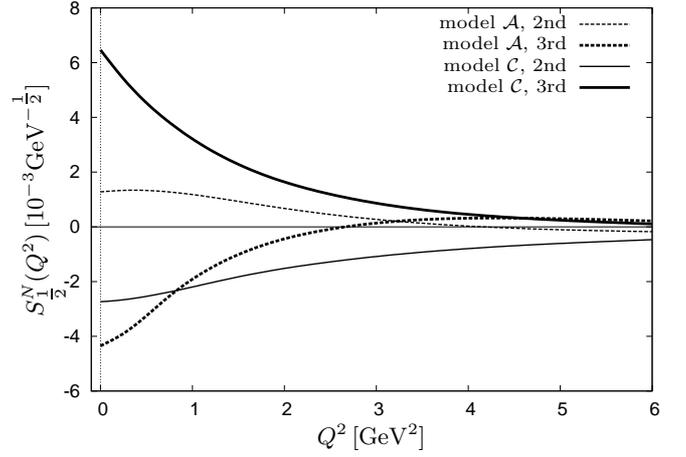}
\caption{
  Prediction of the $P_{33}(1920)$ longitudinal helicity amplitude
  $S_{1/2}^N$ calculated in model $\mathcal C$ (solid lines)
  and model $\mathcal A$ (dashed lines). See also caption to
  Fig.~\ref{TRANSFF_A12_S11_1535}.\label{TRANSFF_S_P33_1920}
}
\end{figure}

For the $P_{33}(1920)$ state with positive parity the
helicity amplitudes are displayed in Figs.~\ref{TRANSFF_AS132_P33_1920}
and~\ref{TRANSFF_S_P33_1920}. In~\cite{Ronniger} it is shown that there exist
several states around 1920 MeV which correspond to the second and third excited
$\Delta_{\nicefrac{3}{2}^+}$ state and which are predicted at 1834 MeV and
at 1912 MeV for model $\mathcal A$ and at 1899 MeV and at 1932 MeV for model
$\mathcal C$\,, respectively.
The transverse amplitudes are in general very small and match the photon decay data
of~\cite{Awaji,Horn,Penner} whereas the data of Anisovich \textit{et al.}~\cite{Anisovich_3}
cannot be reproduced. The predictions for the longitudinal amplitude as well as the
$A^N_{\nicefrac{3}{2}}$ amplitude for the third excitation are effectively zero.

\begin{figure}[ht!]
\centering
\psfrag{x-axis}[c][c]{$Q^2\,[\textrm{GeV}^2]$}
\psfrag{y-axis}[c][c]{$A_{\tfrac{1}{2}}^N/A_{\tfrac{3}{2}}^N(Q^2)\,[10^{-3}\textrm{GeV}^{-\tfrac{1}{2}}]$}
\psfrag{Anisovich-M2}[r][r]{\scriptsize Anisovich~\cite{Anisovich_3}, $A_{1/2}$}
\psfrag{Anisovich-M4}[r][r]{\scriptsize Anisovich~\cite{Anisovich_3}, $A_{3/2}$}
\psfrag{PDG-M2}[r][r]{\scriptsize PDG~\cite{PDG}, $A_{1/2}$}
\psfrag{PDG-M4}[r][r]{\scriptsize PDG~\cite{PDG}, $A_{3/2}$}
\psfrag{Burkert-M2}[r][r]{\scriptsize Burkert~\cite{Burkert}, $A_{1/2}$}
\psfrag{Burkert-M4}[r][r]{\scriptsize Burkert~\cite{Burkert}, $A_{3/2}$}
\psfrag{Aznauryan-M2}[r][r]{\scriptsize Aznauryan~\cite{Aznauryan05_2}, $A_{1/2}$}
\psfrag{Aznauryan-M4}[r][r]{\scriptsize Aznauryan~\cite{Aznauryan05_2}, $A_{3/2}$}
\psfrag{Maid-M2}[r][r]{\scriptsize MAID~\cite{Drechsel,Tiator}, $A_{1/2}$}
\psfrag{Maid-M4}[r][r]{\scriptsize MAID~\cite{Drechsel,Tiator}, $A_{3/2}$}
\psfrag{Tiator-M2}[r][r]{\scriptsize Fit: Tiator~\cite{Tiator2011}, $A_{1/2}$}
\psfrag{Tiator-M4}[r][r]{\scriptsize Fit: Tiator~\cite{Tiator2011}, $A_{3/2}$}
\psfrag{Merten-M2}[r][r]{\scriptsize model $\mathcal A$, $A_{1/2}$}
\psfrag{Merten-M4}[r][r]{\scriptsize model $\mathcal A$, $A_{3/2}$}
\psfrag{Gauss-M2}[r][r]{\scriptsize model $\mathcal C$, $A_{1/2}$}
\psfrag{Gauss-M4}[r][r]{\scriptsize model $\mathcal C$, $A_{3/2}$}
\includegraphics[width=\linewidth]{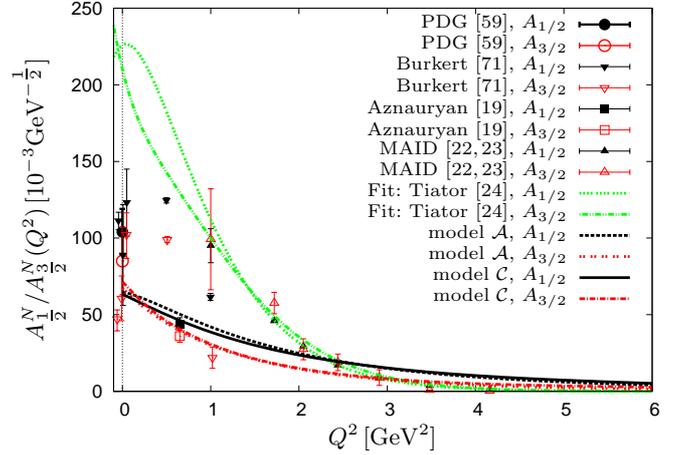}
\caption{
  Comparison of the $D_{33}(1700)$ transverse helicity amplitudes $A_{1/2}^N$
  and $A_{3/2}^N$ calculated in model $\mathcal C$ (solid and dashed-dotted line) and			
  model $\mathcal A$ (dashed lines). See also caption to Fig.~\ref{TRANSFF_A12_S11_1535}.
  \label{TRANSFF_A132_D33_1700}
}
\end{figure}

\begin{figure}[ht!]
\centering
\psfrag{x-axis}[c][c]{$Q^2\,[\textrm{GeV}^2]$}
\psfrag{y-axis}[c][c]{$S_{\frac{1}{2}}^N(Q^2)\,[10^{-3}\textrm{GeV}^{-\tfrac{1}{2}}]$}
\psfrag{Aznauryan}[r][r]{\scriptsize Aznauryan~\cite{Aznauryan05_2}}
\psfrag{Maid}[r][r]{\scriptsize MAID~\cite{Drechsel,Tiator}}
\psfrag{Tiator}[r][r]{\scriptsize Fit: Tiator~\cite{Tiator2011}}
\psfrag{Merten}[r][r]{\scriptsize model $\mathcal A$}
\psfrag{Gauss}[r][r]{\scriptsize model $\mathcal C$}
\includegraphics[width=\linewidth]{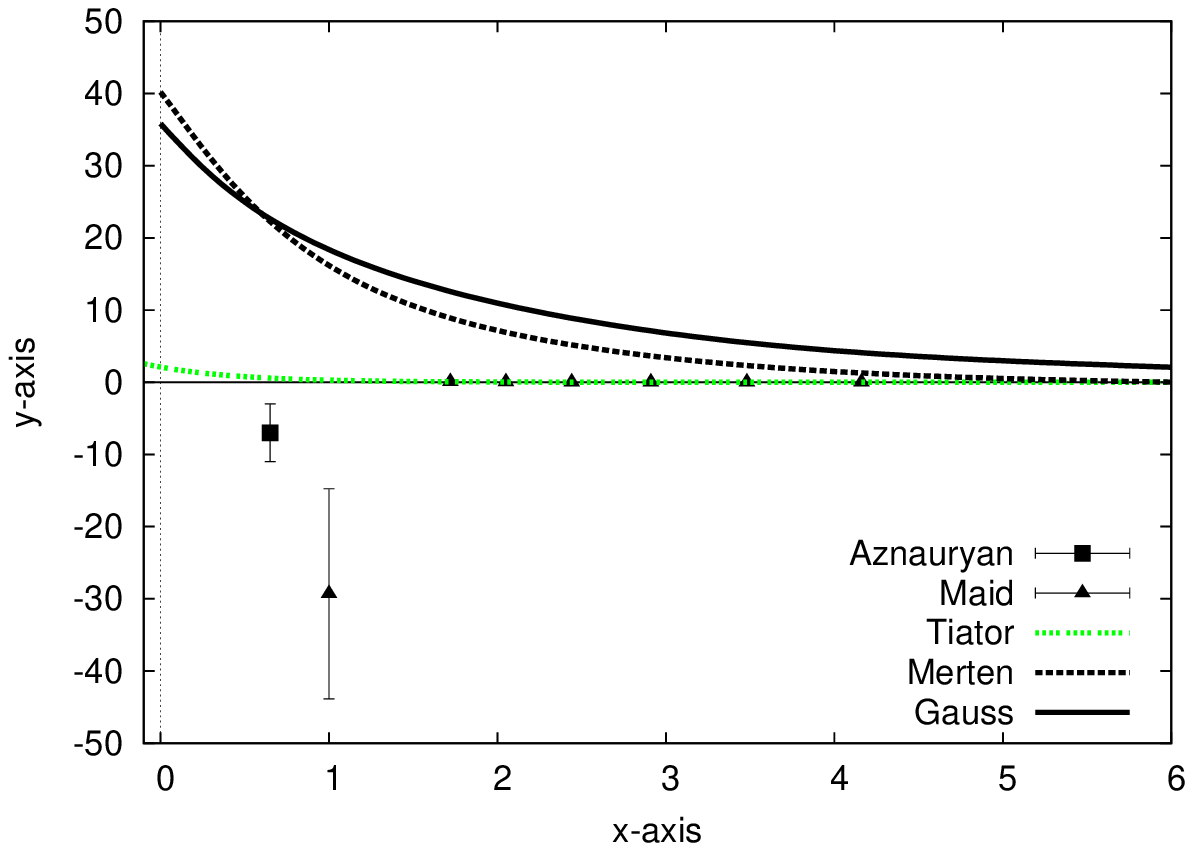}
\caption{
  Comparison of the $D_{33}(1700)$ longitudinal helicity amplitude $S_{1/2}^N$
  of the nucleon calculated in model $\mathcal C$ (solid line) and model
  $\mathcal A$ (dashed line). Note that for the data points of the MAID-analysis
  by Tiator \textit{et al.}~\cite{Tiator} no errors are quoted. See also caption
  to Fig.~\ref{TRANSFF_A12_S11_1535}.
  \label{TRANSFF_S_D33_1700}}
\end{figure}

We now turn to negative parity excited $\Delta$-resonances. For the
$D_{33}(1700)$ transition amplitudes we find that the predictions of both
models are rather close, as displayed in Figs.~\ref{TRANSFF_A132_D33_1700}
and~\ref{TRANSFF_S_D33_1700}. Note that the calculated masses of the
$D_{33}(1700)$-resonance, \textit{viz.}  $M=1594\,\textrm{MeV}$ for model
$\mathcal A$ and $M=1600\,\textrm{MeV}$
for model $\mathcal C$, are about  $100\,\textrm{MeV}$ lower than the
experimental mass at approximately $1700\,\textrm{MeV}$. This of course
affects the pre-factors in Eqs.~(\ref{TransFF_eq4a}) and~(\ref{TransFF_eq4b})
leading to the conclusion that the current matrix elements are calculated to
be too small.
For the transverse amplitude $A_{1/2}^N$ only the single data point from Aznauryan
\textit{et al.}~\cite{Aznauryan05_2} is close to the calculated curves. At the photon point
the calculated values also agree with the PDG data~\cite{PDG}. In contrast the
data from MAID~\cite{Drechsel,Tiator} and Burkert \textit{et
  al.}~\cite{Burkert} cannot be accounted for. Similar observations are made
for the $A_{3/2}^N$-amplitude. The longitudinal $S_{1/2}^N$ amplitude
has a sign opposite to the rare data from Aznauryan
\textit{et al.}~\cite{Aznauryan05_2} and MAID~\cite{Drechsel,Tiator} as shown in
Fig.~\ref{TRANSFF_S_D33_1700}. Note however that the MAID-analysis of Tiator
\textit{et al.}~\cite{Tiator} yields a vanishing $S_{1/2}^N$ amplitude in
contrast to the appreciable amplitudes found in the calculations.

\begin{figure}[ht!]
\centering
\psfrag{x-axis}[c][c]{$Q^2\,[\textrm{GeV}^2]$}
\psfrag{y-axis}[c][c]{$A^N_{\tfrac{1}{2}}/A^N_{\tfrac{3}{2}}\,[10^{-3}\textrm{GeV}^{-\tfrac{1}{2}}]$}
\psfrag{Anisovich-M2}[r][r]{\scriptsize Anisovich~\cite{Anisovich_3}, $A_{1/2}$}
\psfrag{Anisovich-M4}[r][r]{\scriptsize Anisovich~\cite{Anisovich_3}}
\psfrag{Horn-M2}[r][r]{\scriptsize Horn~\cite{Horn}, $A_{1/2}$}
\psfrag{Awaji-M2}[r][r]{\scriptsize Awaji~\cite{Awaji}, $A_{1/2}$}
\psfrag{Horn-M4}[r][r]{\scriptsize Horn~\cite{Horn}, $A_{3/2}$}
\psfrag{Awaji-M4}[r][r]{\scriptsize Awaji~\cite{Awaji}, $A_{3/2}$}
\psfrag{Gauss-M2}[r][r]{\scriptsize model $\mathcal C$, $A_{1/2}$ 2nd}
\psfrag{Gauss-M4}[r][r]{\scriptsize model $\mathcal C$, $A_{3/2}$ 2nd}
\psfrag{Gauss-M2-2nd}[r][r]{\scriptsize model $\mathcal C$, $A_{1/2}$ 3rd}
\psfrag{Gauss-M4-2nd}[r][r]{\scriptsize model $\mathcal C$, $A_{3/2}$ 3rd}
\includegraphics[width=\linewidth]{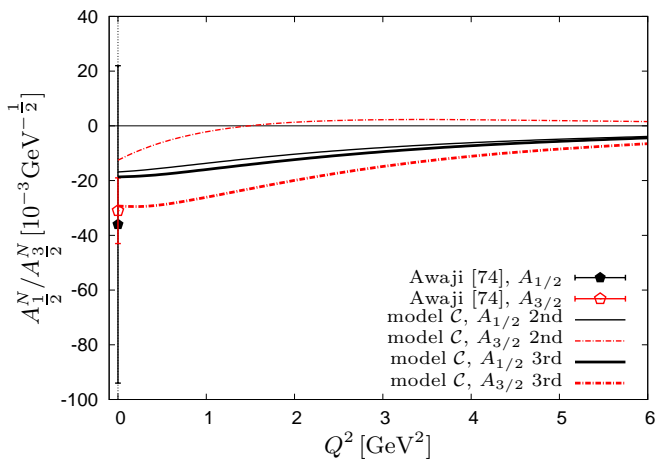}
\caption{
  Comparison of the $D_{33}(1940)$ transverse helicity amplitudes $A_{1/2}^N$
  (solid lines) and $A_{3/2}^N$ (dashed lines) calculated in model $\mathcal C$
  with the data from Awaji \textit{et al.}~\cite{Awaji}. Due to the fact that
  model $\mathcal C$ offers two alternatives for the $D_{33}(1940)$-resonance,
  as shown in~\cite{Ronniger}, both amplitudes, labelled with ''second''
  and ''third'' are displayed. Note that the values at $Q^2=0$ of Horn
  \textit{et al.}~\cite{Horn}, $A^p_{1/2}=(160\pm40)\times10^{-3}\textrm{GeV}^{-1/2}$
  and $A^p_{3/2}=(130\pm30)\times10^{-3}\textrm{GeV}^{-1/2}$,
  are beyond the range displayed. See also caption to Fig.~\ref{TRANSFF_A12_S11_1535}.
  \label{TRANSFF_A12_D33_1940}
}

\end{figure}

\begin{figure}[ht!]
\centering
\psfrag{x-axis}[c][c]{$Q^2\,[\textrm{GeV}^2]$}
\psfrag{y-axis}[c][c]{$S_{\tfrac{1}{2}}^N(Q^2)\,[10^{-3}\textrm{GeV}^{-\tfrac{1}{2}}]$}
\psfrag{Gauss}[r][r]{\scriptsize model $\mathcal C$\,, 2nd}
\psfrag{Gauss-2nd}[r][r]{\scriptsize model $\mathcal C$\,, 3rd}
\includegraphics[width=\linewidth]{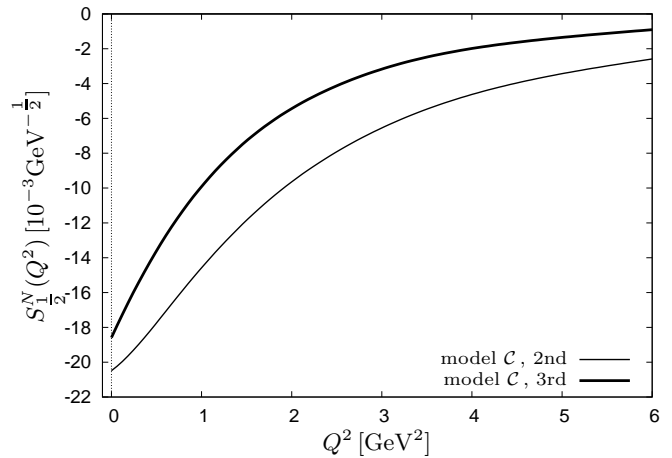}
\caption{
  Prediction of the $D_{33}(1940)$ longitudinal electro-excitation helicity amplitudes
  $S_{1/2}^N$ of the nucleon calculated in model $\mathcal C$\,. Due to
  the fact that model $\mathcal C$ offers two alternatives for the $D_{33}(1940)$-resonance,
  as shown in~\cite{Ronniger}, both amplitudes, labelled with ''first'' and ''second'' are
  displayed. See also caption to Fig.~\ref{TRANSFF_A12_S11_1535}.
  \label{TRANSFF_S_A12_D33_1940}}
\end{figure}

Figs.~\ref{TRANSFF_A12_D33_1940} and~\ref{TRANSFF_S_A12_D33_1940} contain the prediction
for the transverse and longitudinal helicity amplitudes of the $D_{33}(1940)$
resonance in model $\mathcal C$. Note that in this model two resonances with
masses $M=1895\,\textrm{MeV}$ and $M=1959\,\textrm{MeV}$ are predicted in this
energy range, as shown in~\cite{Ronniger}. Accordingly we have displayed two
alternative predictions for the helicity amplitudes. 
The results for the transverse amplitudes, see Fig.~\ref{TRANSFF_A12_D33_1940}\,,
for both resonances are rather similar; the measured photon decay amplitudes
from Horn \textit{et al.}~\cite{Horn} and Awaji \textit{et al.}~\cite{Awaji}
are in conflict, the calculated values favour a small negative value at the
photon point, which agrees with the data from Awaji \textit{et al.}~\cite{Awaji}.
In Fig.~\ref{TRANSFF_S_A12_D33_1940} we also show the corresponding longitudinal
amplitudes.

\paragraph{The $J=5/2$ resonances:}

\begin{figure}[ht!]
\centering
\psfrag{x-axis}[c][c]{$Q^2\,[\textrm{GeV}^2]$}
\psfrag{y-axis}[c][c]{$A_{\tfrac{1}{2}}^N/A_{\tfrac{3}{2}}^N/S_{\tfrac{1}{2}}^N(Q^2)\,[10^{-3}\textrm{GeV}^{-\tfrac{1}{2}}]$}
\psfrag{Anisovich-M2}[r][r]{\scriptsize Anisovich~\cite{Anisovich_3}, $A_{1/2}$}
\psfrag{Anisovich-M4}[r][r]{\scriptsize Anisovich~\cite{Anisovich_3}, $A_{3/2}$}
\psfrag{PDG-M2}[r][r]{\scriptsize PDG~\cite{PDG}, $A_{1/2}$}
\psfrag{PDG-M4}[r][r]{\scriptsize PDG~\cite{PDG}, $A_{3/2}$}
\psfrag{Gauss-M2}[r][r]{\scriptsize model $\mathcal C$, $A_{1/2}$}
\psfrag{Gauss-M4}[r][r]{\scriptsize model $\mathcal C$, $A_{3/2}$}
\psfrag{Gauss-S}[r][r]{\scriptsize model $\mathcal C$, $S_{1/2}$}
\includegraphics[width=\linewidth]{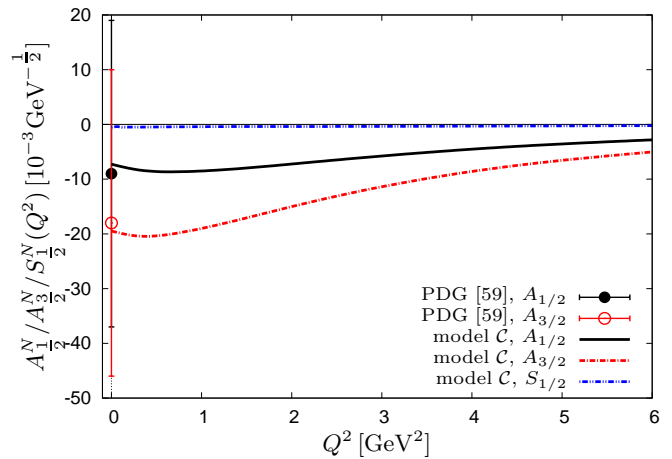}
\caption{
  Comparison of the $D_{35}(1930)$ helicity amplitudes $A_{1/2}^N$
  (black line), $A_{3/2}^N$ (red line) and $S_{1/2}^N$ (blue line) 
  calculated in model $\mathcal C$ with the PDG-data~\cite{PDG}.
  See also caption to Fig.~\ref{TRANSFF_A12_S11_1535}.
  \label{TRANSFF_A_D35_1930}
}
\end{figure}

In Fig.~\ref{TRANSFF_A_D35_1930} we show the $A_{1/2}^N$, $A_{3/2}^N$ and $S_{1/2}^N$
helicity amplitudes
calculated in model $\mathcal C$~\cite{Ronniger} for the $D_{35}(1930)$
resonance. Also displayed is the PDG-data at the photon point~\cite{PDG}, 
where we find that the transverse amplitudes agree well with the experimental
values. The longitudinal amplitude is found to be almost vanishing. Since model
$\mathcal A$ cannot account for a resonance in this energy region no results
are given in this case.

\begin{figure}[ht!]
\centering
\psfrag{x-axis}[c][c]{$Q^2\,[\textrm{GeV}^2]$}
\psfrag{y-axis}[c][c]{$A_{\tfrac{1}{2}}^N/A_{\tfrac{3}{2}}^N/S_{\tfrac{1}{2}}^N(Q^2)\,[10^{-3}\textrm{GeV}^{-\tfrac{1}{2}}]$}
\psfrag{Anisovich-M2}[r][r]{\scriptsize Anisovich~\cite{Anisovich_3}, $A_{1/2}$}
\psfrag{Anisovich-M4}[r][r]{\scriptsize Anisovich~\cite{Anisovich_3}, $A_{3/2}$}
\psfrag{PDG-M2}[r][r]{\scriptsize PDG~\cite{PDG}, $A_{1/2}$}
\psfrag{PDG-M4}[r][r]{\scriptsize PDG~\cite{PDG}, $A_{3/2}$}
\psfrag{Merten-M2}[r][r]{\scriptsize model $\mathcal A$, $A_{1/2}$}
\psfrag{Merten-M4}[r][r]{\scriptsize model $\mathcal A$, $A_{3/2}$}
\psfrag{Merten-S}[r][r]{\scriptsize model $\mathcal A$, $S_{1/2}$}
\psfrag{Gauss-M2}[r][r]{\scriptsize model $\mathcal C$, $A_{1/2}$}
\psfrag{Gauss-M4}[r][r]{\scriptsize model $\mathcal C$, $A_{3/2}$}
\psfrag{Gauss-S}[r][r]{\scriptsize model $\mathcal C$, $S_{1/2}$}
\includegraphics[width=\linewidth]{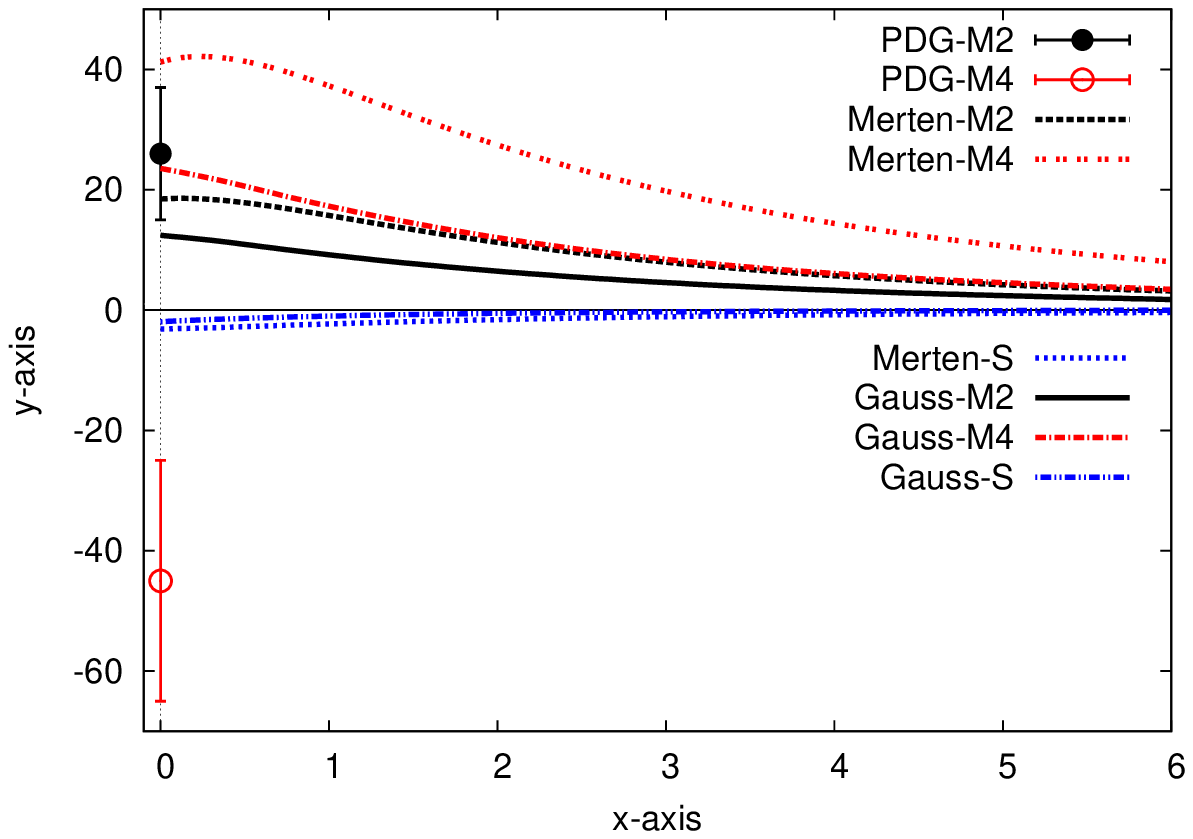}
\caption{
  Comparison of the $F_{35}(1905)$ transverse and longitudinal helicity
  amplitudes $A_{1/2}^N$, $A_{3/2}^N$  and $S_{1/2}^N$ calculated in model
  $\mathcal C$ (solid and dashed-dotted line) and model $\mathcal A$ (dashed lines). See
  also caption to Fig.~\ref{TRANSFF_A12_S11_1535}.
  \label{TRANSFF_A_F35_1905}}
\end{figure}

Both models are able to reproduce the lowest $J=5/2$ $\Delta$ resonance with
positive parity. The prediction of the helicity amplitudes of the
$F_{35}(1905)$ can be found in Fig.~\ref{TRANSFF_A_F35_1905}.
Both models can account very well for the PDG-data at the photon point for
the $A_{1/2}$ transverse amplitudes, but the
$A_{3/2}$ amplitude is found with a sign opposite to that of the data.
As for the previously discussed resonance the results for the longitudinal
amplitudes turn out to be very small.

\begin{figure}[ht!]
\centering
\psfrag{x-axis}[c][c]{$Q^2\,[\textrm{GeV}^2]$}
\psfrag{y-axis}[c][c]{$A_{\tfrac{1}{2}}^N/A_{\tfrac{3}{2}}^N/S_{\tfrac{1}{2}}^N(Q^2)\,[10^{-3}\textrm{GeV}^{-\tfrac{1}{2}}]$}
\psfrag{Anisovich-M2}[r][r]{\scriptsize Anisovich~\cite{Anisovich_3}, $A_{1/2}$}
\psfrag{Anisovich-M4}[r][r]{\scriptsize Anisovich~\cite{Anisovich_3}, $A_{3/2}$}
\psfrag{PDG-M2}[r][r]{\scriptsize PDG~\cite{PDG}, $A_{1/2}$}
\psfrag{PDG-M4}[r][r]{\scriptsize PDG~\cite{PDG}, $A_{3/2}$}
\psfrag{Merten-M2}[r][r]{\scriptsize model $\mathcal A$, $A_{1/2}$}
\psfrag{Merten-M4}[r][r]{\scriptsize model $\mathcal A$, $A_{3/2}$}
\psfrag{Merten-S}[r][r]{\scriptsize model $\mathcal A$, $S_{1/2}$}
\psfrag{Gauss-M2}[r][r]{\scriptsize model $\mathcal C$, $A_{1/2}$}
\psfrag{Gauss-M4}[r][r]{\scriptsize model $\mathcal C$, $A_{3/2}$}
\psfrag{Gauss-S}[r][r]{\scriptsize model $\mathcal C$, $S_{1/2}$}
\includegraphics[width=\linewidth]{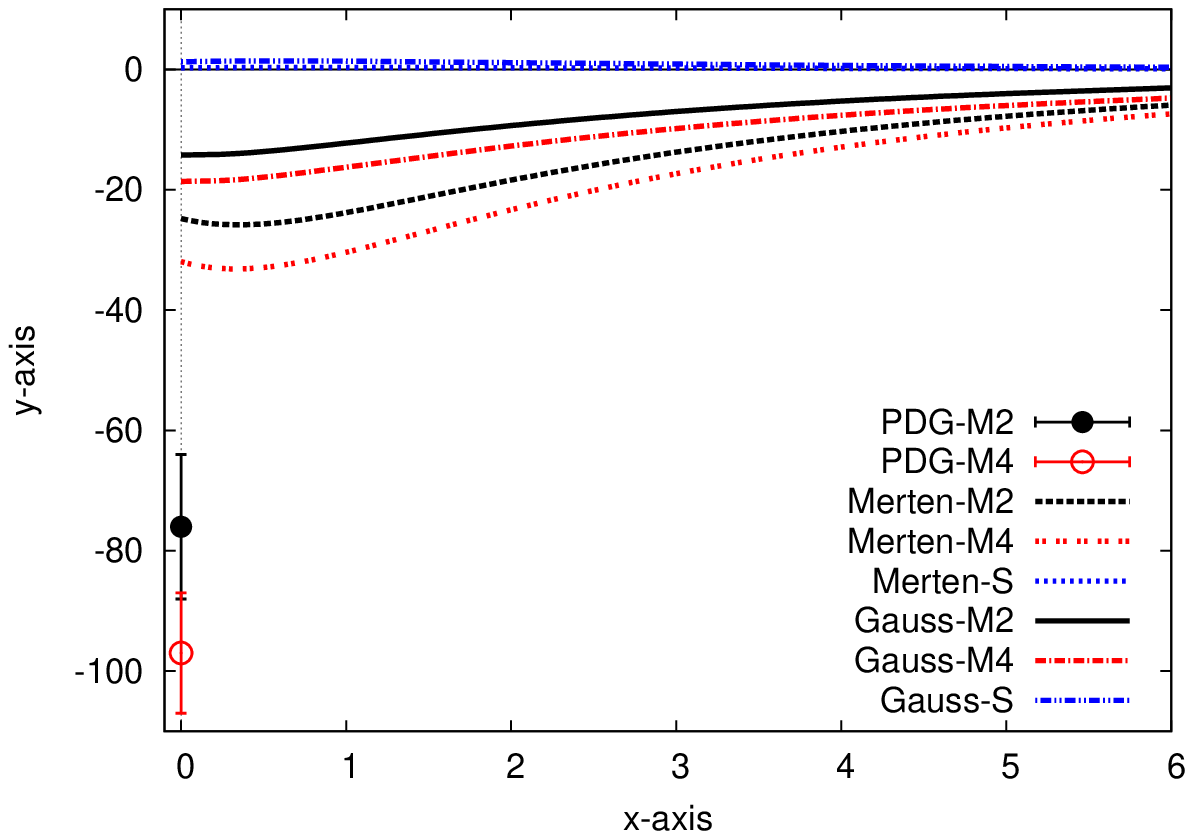}
\caption{
  Comparison of the $F_{37}(1950)$ transverse and longitudinal helicity
  amplitudes $A_{1/2}^N$, $A_{3/2}^N$  and $S_{1/2}^N$ calculated in model
  $\mathcal C$ (solid and dashed-dotted line) and model $\mathcal A$ (dashed lines). See
  also caption to Fig.~\ref{TRANSFF_A12_S11_1535}.
  \label{TRANSFF_A_F37_1950}
}
\end{figure}

\paragraph{The $J=7/2$ resonances:}
For $J=7/2$ there exists only one four star resonance, the $F_{37}(1950)$. The predictions
of the corresponding transverse and longitudinal helicity amplitudes are shown
in Fig.~\ref{TRANSFF_A_F37_1950}\,.
Here the predictions of the transverse amplitudes are much too small in order
to explain the experimental photon couplings. 

\paragraph{The $J=11/2$ resonances:}

\begin{figure}[ht!]
\centering
\psfrag{x-axis}[c][c]{$Q^2\,[\textrm{GeV}^2]$}
\psfrag{y-axis}[c][c]{$A_{\tfrac{1}{2}}^N/A_{\tfrac{3}{2}}^N/S_{\tfrac{1}{2}}^N(Q^2)\,[10^{-3}\textrm{GeV}^{-\tfrac{1}{2}}]$}
\psfrag{Merten-M2}[r][r]{\scriptsize model $\mathcal A$, $A_{1/2}$}
\psfrag{Merten-M4}[r][r]{\scriptsize model $\mathcal A$, $A_{3/2}$}
\psfrag{Merten-S}[r][r]{\scriptsize model $\mathcal A$, $S_{1/2}$}
\psfrag{Gauss-M2}[r][r]{\scriptsize model $\mathcal C$, $A_{1/2}$}
\psfrag{Gauss-M4}[r][r]{\scriptsize model $\mathcal C$, $A_{3/2}$}
\psfrag{Gauss-S}[r][r]{\scriptsize model $\mathcal C$, $S_{1/2}$}
\includegraphics[width=\linewidth]{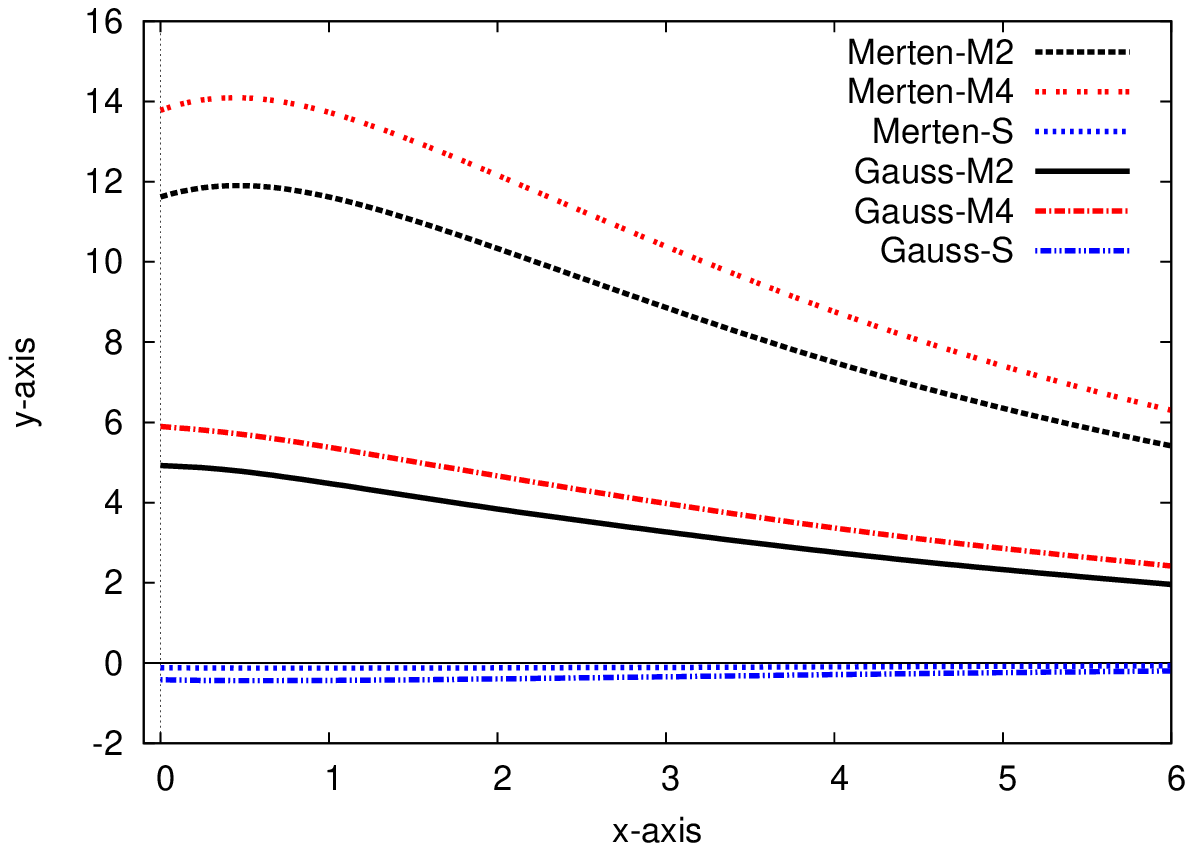}
\caption{
  Prediction of the $H_{3\,11}(2420)$ transverse and longitudinal helicity
  amplitudes $A_{1/2}^N$, $A_{3/2}^N$  and $S_{1/2}^N$ calculated in model
  $\mathcal C$ (solid and dashed-dotted line) and model $\mathcal A$ (dashed lines). See
  also caption to Fig.~\ref{TRANSFF_A12_S11_1535}.
  \label{TRANSFF_H311_2420}
}
\end{figure}
Fig.~\ref{TRANSFF_H311_2420} shows the prediction of the transverse and longitudinal
helicity amplitudes of the $\Delta_{11/2^+}(2420)$ resonance. 
The amplitudes found in model $\mathcal C$ are slightly smaller than those in model
$\mathcal A$. In both cases the longitudinal amplitude virtually vanishes.

\subsection{Photon couplings\label{PhotonCoupl}}

\begin{table*}[ht!]
\centering
\caption{
  Transverse photon couplings calculated for $N\to N^\ast$ transitions in model
  $\mathcal A$ and $\mathcal C$ in comparison to experimental data. 
  All calculated photon couplings were determined by calculating the helicity
  amplitudes at $Q^2=10^{-4}\,\textrm{GeV}^2$ close to the photon point.
  A hyphen indicates that data do not exist. All amplitudes are
  in units of $10^{-3}\textrm{GeV}^{-1/2}$\,, all masses are given in MeV.
  The references~\cite{Penner,Barbour1978} do not quote errors.
\label{tab:PhotonCoupl1}}
\begin{tabular}{lc|cc|c*{4}{r@{\hspace*{5pt}}}|*{2}{c@{\hspace*{5pt}}}r}\toprule
State          &      & \multicolumn{2}{c|}{Mass}& & \multicolumn{2}{c@{\hspace*{5pt}}}{Model $\mathcal A$}  & \multicolumn{2}{c@{\hspace*{5pt}}}{Model $\mathcal C$} &   \multicolumn{2}{|c@{\hspace*{5pt}}}{Exp.} & Ref. \\
               & Rat. & model $\mathcal A$ & model $\mathcal C$ & Ampl.             &  $p$   &  $n$   &  $p$   &  $n$   &  $p$      &  $n$       & \\\midrule
$S_{11}(1535)$ & **** & 1417         & 1475         & $A_{\nicefrac{1}{2}}$ & 111.68 & -74.75 & 85.93  & -54.96 &  90$\pm$30& -46$\pm$27 & \cite{PDG} \\
$S_{11}(1650)$ & **** & 1618         & 1681         & $A_{\nicefrac{1}{2}}$ &   2.55 & -16.03 & -4.56  &  -6.86 &  53$\pm$16& -15$\pm$21 & \cite{PDG} \\
\multirow{2}{*}{$S_{11}(1895)$} & \multirow{2}{*}{**} & 1872 & 1839 & \multirow{2}{*}{$A_{\nicefrac{1}{2}}$}& 43.36&-23.93&52.71&-29.01&\multirow{2}{*}{12$\pm$6}&\multirow{2}{*}{--}&\multirow{2}{*}{\cite{Anisovich_3}} \\
               &      & 1886         & 1882         &                   &  38.95 & -18.44 & 17.18  &  -8.27 &           &            &          \\\midrule
$P_{11}(1440)$ & **** & 1498         & 1430         & $A_{\nicefrac{1}{2}}$ &  33.51 & -18.68 & 33.10  & -17.43 & -60$\pm$4 &  40$\pm$10 & \cite{PDG} \\
$P_{11}(1710)$ & ***  & 1700         & 1712         & $A_{\nicefrac{1}{2}}$ &  58.36 & -30.59 & 30.95  & -13.57 &  24$\pm$10&  -2$\pm$14 & \cite{PDG} \\
$P_{11}(1880)$ & **   & 1905         & 1872         & $A_{\nicefrac{1}{2}}$ &  24.35 & -15.55 & 24.44  & -11.87 &  14$\pm$3 &         -- & \cite{Anisovich_3} \\\midrule
$P_{13}(1720)$ & **** & 1655         & 1690         & $A_{\nicefrac{1}{2}}$ &  81.69 & -33.06 & 50.28  & -22.56 & 18$\pm$30 &   1$\pm$15 & \cite{PDG} \\
               &      &              &              & $A_{\nicefrac{3}{2}}$ & -26.24 &  11.76 &-17.10  &   2.69 & -19$\pm$20& -29$\pm$61 & \cite{PDG} \\
\multirow{2}{*}{$P_{13}(1900)$} & \multirow{2}{*}{***} & 1859 & \multirow{2}{*}{1840} & \multirow{2}{*}{$A_{\nicefrac{1}{2}}$} &  5.06 &   3.17 & \multirow{2}{*}{2.31} &  \multirow{2}{*}{5.17} & \multirow{2}{*}{26$\pm$15/-17} & \multirow{2}{*}{--/-16} & \multirow{2}{*}{\cite{Anisovich_3}/\cite{Penner}} \\
                                &                     & 1894 &                       &                                        & 12.58 & -14.53 &                       &                        &                            &                     & \\
                                &                     &      &                       & \multirow{2}{*}{$A_{\nicefrac{3}{2}}$} &  2.29 &  18.15 & \multirow{2}{*}{4.03} & \multirow{2}{*}{13.79} & \multirow{2}{*}{-65$\pm$30/31} & \multirow{2}{*}{--/-2} & \multirow{2}{*}{\cite{Anisovich_3}/\cite{Penner}} \\
                                &                     &      &                       &                                        &  6.49 & -14.90 &                       &                        &                            &                     & \\\midrule
$D_{13}(1520)$ & **** & 1453         & 1520         & $A_{\nicefrac{1}{2}}$ & -54.80 &   2.47 &-39.39  &   0.65 & -24$\pm$9 & -59$\pm$9  & \cite{PDG} \\
               &      &              &              & $A_{\nicefrac{3}{2}}$ &  48.45 &  -52.27& 32.80  & -31.64 & 150$\pm$15&-139$\pm$11 & \cite{PDG} \\
$D_{13}(1700)$ & ***  & 1573         & 1686         & $A_{\nicefrac{1}{2}}$ & -20.69 &  16.52 &-10.16  &  10.65 & -18$\pm$13&   0$\pm$50 & \cite{PDG} \\
               &      &              &              & $A_{\nicefrac{3}{2}}$ &  -5.45 &  38.89 & -7.08  &  26.42 &  -2$\pm$24&  -3$\pm$44 & \cite{PDG} \\
\multirow{2}{*}{$D_{13}(1875)$} & \multirow{2}{*}{***} & 1896 & 1849 & \multirow{2}{*}{$A_{\nicefrac{1}{2}}$} & 49.87 & -19.04 & 42.29 & -13.71 & 18$\pm$10/-20$\pm$8& \multirow{2}{*}{7$\pm$13} & \cite{Anisovich_3}/\cite{Awaji}/\cite{PDG}/ \\
                                &                     & 1920 & 1921 &                                         &  1.62 &  -6.73 & -3.72 &  -6.76 & 12/26$\pm$52       &                     & \cite{Penner}/\cite{Devenish1974}\\
               &      &              &              & \multirow{2}{*}{$A_{\nicefrac{3}{2}}$} & -20.86  & 13.11 &  -21.46  & 10.17 & -9$\pm$5/17$\pm$11& \multirow{2}{*}{-53$\pm$34} & \cite{Anisovich_3}/\cite{Awaji}/\cite{PDG}/ \\
               &      &              &              &                                        &  -5.78  & -2.38 &    0.64  & -4.27 & -10/128$\pm$57    &                             & \cite{Penner}/\cite{Devenish1974}\\\midrule
$D_{15}(1675)$ & **** & 1623         & 1678         & $A_{\nicefrac{1}{2}}$ &   3.74 & -25.80 &  6.16  & -19.91 &  19$\pm$8 & -43$\pm$12 & \cite{PDG} \\
               &      &              &              & $A_{\nicefrac{3}{2}}$ &   5.39 & -36.41 & -1.36  & -22.98 &  15$\pm$9 & -58$\pm$13 & \cite{PDG} \\
\multirow{2}{*}{$D_{15}(2060)$} & \multirow{2}{*}{**} & 1935 & 1922 & \multirow{2}{*}{$A_{\nicefrac{1}{2}}$} &  50.63 & -28.09 & 26.71 & -16.48 &  \multirow{2}{*}{65$\pm$12} & \multirow{2}{*}{--} & \multirow{2}{*}{\cite{Anisovich_3}} \\
                                &                     & 2063 & 2017 &                                        &   0.83 & -14.53 &  2.74 & -12.84 & \\
                                &                     &      &      & \multirow{2}{*}{$A_{\nicefrac{3}{2}}$} & -17.97 &  10.01 & -8.99 &   2.06 & \multirow{2}{*}{55$^{+15}_{-35}$} & \multirow{2}{*}{--} & \multirow{2}{*}{\cite{Anisovich_3}} \\
                                &                     &      &      &                                        &   1.35 & -20.16 & -2.92 & -17.67 &                            &                     & \\\midrule
$F_{15}(1680)$ & **** & 1695         & 1734         & $A_{\nicefrac{1}{2}}$ & -45.91 &  32.65 &-29.98  &  22.25 & -15$\pm$6 &  29$\pm$10 & \cite{PDG} \\
               &      &              &              & $A_{\nicefrac{3}{2}}$ &  42.16 & -12.85 & 24.10  &  -6.95 & 133$\pm$12& -33$\pm$9  & \cite{PDG} \\
\multirow{2}{*}{$F_{15}(1860)$} & \multirow{2}{*}{**} & 1892 & 1933 & \multirow{2}{*}{$A_{\nicefrac{1}{2}}$} & -9.86 & -11.41 &  1.22 & -13.86 & \multirow{2}{*}{20$\pm$12} &  \multirow{2}{*}{--} & \multirow{2}{*}{\cite{Anisovich_3}} \\
                                &                     & 1918 & 1978 &                                        & -5.33 &  17.12 & -5.41 &   4.31 &                             &                      & \\
                                &                     &      &      & \multirow{2}{*}{$A_{\nicefrac{3}{2}}$} & -0.41 & -23.28 & -0.60 & -11.28 & \multirow{2}{*}{50$\pm$20}  & \multirow{2}{*}{--}  & \multirow{2}{*}{\cite{Anisovich_3}} \\
                                &                     &      &      &                                        & -5.34 &   6.48 & -2.21 &  -2.67 &                             &                      & \\
\multirow{2}{*}{$F_{15}(2000)$} & \multirow{2}{*}{**} & \multirow{2}{*}{2082} & 1978 & \multirow{2}{*}{$A_{\nicefrac{1}{2}}$} & \multirow{2}{*}{-0.05} & \multirow{2}{*}{0.59} & -5.41 &   4.31 & \multirow{2}{*}{35$\pm$15} & \multirow{2}{*}{--} & \multirow{2}{*}{\cite{Anisovich_3}} \\
                                &                     &                       & 2062 &                                        &                        &                       & 32.96 & -21.35 &                            &                     & \\
                                &                     &                       &      & \multirow{2}{*}{$A_{\nicefrac{3}{2}}$} & \multirow{2}{*}{-0.02} & \multirow{2}{*}{0.61} & -2.21 &  -2.70 & \multirow{2}{*}{50$\pm$14} & \multirow{2}{*}{--} & \multirow{2}{*}{\cite{Anisovich_3}} \\
                                &                     &                       &      &                                        &                        &                       &-16.72 &   6.06 &                            &                     & \\\midrule
\multirow{2}{*}{$F_{17}(1990)$} & \multirow{2}{*}{**} & \multirow{2}{*}{1954} & \multirow{2}{*}{1997} & \multirow{2}{*}{$A_{\nicefrac{1}{2}}$} & \multirow{2}{*}{-2.98} &  \multirow{2}{*}{-9.19} &\multirow{2}{*}{-3.94}  &  \multirow{2}{*}{-3.22} & 42$\pm$14/30$\pm$29 &  --/-1 & \cite{Anisovich_3}/\cite{Awaji} \\ 
                                &                     &                       &                       &                                        &                         &                         &                         &                         & 40                  &  -69 & \cite{Barbour1978} \\
                                &                     &                       &                       & \multirow{2}{*}{$A_{\nicefrac{3}{2}}$} &  \multirow{2}{*}{-3.96} & \multirow{2}{*}{-11.81} & \multirow{2}{*}{0.39}  &  \multirow{2}{*}{-5.88} & 58$\pm$12/86$\pm$60& --/-178  & \cite{Anisovich_3}/\cite{Awaji} \\
                                &                     &                       &                       &                                        &                         &                         &                         &                         & 4                   &  -72 & \cite{Barbour1978} \\\midrule
$G_{17}(2190)$ & **** & 1986         & 1980         & $A_{\nicefrac{1}{2}}$ & -27.72 &   8.47 & -12.42 &   2.69 & -65$\pm$8  & -- & \cite{Anisovich_3} \\
               &      &              &              & $A_{\nicefrac{3}{2}}$ &  19.04 & -13.45 &   8.80 &  -6.48 &  35$\pm$17 & -- & \cite{Anisovich_3} \\\midrule
$G_{19}(2250)$ & **** & 2181         & 2169         & $A_{\nicefrac{1}{2}}$ &   1.26 & -11.16 &  2.18  &  -6.65 & $|A^p_{\nicefrac{1}{2}}|<10$ & --         & \cite{Anisovich_3} \\
               &      &              &              & $A_{\nicefrac{3}{2}}$ &   1.64 & -13.70 & -0.27  &  -6.54 & $|A^p_{\nicefrac{1}{2}}|<10$ & --         & \cite{Anisovich_3} \\\midrule
$H_{19}(2220)$ & **** & 2183         & 2159         & $A_{\nicefrac{1}{2}}$ &  22.06 & -13.65 & 10.63  &  -6.93 & $|A^p_{\nicefrac{1}{2}}|<10$ & --         & \cite{Anisovich_3} \\
               &      &              &              & $A_{\nicefrac{3}{2}}$ & -17.40 &   7.04 & -7.78  &   3.05 & $|A^p_{\nicefrac{1}{2}}|<10$ & --         & \cite{Anisovich_3} \\\midrule
$I_{1,11}(2600)$& *** & 2394         & 2342         & $A_{\nicefrac{1}{2}}$ &  14.06 &  -5.45 & -5.56  &  -1.89 & --                           & --         & -- \\
               &      &              &              & $A_{\nicefrac{3}{2}}$ & -10.07 &   5.97 & -4.07  &   2.46 & --                           & --         & -- \\\bottomrule
\end{tabular}
\end{table*}

\begin{table*}[ht!]
\centering
\caption{
  Transverse photon couplings calculated for $N\to\Delta$ transitions in model
  $\mathcal A$ and $\mathcal C$ in comparison to experimental data. 
  All calculated photon couplings were determined by calculating the helicity
  amplitudes at $Q^2=10^{-4}\,\textrm{GeV}^2$ close to the photon point.
  A hyphen indicates that data do not exist. All amplitudes are
  in units of $10^{-3}\textrm{GeV}^{-1/2}$\,, all masses are given in MeV.
  Reference~\cite{Penner} does not quote errors.
\label{tab:PhotonCoupl2}}
\begin{tabular}{lc|cc|c*{2}{r@{\hspace*{5pt}}}|c@{\hspace*{5pt}}r}\toprule
State          &      & \multicolumn{2}{c|}{Mass}& & Model $\mathcal A$  & Model $\mathcal C$ &   Exp. & Ref. \\
               & Rat. & model $\mathcal A$ & model $\mathcal C$ & Ampl.           &        &        &            & \\\midrule
$S_{31}(1620)$ & **** & 1620               & 1636         & $A_{\nicefrac{1}{2}}$ & 16.63  &  15.33 &  27$\pm$11 & \cite{PDG} \\
$S_{31}(1900)$ & **   & --                 & 1956         & $A_{\nicefrac{1}{2}}$ & --     & -1.43  & 59$\pm$16/29$\pm$8/-4$\pm$16 & \cite{Anisovich_3}/\cite{Awaji}/\cite{Crawford} \\\midrule
$P_{31}(1750)$ & *    & --                 & 1765         & $A_{\nicefrac{1}{2}}$ & --     &   6.27 &   53       & \cite{Penner} \\
$P_{31}(1910)$ & **** & 1829/1869          & 1892         & $A_{\nicefrac{1}{2}}$ & 2.38/0.69& 1.98 &   3$\pm$14 & \cite{PDG} \\\midrule 
$P_{33}(1232)$ & **** & 1233               & 1231         & $A_{\nicefrac{1}{2}}$ & -93.23 & -68.08 & -135$\pm$6 & \cite{PDG} \\
               &      &                    &              & $A_{\nicefrac{3}{2}}$ &-158.61 &-122.08 & -250$\pm$8 & \cite{PDG}  \\
$P_{33}(1600)$ & ***  & --                 & 1596         & $A_{\nicefrac{1}{2}}$ & --     & -14.98 & -23$\pm$20 & \cite{PDG} \\
               &      &                    &              & $A_{\nicefrac{3}{2}}$ & --     & -35.24 &  -9$\pm$21 & \cite{PDG}  \\
\multirow{2}{*}{$P_{33}(1920)$} & \multirow{2}{*}{***} & \multirow{2}{*}{1834/1912}& \multirow{2}{*}{1899/1932}& \multirow{2}{*}{$A_{\nicefrac{1}{2}}$} & \multirow{2}{*}{20.89/1.79} & \multirow{2}{*}{14.89/11.90} & 130$^{+30}_{-60}$/40$\pm$14/ & \cite{Anisovich_3}/\cite{Awaji}/ \\
                                &                      &                           &                           &                                        &                              &                              & 22$\pm$8/-7                  & \cite{Horn}/\cite{Penner}\\
               &      &                    &              & \multirow{2}{*}{$A_{\nicefrac{3}{2}}$} & \multirow{2}{*}{-18.56/-0.58} & \multirow{2}{*}{1.36/9.16}& -115$^{+25}_{-50}$/23$\pm$17/ & \cite{Anisovich_3}/\cite{Awaji}/ \\
               &      &                    &              &                                        &                              &                           & 42$\pm$12/-1                  & \cite{Horn}/\cite{Penner} \\\midrule
$D_{33}(1700)$ & **** & 1594               & 1600         & $A_{\nicefrac{1}{2}}$ &  64.99 &  63.39 & 104$\pm$15 & \cite{PDG} \\
               &      &                    &              & $A_{\nicefrac{3}{2}}$ &  67.25 &  71.47 &  85$\pm$22 & \cite{PDG} \\
$D_{33}(1940)$ & **   & --                 & 1895/1959    & $A_{\nicefrac{1}{2}}$ & --     & -16.86/-14.98 & -36$\pm$58/160$\pm$40 & \cite{Awaji}/\cite{Horn} \\
               &      &                    &              & $A_{\nicefrac{3}{2}}$ & --     & -12.56/-27.19 & -31$\pm$12/110$\pm$30 & \cite{Awaji}/\cite{Horn} \\\midrule
$D_{35}(1930)$ & ***  & --                 & 2022         & $A_{\nicefrac{1}{2}}$ & --     &  -7.27 &  -9$\pm$28 & \cite{PDG} \\
               &      &                    &              & $A_{\nicefrac{3}{2}}$ & --     & -19.49 & -18$\pm$28 & \cite{PDG} \\\midrule
$F_{35}(1905)$ & **** & 1860               & 1896         & $A_{\nicefrac{1}{2}}$ &  18.46 &  12.42 &  26$\pm$11 & \cite{PDG} \\
               &      &                    &              & $A_{\nicefrac{3}{2}}$ &  41.22 &  23.54 & -45$\pm$20 & \cite{PDG} \\\midrule
$F_{37}(1950)$ & **** & 1918               & 1934         & $A_{\nicefrac{1}{2}}$ & -24.80 & -14.22 & -76$\pm$12 & \cite{PDG} \\
               &      &                    &              & $A_{\nicefrac{3}{2}}$ & -31.94 & -18.62 & -97$\pm$10 & \cite{PDG} \\\midrule
$H_{39}(2420)$ & **** & 2399               & 2363         & $A_{\nicefrac{1}{2}}$ &  11.62 &   4.92 & --         & -- \\
               &      &                    &              & $A_{\nicefrac{3}{2}}$ &  13.78 &   5.90 & --         & -- \\
\bottomrule
\end{tabular}
\end{table*}
In tables~\ref{tab:PhotonCoupl1} and~\ref{tab:PhotonCoupl2}  we have summarised the results for the photon
decay amplitudes as partially already discussed in 
subsections~\ref{NNHelAmpl} and \ref{NDeltaHelAmpl}\,. This tables also lists
the available experimental data.
Most of the decay amplitudes can be accounted for quite satisfactory. In
general no large differences between both models is found. 
For some amplitudes of resonances with higher angular momentum no
experimental data are available to our knowledge.

\subsection{The nucleon$\,\to\Delta(1232)$ transition form factors\label{MagnTransFF}}

The $N\to\Delta$ electric and magnetic transition form factors between
the ground-state nucleon and the $P_{33}(1232)$ state are related to
the helicity amplitudes by Eqs.~(\ref{TransFF_eq6a}) and~(\ref{TransFF_eq6b}) 
\begin{subequations}
\begin{align}
  G^{\ast}_E(Q^2) 
  := 
  & 
  F(Q^2)\left(\tfrac{1}{\sqrt{3}}A^N_{\nicefrac{3}{2}}-A^N_{\nicefrac{1}{2}}\right)\label{TransFF_eq6a},\!\!
  \\
  G^{\ast}_M(Q^2) 
  := 
  & 
  F(Q^2)\left(\sqrt{3}A^N_{\nicefrac{3}{2}}+A^N_{\nicefrac{1}{2}}\right)\label{TransFF_eq6b},
\end{align}
\end{subequations}
respectively, where $F(Q^2)$ is a kinematical pre-factor defined as
\begin{align}
  F(Q^2)   =
  -\sqrt{\frac{M_N}{4\pi\alpha}\frac{M_\Delta^2-M_N^2}{2M_\Delta^2}}\frac{M_N}{|\vec
    k|}	
  \label{TransFF_eq7}
\end{align}
in the notation of Ash \textit{et al.}~\cite{Ash1967}.
Furthermore, for the sake of completeness the Coulomb-transition form factor is
given by 
\begin{align}
G^{\ast}_C(Q^2)=-2\frac{M_\Delta}{|\vec k|}F(Q^2)\sqrt{2}\,S_{\nicefrac{1}{2}}^N\,. \label{TransFF_eq8}
\end{align}

\begin{figure}[ht!]
\centering
\psfrag{x-axis}[c][c]{$Q^2\,[\textrm{GeV}^2]$}
\psfrag{y-axis}[c][c]{$\frac{G^{\ast p}_M}{3G_D}(Q^2)$}
\psfrag{Anisovich}[r][r]{\scriptsize Anisovich~\cite{Anisovich_3}}
\psfrag{PDG}[r][r]{\scriptsize PDG~\cite{PDG}}
\psfrag{Bartel}[r][r]{\scriptsize Bartel~\cite{Bartel}}
\psfrag{Stein}[r][r]{\scriptsize Stein~\cite{Stein}}
\psfrag{Frolov}[r][r]{\scriptsize Frolov~\cite{Frolov}}
\psfrag{Foster}[r][r]{\scriptsize Foster~\cite{Foster}}
\psfrag{Maid}[r][r]{\scriptsize MAID~\cite{Drechsel,Tiator}}
\psfrag{Villano}[r][r]{\scriptsize Villano~\cite{Villano2009}}
\psfrag{Tiator}[r][r]{\scriptsize Fit: Tiator~\cite{Tiator2011}}
\psfrag{Merten}[r][r]{\scriptsize model $\mathcal A$}
\psfrag{Gauss}[r][r]{\scriptsize model $\mathcal C$}
\includegraphics[width=\linewidth]{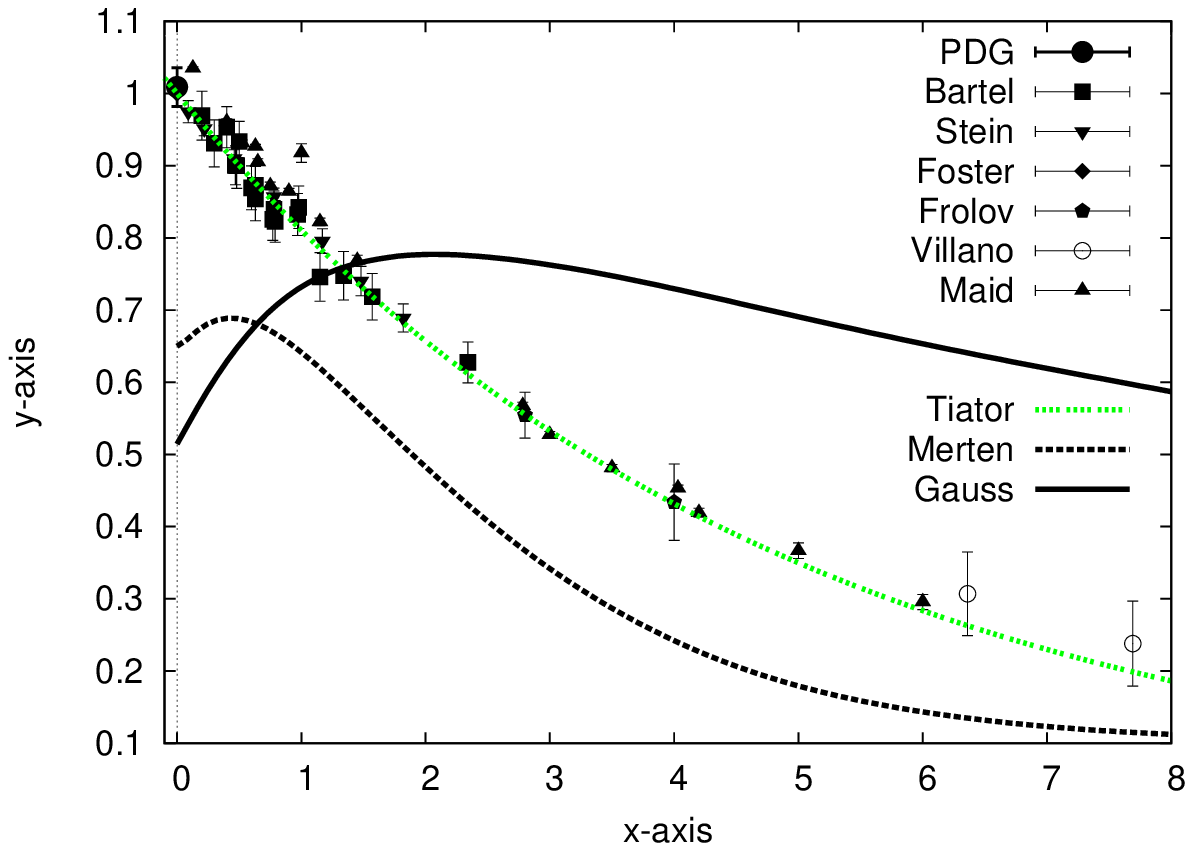}
\caption{Comparison of $\Delta(1232)$ magnetic transition form factor
  $G^{\ast p}_M$ calculated within model $\mathcal C$ (solid line)
  and model $\mathcal A$ (dashed line). See also caption
  to Fig.~\ref{TRANSFF_A12_S11_1535}.\label{MagneticTRANSFF}}
\end{figure}
\begin{figure}[ht!]
\centering
\psfrag{x-axis}[c][c]{$Q^2\,[\textrm{GeV}^2]$}
\psfrag{y-axis}[c][c]{$G^{\ast p}_E(Q^2)$}
\psfrag{Anisovich}[r][r]{\scriptsize Anisovich~\cite{Anisovich_3}}
\psfrag{PDG}[r][r]{\scriptsize PDG~\cite{PDG}}
\psfrag{Maid}[r][r]{\scriptsize MAID~\cite{Drechsel,Tiator}}
\psfrag{Tiator}[r][r]{\scriptsize Fit: Tiator~\cite{Tiator2011}}
\psfrag{Merten}[r][r]{\scriptsize model $\mathcal A$}
\psfrag{Gauss}[r][r]{\scriptsize model $\mathcal C$}
\includegraphics[width=\linewidth]{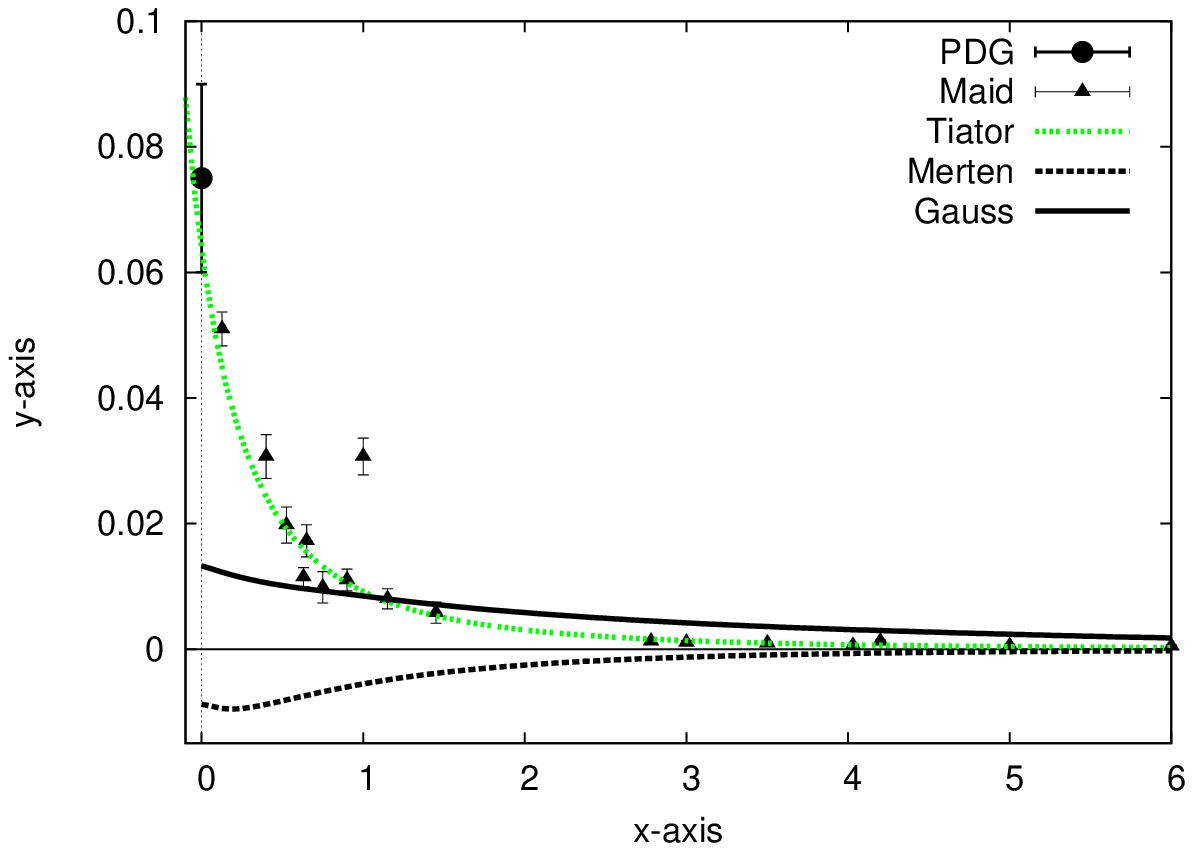}
\caption{Comparison of $\Delta(1232)$ electric transition form factor
  $G^{\ast p}_E$ calculated within model $\mathcal C$ (solid line) and
  model $\mathcal A$ (dashed line). See also caption to
  Fig.~\ref{TRANSFF_A12_S11_1535}.\label{ElectricTRANSFF}}
\end{figure}
\begin{figure}[ht!]
\centering
\psfrag{x-axis}[c][c]{$Q^2\,[\textrm{GeV}^2]$}
\psfrag{y-axis}[c][c]{$G^{\ast p}_C(Q^2)$}
\psfrag{Maid}[r][r]{\scriptsize MAID~\cite{Drechsel,Tiator}}
\psfrag{Tiator}[r][r]{\scriptsize Fit: Tiator~\cite{Tiator2011}}
\psfrag{Merten}[r][r]{\scriptsize model $\mathcal A$}
\psfrag{Gauss}[r][r]{\scriptsize model $\mathcal C$}
\includegraphics[width=\linewidth]{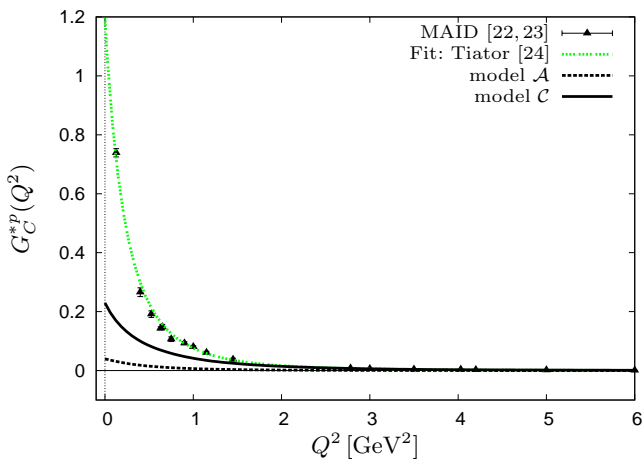}
\caption{Comparison of $\Delta(1232)$ Coulomb transition form factor 
  $G^{\ast p}_C$ calculated within model $\mathcal C$ (solid line)
  and model $\mathcal A$ (dashed line). See also caption to
  Fig.~\ref{TRANSFF_A12_S11_1535}.\label{CoulombTRANSFF}}
\end{figure}
In Fig.~\ref{MagneticTRANSFF} the calculated magnetic transition form
factor divided by thrice the standard dipole form factor is compared to
experimental data and analyses. This representation enhances the discrepancies
between the calculated and experimental results: Although model $\mathcal{A}$
still gives a fair description at larger momentum transfers, albeit in general
too small, model $\mathcal{C}$ yields too large values in this regime.
In both models the values at low momenta are too small, a
discrepancy which this calculation shares with virtually all calculations within
a constituent quark model. Usually this is regarded to be an indication of
effects due to the coupling to pions. In Fig.~\ref{ElectricTRANSFF} we also present
to corresponding electric transition form factor. Only model $\mathcal C$ agrees
with the PDG-data~\cite{PDG} of the MAID-analysis~\cite{Drechsel,Tiator},
whereas model $\mathcal A$ even has the wrong sign.
In model $\mathcal{A}$ we recalculated the form factor with a higher numerical
accuracy than was done by Merten \textit{et al.} in~\cite{Merten}.
The Coulomb transition form factor is displayed in
Fig.~\ref{CoulombTRANSFF}\,. Although the calculated result in model
$\mathcal{C}$ is significantly larger than in model $\mathcal{A}$\,, both are too
small to account for the data from the
MAID-analysis~\cite{Drechsel,Tiator,Tiator2011}\,. 

\section{Summary and conclusion\label{Summary}}

In this paper we supplement the investigation of a novel spin-flavour dependent
interaction within the framework of a relativistically covariant constituent
quark model for the structure of baryon resonances, as presented in~\cite{Ronniger}
by a calculation of helicity amplitudes for electro-excitation of the
nucleon. The calculational framework for the computation of current-matrix
elements is the same as presented by Merten \textit{et al.}~\cite{Merten}. 
In the current contribution the Salpeter-amplitudes were obtained in a calculation of the
baryon mass spectra, see~\cite{Ronniger}, where, in addition to confinement and
a spin-flavour dependent interaction motivated by instanton effects as used in
an older version called model $\mathcal A$\,, a phenomenological short ranged
spin-flavour dependent interaction was introduced in order to improve in particular upon
the description of some excited negative parity $\Delta$ resonances at
$\approx$ 1.9 GeV; this novel version of the model is called $\mathcal{C}$\,.
On the basis of these amplitudes the current-matrix elements relevant for the
helicity amplitudes were calculated in lowest order without introducing any
additional parameters. In the course of the investigations within the novel
version of the relativistic quark model $\mathcal{C}$ an improved parameter
set was found as discussed in sec.~\ref{ImprovedModel}\,. The modification 
affects mainly the neutron form factor which now rather accurately reproduces
the experimental data.

The calculated results are compared to experimental data as far as available
for resonances with a three or four star rating according to the
PDG~\cite{PDG}\,. The experimental data comprise the couplings at the photon
point from PDG~\cite{PDG} and~\cite{Anisovich_3} as well as recent determinations
of transverse and longitudinal amplitudes as reported by
Aznauryan~\cite{Aznauryan05_1,Aznauryan05_2,Aznauryan2009,Aznauryan2012} and
in the MAID-analysis~\cite{Drechsel,Tiator}, see also~\cite{Tiator2011}\,.
The results for the helicity amplitudes of nucleon resonances can be summarised
as follows:
\begin{itemize}
\item
A satisfactory description of data for the $S_{11}(1535)$, $P_{11}(1440)$,
$D_{13}(1520)$ and $F_{15}(1680)$ data was found. Exceptions are: A
node in the  transverse $P_{11}(1440)$-amplitude as found experimentally was
not reproduced by the calculations; we also do not find the observed minimum
in the longitudinal $S_{11}(1535)$-amplitude and the calculations
underestimate the transverse\linebreak $S_{11}(1650)$ as well as the longitudinal
$F_{15}(1680)$-ampli\-tude for low momentum transfers. Also the amplitudes of
the $D_{13}(1520)$ resonances are slightly too small in model $\mathcal C$.
Furthermore predictions of helicity amplitudes are given for higher excited resonances
for both models. Some of these were recently found by~\cite{Anisovich_1,Anisovich_2,Anisovich_3},
\textit{e.g.} the $N_{1/2}^+(1880)$ and $N_{1/2}^-(1895)$ resonance. 
\end{itemize}

For the nucleon-$\Delta$ transitions helicity amplitudes, as discussed in
sec.~\ref{NDeltaHelAmpl}, we note that: 
\begin{itemize}
\item
There exists agreement with the scarce data for the $S_{31}(1620)$ amplitude
if we disregard two data points for the longitudinal amplitude. There is an
indication for a sign disagreement between the data of Aznauryan \textit{et al.}~\cite{Aznauryan05_2}
and that for the MAID-analysis~\cite{Drechsel,Tiator} or alternatively a node
in the amplitude exists which is not reproduced by both models in this case.
\item
The $P_{33}(1232)$ helicity amplitudes are generally underestimated by both
models, slightly more so in model $\mathcal C$\,. For the longitudinal
amplitude in particular we find a maximum in the theoretical curves for which
there exists no experimental evidence.
\item
Predictions of the negative parity excited $\Delta^{\ast}(1900,$ $1940,$ $1930)$ helicity amplitudes
can be made in model $\mathcal C$. The position of these states cannot be
reproduced in the original model $\mathcal{A}$ and was the main motivation to
supplement the dynamics of the model by an additional short-ranged
spin-flavour dependent interaction. It is rewarding that the calculated photon
decay amplitudes agree reasonably well with the PDG data for these three
resonances.
\end{itemize}
In addition we presented predictions for helicity amplitudes of some lower
rated resonances, such as $P_{31}(1750)$ and $D_{33}(1940)$
as well as predictions to some photon decay amplitudes analysed by the CB-ELSA
collaboration \textit{et al.}~\cite{Anisovich_3}. The corresponding photon decay
amplitude data from the CB-ELSA collaboration are presently mostly included 
in the new PDG-data and thus only occasionally listed explicitly
in table~\ref{tab:PhotonCoupl1} and~\ref{tab:PhotonCoupl2}. For nucleon photon
decay amplitudes displayed in table~\ref{tab:PhotonCoupl1} we found that model
$\mathcal C$ reproduces the data of $A_{1/2}^p$ slightly better than model
$\mathcal A$\,. The $A_{3/2}^p$ decay amplitudes in both models
are too small in general. Analysing table~\ref{tab:PhotonCoupl2} for
$\Delta$-transition amplitudes we find a slightly better agreement for
model $\mathcal A$ than for model $\mathcal C$\,.

For the magnetic form factor of the $\Delta(1232)$-$N$ transition we have
found that both models cannot accurately account for the data, the
calculated values being too small in model $\mathcal A$
generally and in model $\mathcal C$ for $Q^2 < 1.5\, \textnormal{GeV}^2$.
With the MAID-data~\cite{Drechsel,Tiator} and the fits reported
in~\cite{Tiator2011} it becomes also possible to make 
a statement on the electric transition form factor which is 
better reproduced by model $\mathcal C$\,. Model $\mathcal A$ even produces
a wrong sign compared to the MAID-data. The momentum dependence of the
Coulomb transition form factor is well described by model $\mathcal C$, but
too small in magnitude by more than a factor 3. Model
$\mathcal A$ yields almost vanishing values in this case.
The corresponding elastic nucleon form factors calculations were already
presented in~\cite{Ronniger}. 

Although the overall agreement of calculated and experimental helicity data
in both versions $\mathcal A$ and $\mathcal C$ of the relativistic quark models
are of similar quality, the new model $\mathcal C$ apart from accounting better
for the baryon mass spectrum also does improve on specific observables such as
the ground state form factors.


\section*{Acknowledgements}
We gratefully acknowledge the assistance of Simon T\"olle in optimising the
program-code. We are indebted to the referee for a multitude of valuable
comments and suggestions.



\begin{thebibliography}{12}
\bibitem{Ronniger}
  M.~Ronniger, B.C.~Metsch,
  Eur. Phys. J. A \textbf{47}, 162 (2011).
\bibitem{LoeMePe1}     
  U. L\"oring, K. Kretzschmar, B.C. Metsch, H.R. Petry,
  Eur. Phys. J. A \textbf{10}, 309-346 (2001).
\bibitem{LoeMePe2}     
  U. L\"oring, B.C. Metsch, H.R. Petry, 
  Eur. Phys. J. A \textbf{10}, 395-446 (2001).
\bibitem{LoeMePe3}     
  U. L\"oring, B.C. Metsch, H.R. Petry,
  Eur. Phys. J. A \textbf{10}, 447-486 (2001).
\bibitem{Merten}      
  D. Merten, U. L\"oring, K. Kretzschmar, B.C. Metsch, H.R. Petry,
  Eur. Phys. J. A \textbf{14}, 477 (2002).
\bibitem{Glozman1996}
 L. Ya. Glozman, D. O. Riska,
 Phys. Rep. \textbf{268}, 263-303 (1996).
\bibitem{Glozman1997}
 L. Ya. Glozman, Z. Papp, W. Plessas, K. Varga and R. F. Wagenbrunn,
 Nucl. Phys. A \textbf{623}, 90-99 (1997).
\bibitem{Glozman1998_1}
 L. Ya. Glozman, Z. Papp, W. Plessas, K. Varga and R. F. Wagenbrunn,
 Phys. Rev. C \textbf{57}, 3406 (1998).
\bibitem{Glozman1998_2}
 L. Ya. Glozman, W. Plessas, K. Varga and R. F. Wagenbrunn,
 Phys. Rev. D \textbf{58}, 094030 (1998).
\bibitem{Theussl}
 L. Theu\ss{}l, R. F. Wagenbrunn, B. Desplanques and W. Plessas,
 Eur. Phys. J. A \textbf{12}, 91 (2001). 
\bibitem{Glantschnig}
 K. Glantschnig, R. Kainhofer, W. Plessas , B. Sengl and R. F. Wagenbrunn,
 Eur. Phys. J. A \textbf{23}, 507-515 (2005).
\bibitem{Melde07}
 T. Melde, K. Berger, L. Canton, W. Plessas and R. F. Wagenbrunn
 Phys. Rev. D \textbf{76}, 074020 (2007).
\bibitem{Melde08}
 T. Melde, W. Plessas, and B. Sengl,
 Phys. Rev. D \textbf{77}, 114002 (2008).
\bibitem{Plessas}
 W. Plessas and T. Melde,
 AIP Conf. Proc. \textbf{1056}, 15-22 (2008).
\bibitem{Anisovich_1}
 A.V. Anisovich, E. Klempt, V.A. Nikonov, A.V. Sarantsev and U. Thoma,
 Eur. Phys. J. A \textbf{47}, 27 (2011).
\bibitem{Anisovich_2}
 A.V. Anisovich, V.A. Nikonov, A.V. Sarantsev, U. Thoma and E. Klempt,
 Eur. Phys. J. A \textbf{47}, 153 (2011).
\bibitem{Anisovich_3}
 A.V. Anisovich, R. Beck, E. Klempt, V.A. Nikonov, A.V. Sarantsev and U. Thoma,
 Eur. Phys. J. A \textbf{48}, 15 (2012).
\bibitem{Aznauryan05_1}
 I. G. Aznauryan, V. D. Burkert, H. Egiyan, K. Joo, R. Minehart, and L. C. Smith,
 Phys. Rev. C \textbf{71}, 015201 (2005).
\bibitem{Aznauryan05_2}
 I. G. Aznauryan, V. D. Burkert, G. V. Fedotov, B. S. Ishkhanov, and V. I. Mokeev,
 Phys. Rev. C \textbf{72}, 045201 (2005).
\bibitem{Aznauryan2009}
 I. G. Aznauryan, V. D. Burkert,
 Phys. Rev. C \textbf{80}, 055203 (2009).
\bibitem{Aznauryan2012}
 I. G. Aznauryan, V. D. Burkert,
 Prog. Part. Nucl. Phys. 67, 1 (2012).
\bibitem{Drechsel}
 D. Drechsel, S. S. Kamalov, and L. Tiator,
 Eur. Phys. J. A \textbf{34}, 69 (2007).
\bibitem{Tiator}
 L. Tiator, D. Drechsel, S. S. Kamalov, M. Vanderhaeghen,
 Chinese Phys. C \textbf{33}, 1069 (2009).
\bibitem{Tiator2011}
 L. Tiator, D. Drechsel, S. S. Kamalov and M. Vanderhaeghen,
 Eur. Phys. J. Special Topics \textbf{198}, 141 (2011).
\bibitem{Kreuzer}
 S. Kreuzer, \textit{Ein masseabh\"angiges Confinement-Potential f\"ur das Bethe-Salpeter-Modell},
 Diploma thesis, University of Bonn, (2006).
\bibitem{PhDMerten}      
  D. Merten, \textit{'Hadron Form factors and Decays'}
  PhD thesis, University of Bonn (2002).
\bibitem{Mergell}
 P. Mergell, U.-G. Mei\ss ner and D. Drechsel,
 Nucl. Phys. A \textbf{596}, 367 (1996).
\bibitem{Bodek}
 A. Bodek, S. Avvakumov, R. Bradford, and H. Budd,
 J. Phys. Conf. Ser. \textbf{110}, 082004 (2008).
\bibitem{Christy}
 M.E. Christy \emph{et al.},
 Phys. Rev. C \textbf{70}, 015206 (2004).
\bibitem{Qattan}
 I. A. Qattan \emph{et al.},
 Phys. Rev. Lett. \textbf{94}, 142301 (2005).
\bibitem{Eden}
 T. Eden \emph{et al.},
 Phys. Rev. C \textbf{50}, 1749 (1994).
\bibitem{Herberg}
 C. Herberg \emph{et al.},
 Eur. Phys. J. A \textbf{5}, 131 (1999).
\bibitem{Ostrick}
 M. Ostrick \emph{et al.},
 Phys. Rev. Lett. \textbf{83}, 276 (1999).
\bibitem{Passchier}
 I. Passchier \emph{et al.},
 Phys. Rev. Lett. \textbf{82}, 4988 (1999).
\bibitem{Schiavilla}
 R. Schiavilla \emph{et al.},
 Phys. Rev. C \textbf{64}, 041002 (2001).
\bibitem{Rohe}
 D. Rohe \emph{et al.},
 Phys. Rev. Lett. \textbf{83}, 21 (1999).
\bibitem{Golak}
 J. Golak \emph{et al.},
 Phys. Rev. C \textbf{63}, 034006 (2001).
\bibitem{Zhu}
 H. Zhu \emph{et al.},
 Phys. Rev. Lett. \textbf{87}, 081801 (2001).
\bibitem{Madey}
 R. Madey \emph{et al.},
 Phys. Rev. Lett. \textbf{91}, 122002 (2003).
\bibitem{Warren}
 G. Warren \emph{et al.},
 Phys. Rev. Lett. \textbf{92}, 042301 (2004).
\bibitem{Glazier}
 D. I. Glazier \emph{et al.},
 Eur. Phys. J. A \textbf{24}, 101 (2005).
\bibitem{Alarcon}
 R. Alarcon \emph{et al.},
 Eur. Phys. J. A \textbf{31}, 588 (2007).
\bibitem{Bartel}
 T. Bartel \emph{et al.},
 Nucl. Phys. B \textbf{58}, 469 (1973).
\bibitem{Anklin}
 H. Anklin \emph{et al.},
 Phys. Lett. B \textbf{428}, 248 (1998).
\bibitem{Xu}
 W. Xu \emph{et al.},
 Phys. Rev. Lett. \textbf{85} (2000).
\bibitem{Kubon}
 G. Kubon \emph{et al.},
 Phys. Lett. B \textbf{524}, 26 (2002).
\bibitem{Amaldi}
 E. Amaldi \emph{et al.},
 Phys. Lett. B \textbf{41}, 216 (1972).
\bibitem{Brauel}
 P. Brauel \emph{et al.},
 Phys. Lett. B \textbf{45}, 389 (1973).
\bibitem{Bloom}
 E. D. Bloom \emph{et al.},
 Phys. Rev. Lett. \textbf{30}, 1186 (1973).
\bibitem{DGuerra}
 A. Del Guerra \emph{et al.},
 Nucl. Phys. B \textbf{99}, 253 (1975).
\bibitem{Joos}
 P. Joos \emph{et al.},
 Phys. Lett. B \textbf{62}, 230 (1976).
\bibitem{Baker}
 N. J. Baker \emph{et al.},
 Phys. Rev. D \textbf{23}, 2499 (1981).
\bibitem{Miller}
 K. L. Miller \emph{et al.},
 Phys. Rev. D \textbf{26}, 537 (1982).
\bibitem{Kitagaki83}
 T. Kitagaki \emph{et al.},
 Phys. Rev. D \textbf{28}, 436 (1983).
\bibitem{Kitagaki90}
 T. Kitagaki \emph{et al.},
 Phys. Rev. D \textbf{42}, 1331 (1990).
\bibitem{Allasia}
 D. Allasia \emph{et al.},
 Nucl. Phys. B \textbf{343}, 285 (1990).
\bibitem{Bernard}
 V. Bernard, L. Elouadrhiri, and U.-G. Mei\ss ner,
 J. Phys. G \textbf{28}, 1-35 (2002).
\bibitem{Warns1990}
 M. Warns \textit{et al.},
 Z. Phys. C \textbf{45}, 627 (1990).
\bibitem{PDG}
 J. Beringer \emph{et al.} (Particle Data Group),
 Phys. Rev. D \textbf{86}, 010001 (2012)
\bibitem{Kummer}
 P. Kummer \emph{et al.},
 Phys. Rev. Lett. \textbf{30}, 873 (1973).
\bibitem{Beck}
 U. Beck \emph{et al.},
 Phys. Lett. B \textbf{51}, 103 (1974).
\bibitem{Alder}
 J. Alder \emph{et al.},
 Nucl. Phys. B \textbf{91}, 386 (1975).
\bibitem{Breuker}
 H. Breuker \emph{et al.},
 Phys. Lett. B \textbf{74}, 409 (1978).
\bibitem{Brasse78}
 F. W. Brasse \emph{et al.},
 Nucl. Phys. B \textbf{139}, 37 (1978).
\bibitem{Benmerrouche}
 M. Benmerrouche \emph{et al.},
 Phys. Rev. Lett. \textbf{67}, 1070 (1991).
\bibitem{Krusche}
 B. Krusche \emph{et al.},
 Phys. Rev. Lett. \textbf{74},3736 (1995) .
\bibitem{Armstrong}
 C. S. Armstrong \emph{et al.},
 Phys. Rev. D \textbf{60}, 052004 (1999).
\bibitem{Thompson}
 R. Thompson \emph{et al.},
 Phys. Rev. Lett. \textbf{86}, 1702 (2001).
\bibitem{Keister}
 B. D. Keister and S. Capstick, N* PHYSICS, edited by T.-S. H. Lee and W. Roberts,
 341 World Scientific, Singapore (1997).
\bibitem{Capstick1995}
 S. Capstick and B. D. Keister,
 Phys. Rev. D \textbf{51}, 3598 (1995).
\bibitem{Burkert} 
 V. Burkert, \textit{Reserach Program at CEBAF II},
 edited by V. Burkert et al., CEBAF (USA), 161 (1986).
\bibitem{Gerhardt}
 C. Gerhardt,
 Z. Phys. C \textbf{4}, 311 (1980).
\bibitem{Ahrens}
 J. Ahrens \emph{et al.},
 Phys. Rev. Lett. \textbf{88}, 232002 (2002).
\bibitem{Awaji}
 N. Awaji \textit{et al.},
 DPNU-29-81, (1981) 11.
\bibitem{Crawford}
 C. B. Crawford \emph{et al.},
 Phys. Rev. Lett. \textbf{98}, 052301 (2007).
\bibitem{Penner}
 G. Penner, U. Mosel,
 Phys. Rev. C \textbf{66}, 055212 (2002).
\bibitem{Horn}
 I. Horn \textit{et al.},
 Eur. Phys. J. A \textbf{38}, 173 (2008).
\bibitem{Barbour1978}
 I. M. Barbour, R. L. Crawford and N. H. Parsons,
 Nucl. Phys. B \textbf{141}, 253 (1978).
\bibitem{Devenish1974}
 R. C. E. Devenish  D. H. Lyth and  W. A. Rankin,
 Phys. Lett.  B \textbf{52}, 227 (1974).
\bibitem{Ash1967}
 W. W. Ash, K. Berkelman, C. A. Lichtenstein, A. Ramanaukas and R. H. Siemann
 Phys. Lett. B \textbf{24}, 165 (1967).
\bibitem{Stein}
 S. Stein \emph{et al.},
 Phys. Rev. D \textbf{12}, 1884 (1975).
\bibitem{Foster}
 F. Foster and G. Hughes,
 Rep. Prog. Phys. \textbf{46}, 1445 (1983).
\bibitem{Frolov}
 V. V. Frolov \emph{et al.},
 Phys. Rev. Lett. \textbf{82}, 45 (1999).
\bibitem{Villano2009}
 A. N. Villano \textit{et al.},
 Phys. Rev. C \textbf{80}, 035203 (2009).
\end{thebibliography}
\end{document}